\pdfoutput=1
\documentclass[11pt,twoside,a4paper,cmspaper,final,collab]{cms-tdr}
\def\svnVersion{d5de271}\def\svnDate{2022/07/02}\def\cmsCernNoTag{CERN-EP-2022-004}\def\cmsCernDate{\today}\def\cmsMessage{Published in Physical Review Letters as \href{http://dx.doi.org/10.1103/PhysRevLett.129.081802}{\doi{10.1103/PhysRevLett.129.081802}.}}
\begin{document}\cmsNoteHeader{HIG-20-005}

\newcommand{\fb}{\mbox{\ensuremath{\,\text{fb}}}\xspace}
\newcommand{\HH}{\ensuremath{\PH\PH}\xspace}
\newcommand{\SM}{\ensuremath{\text{SM}}\xspace}
\newcommand{\ZZ}{\ensuremath{\PZ\PZ}\xspace}
\newcommand{\bb}{\ensuremath{\PQb\PQb}\xspace} 
\newcommand{\JJ}{\ensuremath{\rm jj}\xspace} 
\newcommand{\bbbb}{\ensuremath{\PQb\PAQb\PQb\PAQb}\xspace}
\newcommand{\bbtt}{\ensuremath{\PQb\PAQb\Pgt^{+}\Pgt^{-}}\xspace}
\newcommand{\bbVV}{\ensuremath{\PQb\PAQb\text{VV}^{*}}\xspace}
\newcommand{\bbgg}{\ensuremath{\PQb\PAQb\Pgg\Pgg}\xspace}
\newcommand{\bbWW}{\ensuremath{\PQb\PAQb\PW^{\pm}\PW^{\mp*}}\xspace}
\newcommand{\WWWW}{\ensuremath{\PW^{\pm}\PW^{\mp*}\PW^{\pm}\PW^{\mp*}}\xspace}
\newcommand{\WWgg}{\ensuremath{\PW^{\pm}\PW^{\mp*}\Pgg\Pgg}\xspace}
\newcommand{\lambdahhh}{\ensuremath{\lambda}\xspace} 
\newcommand{\klambda}{\ensuremath{\kappa_{\lambda}}\xspace} 
\newcommand{\ktop}{\ensuremath{\kappa_{\PQt}}\xspace} 
\newcommand{\CV}{\ensuremath{\kappa_\text{V}}\xspace} 
\newcommand{\CVV}{\ensuremath{\kappa_\text{2V}}\xspace} 
\newcommand{\cg}{\ensuremath{\text{c}_\text{g}}\xspace} 
\newcommand{\cgg}{\ensuremath{\text{c}_{2\text{g}}}\xspace} 
\newcommand{\ctwo}{\ensuremath{\text{c}_2}\xspace} 
\newcommand{\GGF}{\ensuremath{\Pg\Pg\text{F}}\xspace}
\newcommand{\VBF}{\ensuremath{\text{VBF}}\xspace}
\newcommand{\Hfirst}{\ensuremath{{\PH}_1}\xspace}
\newcommand{\Hsecond}{\ensuremath{{\PH}_2}\xspace}
\newcommand{\mperp}{\ensuremath{\text{m}_\perp}\xspace}
\newcommand{\mparall}{\ensuremath{\text{m}_\parallel}\xspace}
\newcommand{\Jfirst}{\ensuremath{\text{j}_1}\xspace}
\newcommand{\Jsecond}{\ensuremath{\text{j}_2}\xspace}
\newcommand{\ASR}{\ensuremath{A_\text{SR}}\xspace}
\newcommand{\ACR}{\ensuremath{A_\text{CR}}\xspace}
\newcommand{\VSR}{\ensuremath{V_\text{SR}}\xspace}
\newcommand{\VCR}{\ensuremath{V_\text{CR}}\xspace}
\newcommand{\ASRthreeb}{\ensuremath{A_\text{SR}^{3\PQb}}\xspace}
\newcommand{\ASRfourb}{\ensuremath{A_\text{SR}^{4\PQb}}\xspace}
\newcommand{\ACRthreeb}{\ensuremath{A_\text{CR}^{3\PQb}}\xspace}
\newcommand{\ACRfourb}{\ensuremath{A_\text{CR}^{4\PQb}}\xspace}
\newcommand{\VSRthreeb}{\ensuremath{V_\text{SR}^{3\PQb}}\xspace}
\newcommand{\VSRfourb}{\ensuremath{V_\text{SR}^{4\PQb}}\xspace}
\newcommand{\VCRthreeb}{\ensuremath{V_\text{CR}^{3\PQb}}\xspace}
\newcommand{\VCRfourb}{\ensuremath{V_\text{CR}^{4\PQb}}\xspace}

\newlength\cmsTwoWide\setlength{\cmsTwoWide}{0.49\textwidth}
\newlength\cmsThreeWide\setlength{\cmsThreeWide}{0.32\textwidth}
\ifthenelse{\boolean{cms@external}}{\providecommand{\cmsLeft}{upper\xspace}}{\providecommand{\cmsLeft}{left\xspace}}
\ifthenelse{\boolean{cms@external}}{\providecommand{\cmsRight}{lower\xspace}}{\providecommand{\cmsRight}{right\xspace}}

\ifthenelse{\boolean{cms@external}}{\providecommand{\suppMaterial}{Supplemental Material~\cite{supp}}}{\providecommand{\suppMaterial}{Appendix~\ref{app:suppMat}}}

\cmsNoteHeader{HIG-20-005} 
\title{Search for Higgs boson pair production in the four \texorpdfstring{\PQb}{b} quark final state in proton-proton collisions at \texorpdfstring{$\sqrt{s} = 13\TeV$}{sqrt(s) = 13 TeV}}

\author*[UF]{Daniel Guerrero}

\date{\today}

\abstract{
A search for pairs of Higgs bosons produced via gluon and vector boson fusion is presented, focusing on the four \PQb quark final state. The data sample consists of proton-proton collisions at a center-of-mass energy of 13\TeV, collected with the CMS detector at the LHC, and corresponds to an integrated luminosity of 138\fbinv. No deviation from the background-only hypothesis is observed. A 95\% confidence level upper limit on the Higgs boson pair production cross section is observed at 3.9 times the standard model prediction for an expected value of 7.8. Constraints are also set on the modifiers of the Higgs field self-coupling, $\klambda$, and of the coupling of two Higgs bosons to two vector bosons, $\CVV$. The observed (expected) allowed intervals at the 95\% confidence level are $-2.3 < \klambda < 9.4$ ($-5.0 < \klambda < 12.0$) and $-0.1 < \CVV < 2.2$ ($-0.4 < \CVV < 2.5$). These are the most stringent observed constraints to date on the $\PH\PH$ production cross section and on the \CVV coupling.
}

\hypersetup{
pdfauthor={CMS Collaboration},
pdftitle={Search for Higgs boson pair production in the four b quark final state in proton-proton collisions at sqrt(s) = 13 TeV},
pdfsubject={CMS},
pdfkeywords={CMS, HH, bbbb, self-coupling, VBF, Higgs}}

\maketitle

The discovery of the Higgs boson (\PH) by the ATLAS and CMS Collaborations~\cite{ATLASdiscovery,CMSdiscovery,Chatrchyan:2013lba} proves the existence of a fundamental scalar sector of the standard model of particle physics (SM), but the experimental confirmation of the Brout--Englert--Higgs mechanism~\cite{PhysRevLett.13.321,PhysRevLett.13.508,PhysRevLett.13.585} requires the determination of the shape of the postulated scalar potential. This shape is governed by a parameter, $\lambda$, that drives the strength of the Higgs boson self-couplings and can thus be determined experimentally with a measurement of Higgs boson pair (\HH) production.

At the CERN Large Hadron Collider (LHC), at the energy of $\sqrt{s} = 13\TeV$, the dominant \HH production mode in the SM is through the gluon fusion mechanism (\GGF), with a cross section of $31.1^{+2.1}_{-7.2}\fb$ \cite{Dawson:1998py,Borowka:2016ehy,Baglio:2018lrj,deFlorian:2013jea,Shao:2013bz,deFlorian:2015moa,Grazzini:2018bsd,Baglio:2020wgt}, followed by the vector boson fusion process (\VBF), with a cross section of $1.726\pm 0.036\fb$ \cite{Dreyer:2018qbw} and characterized by the presence of two additional hadronic jets, j, giving a $\bbbb\JJ$ final state. Variations of the Higgs boson self-coupling with respect to the SM prediction are parameterized by the modifier $\klambda = \lambdahhh / \lambdahhh^\text{SM}$ and affect the \GGF and \VBF production modes. The \VBF production mode also depends on the strength of the interaction of pairs of vector bosons $\text{V}$ ($=\PW,\PZ$) with a single ($\text{VV}\PH$) and a pair ($\text{VV}\HH$) of Higgs bosons, whose values with respect to the SM prediction are parameterized by the modifiers \CV and \CVV, respectively. Departures from the relation $\CVV = \CV^2$ predicted in the Brout--Englert--Higgs mechanism are possible in models of physics beyond the SM where the Higgs boson is a composite state emerging from the presence of new strong dynamics at the TeV scale~\cite{Bishara:2016kjn}.

The ATLAS and CMS Collaborations have searched for \GGF \HH production with a data set corresponding to an integrated luminosity of about 36\fbinv in a variety of final states~\cite{Aaboud:2018knk, Sirunyan:2018tki, Aaboud:2018sfw, Sirunyan:2017djm, Aaboud:2018zhh, Sirunyan:2017guj, Aaboud:2018ftw, Sirunyan:2018iwt, Aaboud:2018ewm, Aaboud:2018ksn}, whose combinations~\cite{Aad:2019uzh, Sirunyan:2018ayu} set an observed (expected) upper limit at the 95\% confidence level (\CL) on the SM production cross section of 7 (10) and 22 (13) times theoretical prediction, respectively.
Updated searches have been performed with an integrated luminosity of about 140\fbinv~\cite{Aad:2019yxi, Sirunyan:2020xok, ATLAS:2021ifb}, and the most stringent observed limits are from the ATLAS search in the \bbgg final state and correspond to 4.2 times the SM prediction, with a value of \klambda between $-1.5$ and 6.7 at the 95\% \CL.
The \VBF \HH production process has been studied in the \bbbb~\cite{Aad:2020kub} and \bbgg~\cite{Sirunyan:2020xok} final states by the ATLAS and CMS Collaborations, respectively, and the most stringent observed constraints at the 95\% \CL correspond to $-0.43 < \CVV < 2.56$ from \bbbb and 225 times the SM cross section prediction from \bbgg.

This Letter reports on searches for both the \GGF and \VBF \HH production mechanisms in the \bbbb decay channel. In the SM, this decay mode is characterized by a combined branching fraction of $0.339 \pm 0.008$ for $m_{\PH} = 125\GeV$~\cite{deFlorian:2016spz}. The analysis uses a sample of proton-proton collision (pp) events at $\sqrt{s} = 13\TeV$ recorded between 2016 and 2018 with the CMS detector, corresponding to an integrated luminosity of 138\fbinv.

The CMS apparatus~\cite{Chatrchyan:2008zzk} is a multipurpose, nearly hermetic detector, designed to trigger on~\cite{CMS:2020cmk,Khachatryan:2016bia} and identify electrons, muons, photons, and hadrons~\cite{CMS:2015xaf,CMS:2020uim,CMS:2018rym,CMS:2015myp,CMS:2014pgm}. A global event reconstruction ``particle-flow" (PF) algorithm~\cite{Sirunyan:2017ulk} combines the information provided by the all-silicon inner tracker and by the crystal electromagnetic and brass-scintillator hadron calorimeters, operating inside a 3.8\unit{T} superconducting solenoid, with data from gas-ionization muon detectors embedded in the solenoid iron return yoke, to build \PGt leptons, jets, missing transverse momentum, and other physics objects~\cite{CMS:2018jrd,Khachatryan:2016kdb,CMS:2019ctu}.

Hadronic jets are clustered from the PF objects using the anti-\kt algorithm~\cite{Cacciari:2008gp, Cacciari:2011ma} with a distance parameter of 0.4.
Jet energy corrections are derived from simulation studies and corrected with in situ measurements to match the energy scale in data and in simulation~\cite{Khachatryan:2016kdb}.
Jets originating from \PQb quarks are identified using as a discriminant the output of a deep neural network algorithm (\textsc{DeepJet})~\cite{Sirunyan:2017ezt,Bols:2020bkb},
trained using as input information the properties of the PF constituents of the jets and of the secondary vertices associated with them.
For the jets in this search, two working points (WPs) of the \textsc{DeepJet} discriminant are considered: the medium WP, which yields a \PQb jet identification efficiency of 75\% with a corresponding misidentification rate of light flavor and gluon (charm) jets of about 1 (10)\%, and the tight WP, which corresponds to a \PQb jet identification efficiency of 58\% and to a misidentification rate of about 0.1\,(2)\%.

{\tolerance=800
Signal processes from \GGF \HH production are simulated at next-to-leading order (NLO) accuracy in quantum chromodynamics (QCD) with \POWHEG~2.0~\cite{Alioli:2008gx, Nason:2004rx, Frixione:2007vw}, and samples for \VBF \HH production are generated at leading order (LO) accuracy in QCD using \MGvATNLO~2.6.5~\cite{Alwall:2014hca} for various combinations of couplings.
The distributions are scaled by functions of the couplings defined according to the known dependence of the theoretical cross section~\cite{Heinrich:2019bkc} and added together to model arbitrary coupling combinations, and the total predictions are normalized to the corresponding next-to-NLO (NNLO) cross section~\cite{Grazzini:2018bsd} for \GGF and by the ratio of the next-to-NNLO~\cite{Dreyer:2018qbw} to LO SM cross sections for \VBF.
Although not used to model the background, simulated samples for the QCD multijet, \ttbar, and \ZZ backgrounds are used for the optimization of the analysis.
For all simulations, the generators are interfaced with \PYTHIA~8.226 (2016) and 8.230 (2017--2018)~\cite{Sjostrand:2014zea}, and the CMS detector response is modeled with \GEANTfour~\cite{Agostinelli:2002hh}.
The simulated events are weighted to match the distribution of additional pp interactions (pileup) within the same or nearby bunch crossings, relative to the collision of interest, to the one observed in data.
See \suppMaterial{} for further details on the simulated samples.
\par}

The trigger selection for events collected in 2016 requires the presence of four jets with transverse momentum $\pt > 45\GeV$ or of two jets with $\pt > 30\GeV$ and two jets with $\pt > 90\GeV$.
For 2017 (2018) data, the presence of four jets above the \pt thresholds of 40, 45, 60, and 75\GeV is required together with $\HT > 300\,(330)\GeV$, respectively, where \HT denotes the scalar sum of the transverse momentum of the jets reconstructed in the event.
As a consequence of the change in jet trigger thresholds, data collected in 2017 and 2018 are analyzed separately from those collected in 2016.

Offline, events are required to contain at least four jets with pseudorapidity $\abs{\eta} < 2.4\,(2.5)$ and $\pt > 30\,(40)\GeV$ for the 2016 (2017--2018) data, respectively.
Jets are required to satisfy the tight WP of the PF jet identification algorithm~\cite{Sirunyan:2020foa,CMS:2017wyc}, corresponding to an efficiency larger than 99\%.
If their \pt is below 50\GeV, the medium WP of the pileup discriminant~\cite{Sirunyan:2020foa} is also required, for a signal jet efficiency of about 90\%.
If more than four jets satisfy these criteria, the four objects with the largest \textsc{DeepJet} output are selected.
The \pt of these four jets are corrected with a multivariate regression method developed for \PQb jets that improves the determination of the momentum by up to 15\% and simultaneously estimates the per jet resolution achieved~\cite{Sirunyan:2019wwa}.
After the application of this method, the resolution on the dijet invariant mass for $\PH\to\PQb\PAQb$ events reconstructed in this analysis ranges between 11 and 14\%.
At least three of the selected jets are required to satisfy the medium WP of the \textsc{DeepJet} discriminant.

Events are rejected if they contain an electron or a muon with $\pt > 15$ and 10\GeV, respectively, and $\abs{\eta} < 2.4$, where these objects must satisfy identification discriminants and criteria that include isolation and impact parameter with respect to the primary interaction vertex. 
This selection suppresses background events containing leptonic top quark decays.

The two Higgs boson candidates are formed by pairing the four jets.
There are three possible pairings of jets, and in each the two Higgs boson candidates, denoted as \Hfirst and \Hsecond, are defined by the relation $\pt(\Hfirst) > \pt(\Hsecond)$.
The $(\Hfirst,\Hsecond)$ pairings are ordered according to the increasing value of a distance parameter $d = \abs{m_{\Hfirst} - k\mkern1mu m_{\Hsecond}}/\sqrt{\smash[b]{1+k^2}}$.
The constant $k$ is the ratio of the expected peak positions of the reconstructed Higgs boson masses for events that are correctly paired, $k = c_1/c_2 = (125\GeV)/(120\GeV) = 1.04$.
Its value differs from 1 because of the residual jet momentum dependence of the multivariate energy regression that more strongly impacts the softer \PH candidate.
If the difference in the distance parameter of the first and second pairing, $\Delta d$, is larger than 30\GeV, corresponding to about two times the resolution on the Higgs boson mass, the pairing with the smallest $d$ is chosen. Conversely, if $\Delta d \leq 30\GeV$, the experimental resolution limits the capability to identify the correct pairing based on the invariant masses, and a choice is made between the first and second pairing as the one that maximizes the \pt of the two Higgs boson candidates in the four-jet center-of-mass reference frame.
This procedure results in a correct jet pairing of about 96\% of the selected events in a \GGF SM \HH sample,
and amounts to 82--96 (91--98)\% for the different couplings studied in \GGF (\VBF) signal events.

The two non-\PQb jets in the \VBF production events are selected with $\pt > 25\GeV$ and $\abs{\eta} < 4.7$, and they must satisfy the tight WP of the jet identification algorithm and the medium WP of the pileup discriminant if $\pt < 50\GeV$. For the 2017 data, affected by large noise in the endcaps of the electromagnetic calorimeter (ECAL), jets in the region $2.6 < \abs{\eta} < 3.1$ are additionally required to satisfy the tight WP of the pileup discriminant to mitigate the noise effects. The two \VBF jet candidates \Jfirst and \Jsecond are chosen as the highest \pt jet and the second-highest \pt jet that has an opposite $\eta$ sign with respect to the former.

{\tolerance=800
Events that do not contain such a VBF jet pair are assigned to the \GGF category.
About 26--28\% of \GGF events contain additional jets that satisfy the above requirements on the VBF jet candidates, and in order to correctly classify them,
a boosted decision tree (BDT) discriminant is trained to separate \GGF and \VBF \HH signal events.
The discriminant uses $\pt(\Hfirst)$, $\pt(\Hsecond)$, $\pt(\Jfirst)$, $\pt(\Jsecond)$,
the invariant mass and absolute value of pseudorapidity of the jj system,
the angular separation $\Delta R = \sqrt{\smash[b]{(\Delta \eta)^2 + (\Delta \phi)^2}}$, where $\phi$ is the azimuthal angle, between the two \PH candidates and between each \PH and \VBF jet, the absolute value of the polar angles with respect to the beam direction of the two \VBF jets in the center-of-mass frame of the six selected jets and the product of the two Higgs boson centralities 
$\exp{[-([\eta(\Hfirst)-\eta_\text{avg}]/\Delta\eta)^{2} - ([\eta(\Hsecond)-\eta_\text{avg}]/{\Delta\eta})^{2}]}$,
where $\Delta\eta = \eta(\Jfirst) - \eta(\Jsecond)$ and $\eta_\text{avg} = [\eta(\Jfirst) + \eta(\Jsecond)]/2$.
The discriminant is trained to separate the SM \GGF \HH signal from the $\CVV = 2$ \VBF signal, in order to optimize both the sensitivity to the anomalous \CVV coupling hypotheses and the correct classification of SM \GGF signal events.
The value $\CVV = 2$ is chosen because it is representative of the event kinematics in the presence of anomalous couplings, characterized by the large invariant mass of the \JJ and \HH systems.
These signals are associated with a large increase of the total cross section that would make them detectable with the available data set.
A threshold on the BDT output is chosen to assign events to either the \GGF or \VBF category. It results in the correct assignment of about 97\% of all \GGF \HH signal events to the \GGF category and of about 60 (80)\% of SM ($\CVV = 2$) \VBF \HH events that contain the additional jets to the \VBF category.
\par}

Events classified as \GGF or \VBF signal are further divided into subcategories to optimize the sensitivity of the search for anomalous coupling hypotheses.
Events in the \GGF category are divided into a low- and high-mass category if the reconstructed invariant mass of the \HH system, $m_{\HH}$, is below or above 450\GeV, where the boundary is defined according to the kinematic properties of the signal.
The latter category efficiently collects SM \GGF \HH events, while the former increases the acceptance to signals with anomalous \klambda values.
Events in the \VBF category are instead divided into a ``SM-like'' and an ``anomalous \CVV-like'' category depending on the value of the discriminant trained to separate \GGF and \VBF.
The categorization thresholds were chosen to maximize the expected sensitivity to \VBF \HH signals, and result in the assignment of about 25--30\% of the \VBF $\CVV = 2$ events to the anomalous \CVV-like category and 95\% of SM \VBF events to the SM-like category.

The large multijet background that originates from QCD and \ttbar hadronic processes is estimated from the data using background-dominated regions.
Analysis signal ($A_\text{SR}$) and control ($A_\text{CR}$) regions are defined by requiring $\chi < 25 \GeV$ and $25 \leq \chi < 50 \GeV$, respectively, where $\chi$ is the distance from the expected peak position of the two Higgs boson candidates' invariant masses
and is defined as
$\chi = \sqrt{\smash[b]{(m_{\Hfirst} - c_1)^2  + (m_{\Hsecond} - c_2)^2 }}$, where $c_1$ and $c_2$ are as defined for the pairing of the four jets.
Both \ASR and \ACR are divided into a four \PQb jet (4\PQb) and three \PQb jet (3\PQb) region by requiring the \PQb jet candidate with the lowest \textsc{DeepJet} output to satisfy or fail the medium WP of the discriminant, respectively.
There are between 5.5 and 11 times more events in the 3\PQb region than in the 4\PQb region, depending on the topological category and data-taking year considered.
The overall efficiency for both \GGF and \VBF signal events to be selected in the \ASRfourb region ranges from 0.3 to 3\% depending on the couplings considered and is minimal for the SM \VBF production and for the \GGF production with $\klambda\approx5$ due to the interference effects in the \HH production that result in low momenta of the Higgs bosons. The signal acceptance is mostly limited by the trigger acceptance and the jet \PQb tagging efficiency.

Background events in the \ASRfourb region are modeled from events in the \ASRthreeb region.
The former represents the sensitive region of the analysis, while the latter provides a sample enriched in multijet background events with similar kinematic properties.
Events in \ASRfourb were analyzed only after all the methods were defined and validated.
The normalization is determined by scaling the observed number of events in \ASRthreeb by a transfer factor computed as the ratio of the number of events in the \ACRfourb and \ACRthreeb regions.
Variations of the transfer factor depending on the position in the $(m_{\Hfirst},m_{\Hsecond})$ plane are accounted for by measuring it as a function of \mparall, defined as the projection of the point in the plane on the line $m_{\Hfirst} = (c_1/c_2)\,m_{\Hsecond}$ that is used for the \PH candidate reconstruction.
Higher values of \mparall are correlated with a higher average \pt of the selected jets.

Differences in the distributions of several variables between the 3\PQb and the 4\PQb regions are addressed with the BDT-based reweighting method described in Ref.~\cite{Rogozhnikov:2016bdp}, which uses a dedicated metric to identify the phase space regions with the largest differences in the distributions and compute an event weight to correct for them.
This method accurately models multiple variables and their correlations, while minimizing issues related to the statistical uncertainties arising from the limited number of events in the two regions. The BDT is trained in the \ACRfourb and \ACRthreeb regions, and applied to events in \ASRthreeb to model \ASRfourb.

Trainings of this BDT are performed separately for each \GGF and \VBF category.
All trainings use as inputs the following ten variables: \pt of the four \PQb jets, $m_{\HH}$, the invariant masses and \pt of the \Hfirst and \Hsecond candidates, and the absolute value of their pseudorapidity difference ($\abs{\Delta\eta(\Hfirst, \Hsecond)}$).
In the \GGF category, ten additional variables are used: the magnitude of the scalar ($\sum\pt$) and vector ($\pt(\HH)$) sums of the \pt of the four \PQb jets, the angular $\Delta R$ separations between the two jets that constitute \Hfirst and \Hsecond ($\Delta R^{\Hfirst}(\PQb\PQb)$, $\Delta R^{\Hsecond}(\PQb\PQb)$), the minimal $\Delta R$ ($\Delta R^\text{min}$) and the maximal $\abs{\Delta\eta}$ ($\abs{\Delta\eta}^\text{max}$) between all the possible \PQb jet pairs, the absolute value of the angle with respect to the beamline of one Higgs boson in the four-jet reference frame ($\abs{\cos\theta^*}$) and of one jet of \Hfirst in the \Hfirst candidate reference frame ($\abs{\cos\theta_\PQb^{\Hfirst}}$), the sum of the resolution estimators of the three \PQb-tagged jets with the best \textsc{DeepJet} value ($\sum R_e$), and the number of these three jets that satisfy the tight \textsc{DeepJet} WP ($N_\PQb^T$).
In the \VBF category, four additional variables are used: the absolute value of the $\phi$ separation between the two Higgs bosons, the \VBF jets invariant mass and absolute value of the $\eta$ separation, and the output of the production mode BDT discriminant.
These variables are chosen as those that best represent the kinematic properties of the events in the 3\PQb and the 4\PQb regions and that provide separation between signal and background and are used in subsequent steps of the analysis. 

{\tolerance=800 The training parameters are optimized with a two-step procedure. First, a Kolmogorov--Smirnov distance test is used to ensure that the distributions of the BDT input variables in the target \ACRfourb region are compatible with the ones in the reweighted \ACRthreeb region. Once that is verified, a BDT is trained to separate the \ACRfourb and reweighted \ACRthreeb data, thus testing also the correlations of variables. All the training configurations that are chosen are required to have an area of 0.5 under the receiver operating curve of the discriminant, corresponding to no separation.\par}

The procedure is validated by applying it to a signal-depleted region, defined by shifting the signal and control regions according to the definition of $\chi$, using as values of the center position $c_1 = 179\GeV$ and $c_2 = 172\GeV$. The center of this validation region is chosen to be along the $m_{\Hfirst} = 1.04\,m_{\Hsecond}$ line used in the reconstruction of the \PH candidates to provide an accurate proxy of the analysis region. In analogy to the analysis regions, signal and control validation regions, \VSR and \VCR, are defined as $\chi < 25 \GeV$ and $25 \leq \chi < 50 \GeV$, respectively. After training and applying the reweighting BDT in these regions and computing the normalization transfer factors, the data in $\VSRfourb$ were found to be compatible within uncertainties with the predicted background, validating the modeling method. The agreement is quantified with a goodness-of-fit test based on a saturated model~\cite{BAKER1984437} performed on the observables used in the analysis. For a fit under the background-only hypothesis, a $p$-value of 53\% is observed, ranging between 12 and 83\% for the individual categories.

The impact on the estimated background from the presence of signal events in the \ASRthreeb region due to jets failing the \PQb tagging requirement is estimated by generating pseudo-data in \ASRfourb according to the modeled background plus simulated \HH signal, and fitting them under a different background hypothesis that includes the contribution from signal events in \ASRthreeb weighted as done for background events.
This study is repeated for signal yields up to five times larger than the expected sensitivity of this search, and in all cases a signal yield compatible with the true one is observed.
We conclude that signal events in \ASRthreeb do not have any significant impact on the background model and on the results.

For the background model, systematic uncertainties are considered for the limited number of events in the \ASRthreeb.
These uncertainties are uncorrelated across the individual bins of the background templates used for the statistical analysis, and correspond to the propagation of a bin-by-bin Poisson uncertainty from the \ASRthreeb to the \ASRfourb region.
The uncertainty in the estimation of the transfer factor from the 3\PQb to the 4\PQb region is computed from the statistical uncertainty in \ACRthreeb and \ACRfourb and is 1--2\% for the \GGF categories, 2-3\% for the SM-like \VBF category and 18--32\% for the anomalous $\CVV$-like \VBF category.
An uncertainty is also considered for the limited number of events in the validation region, in some cases lower than the number of events in the analysis region.
It is large (30--33\%) for the anomalous \CVV-like \VBF category while it is about 2-3\% and below 1\% for the other \VBF category and the \GGF categories, respectively, and represents the inherent limitation on the capability to validate the performance of the background model.
For analysis categories where the agreement between the observed and predicted background yields in the validation region differs by more than one standard deviation, an additional uncertainty is included and ranges between 1.5 and 4.7\%, depending on the category and year.
Finally, the uncertainty in the performance of the reweighting method to interpolate the kinematics from \ACR into \ASR is estimated by performing alternative trainings in two regions of \ACR.
The two regions are defined by requiring the product of \mperp and \mparall to be either positive or negative, where \mperp is defined as the projection of the point in the $(m_{\Hfirst},m_{\Hsecond})$ plane onto the axis perpendicular to the one corresponding to \mparall previously defined.
The two regions correspond to four quadrants in the $(m_{\Hfirst},m_{\Hsecond})$ plane, and allow for tests of the capability of the reweighting method to model \ASR, starting from events with kinematic properties that are either similar ($\mperp \mparall < 0$) or that are harder or softer ($\mperp \mparall > 0$) compared to \ASR, thus testing the capability of the model to interpolate across different learning domains.
The alternative background templates obtained from trainings in these regions represent the uncertainty on the shape of the predicted background distribution.
All the uncertainties are independent between the 2016 and 2017--2018 background models.
The dominant uncertainties in this search are those associated to the background modeling, and in particular the bin-by-bin and the normalization uncertainties due to the limited number of events in \ASRthreeb, \ACRthreeb and \ACRfourb.

The effects of the imperfect modeling of the detector response and the inaccurate simulation of signal processes are accounted for as systematic uncertainties.
The most important sources of systematic uncertainty are the total integrated luminosity, the jet energy scale and resolution, the efficiency of the trigger and of the \PQb-tagging requirements, the modeling of the pileup distribution, the $\HH\to\bbbb$ branching fraction, and the parameters used for the generators.
A specific uncertainty on the parton shower is considered for the \VBF production mode.
Uncertainties on the theoretically determined \HH cross section are considered only when quoting a limit on the \HH signal strength ($\mu$), defined as the ratio of the value of the cross section limit relative to the theoretical cross section expectation in the SM ($\sigma_{\HH}/\sigma_{\HH}^{\SM}$). These uncertainties are negligible in comparison to the background uncertainties. See \suppMaterial{} for more details.

\begin{figure*}[!htb]
\centering
{\includegraphics[width=0.66\cmsTwoWide] {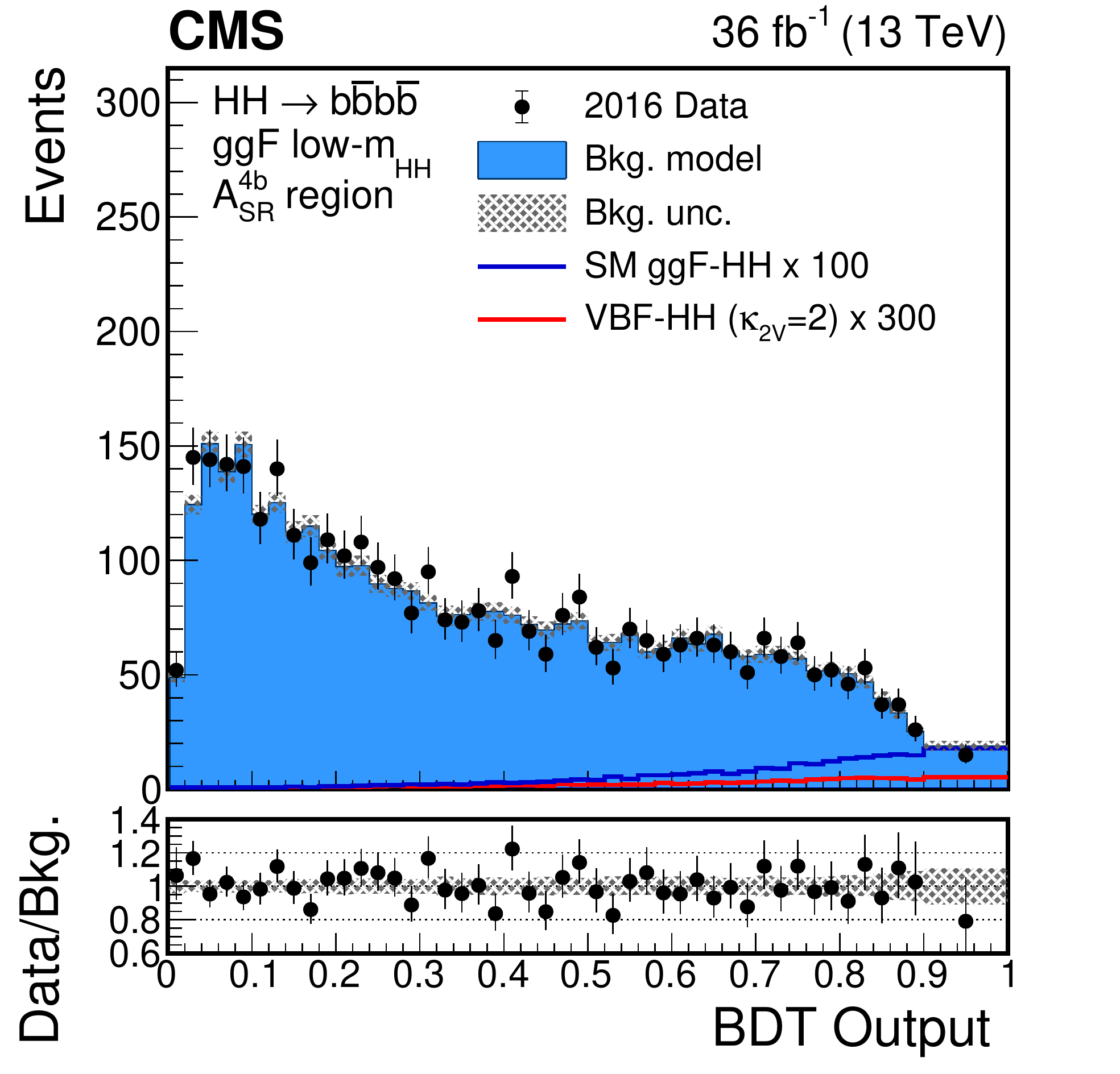}}\hspace{\fill}
{\includegraphics[width=0.66\cmsTwoWide] {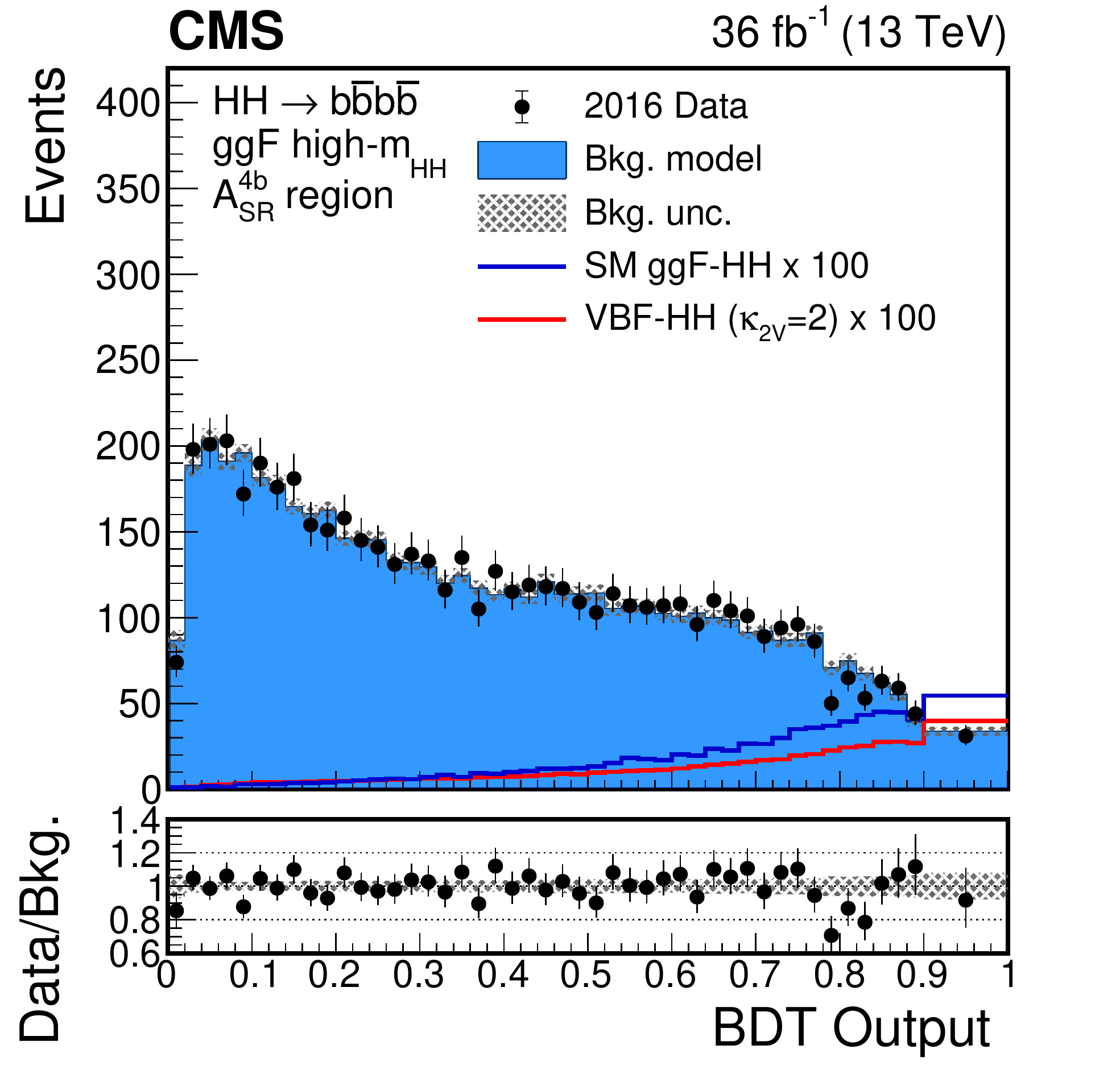}}\hspace{\fill}
{\includegraphics[width=0.66\cmsTwoWide] {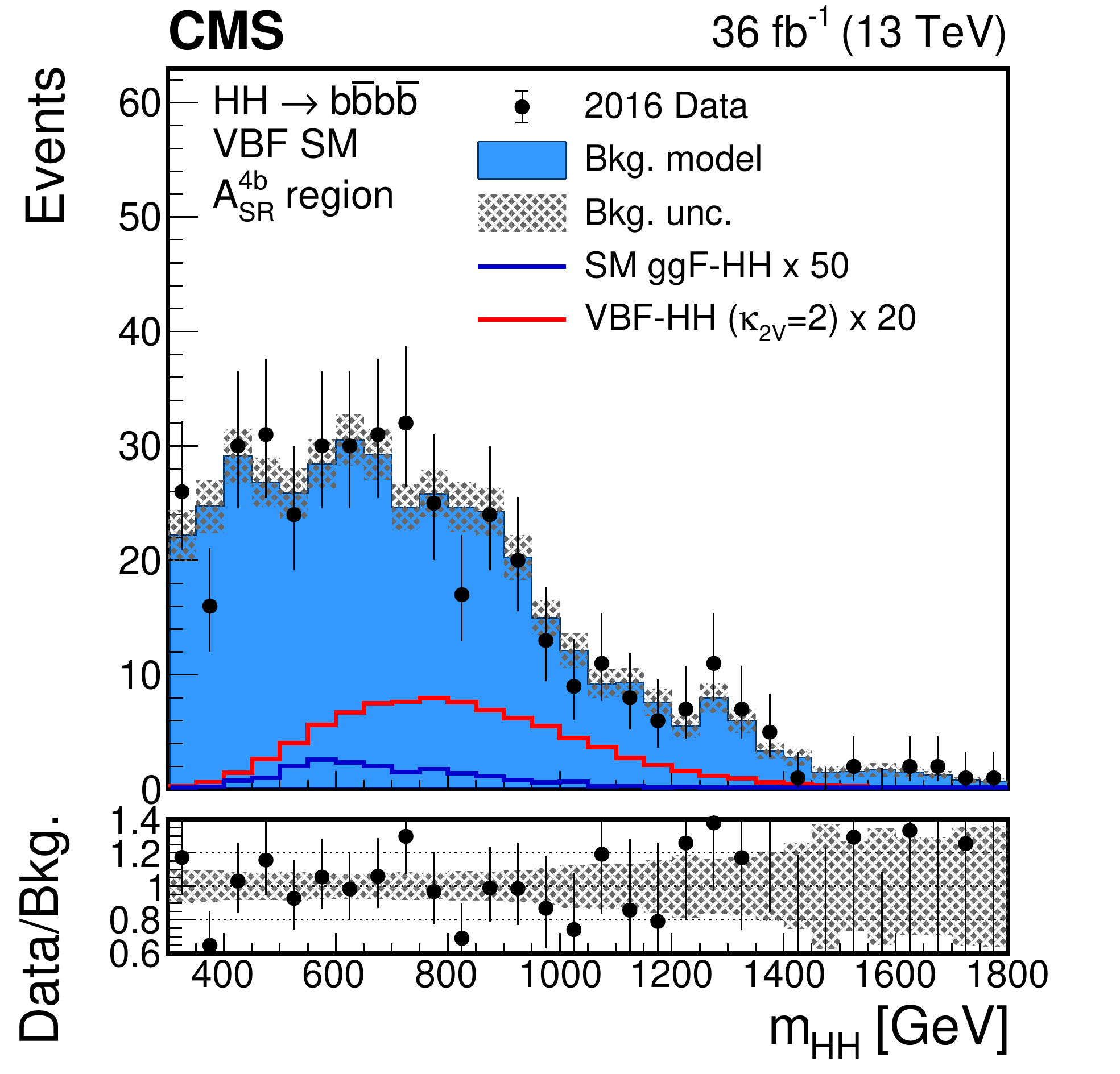}}\\
{\includegraphics[width=0.66\cmsTwoWide] {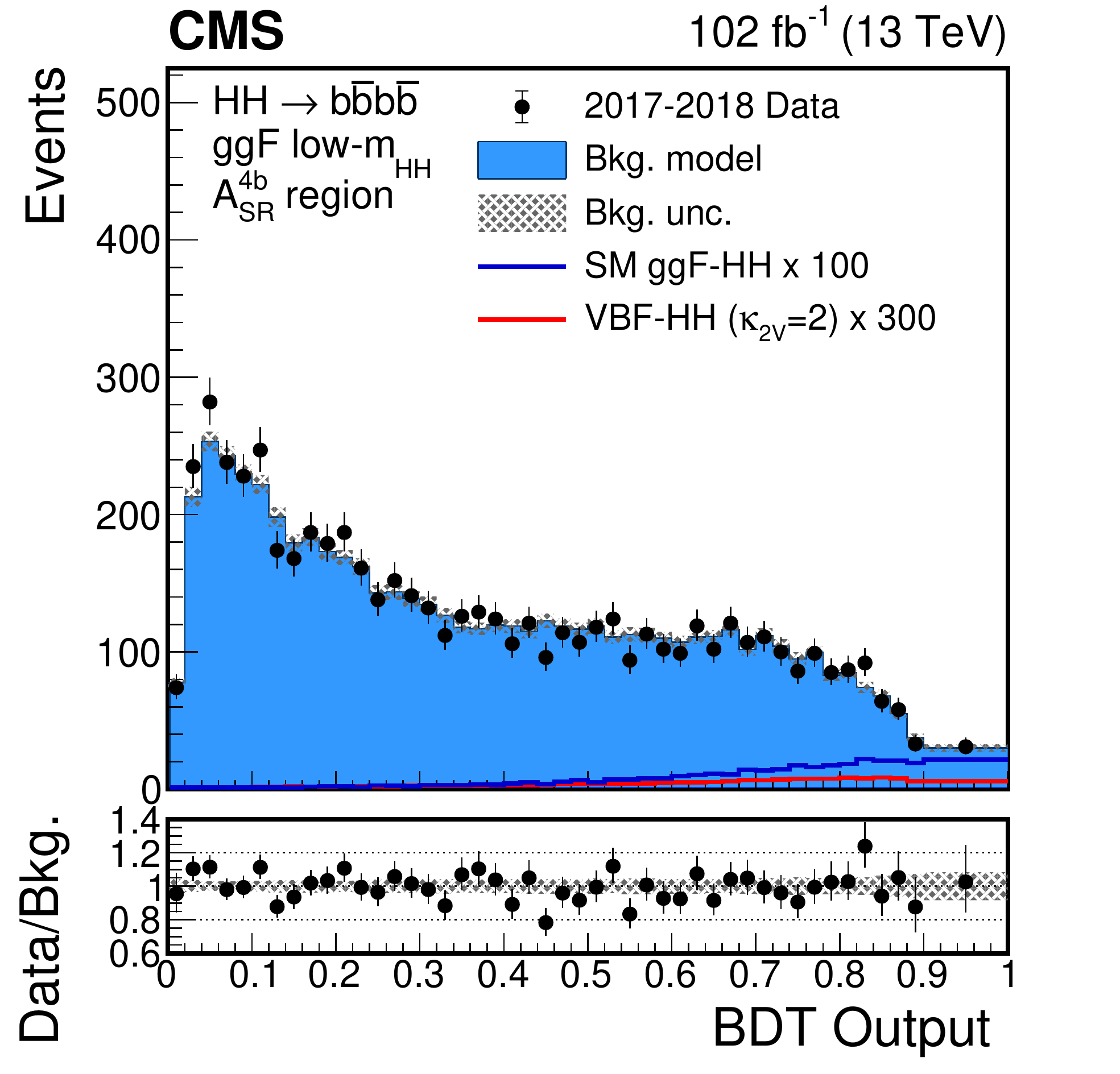}}\hspace{\fill}
{\includegraphics[width=0.66\cmsTwoWide] {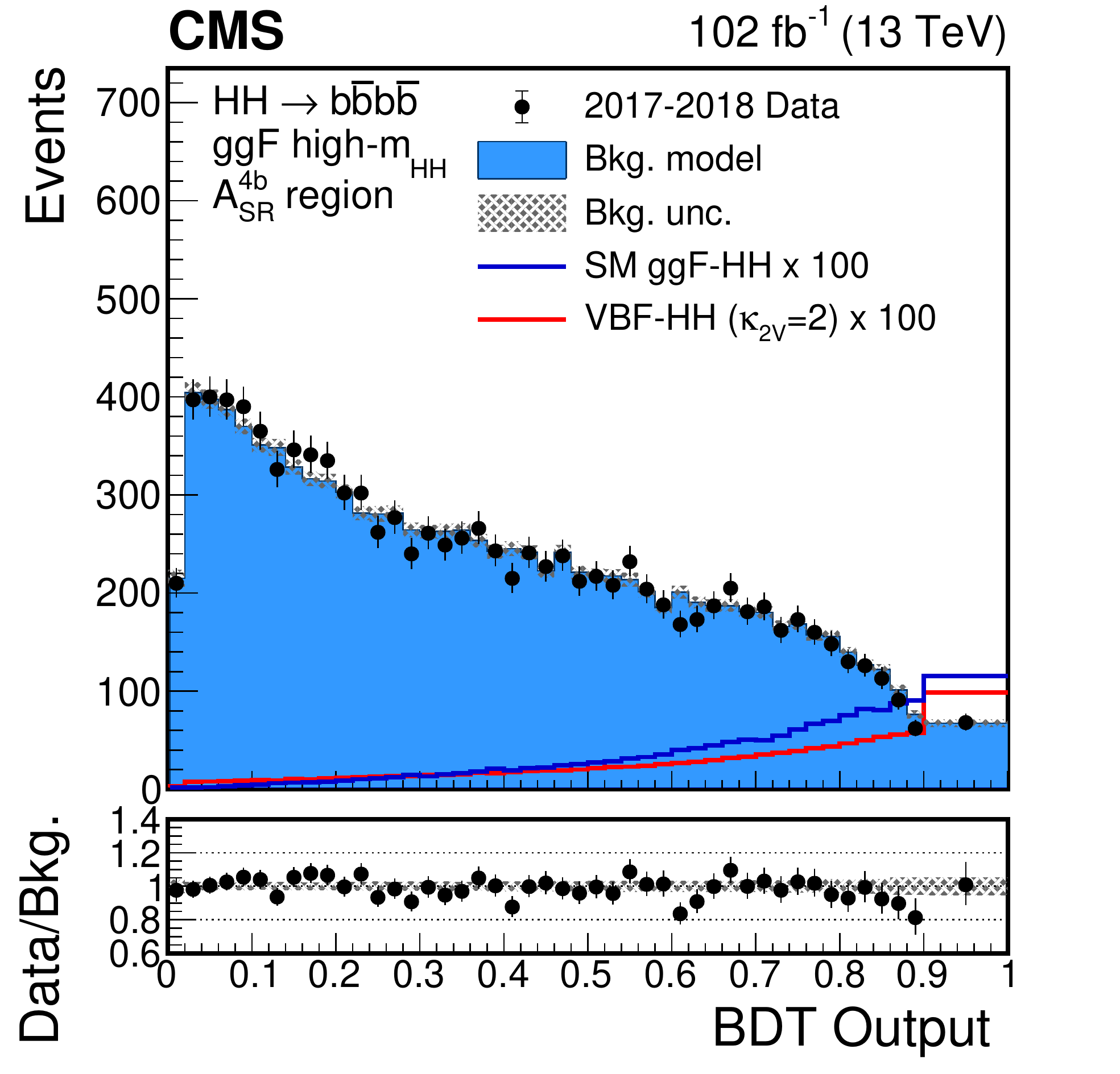}}\hspace{\fill}
{\includegraphics[width=0.66\cmsTwoWide] {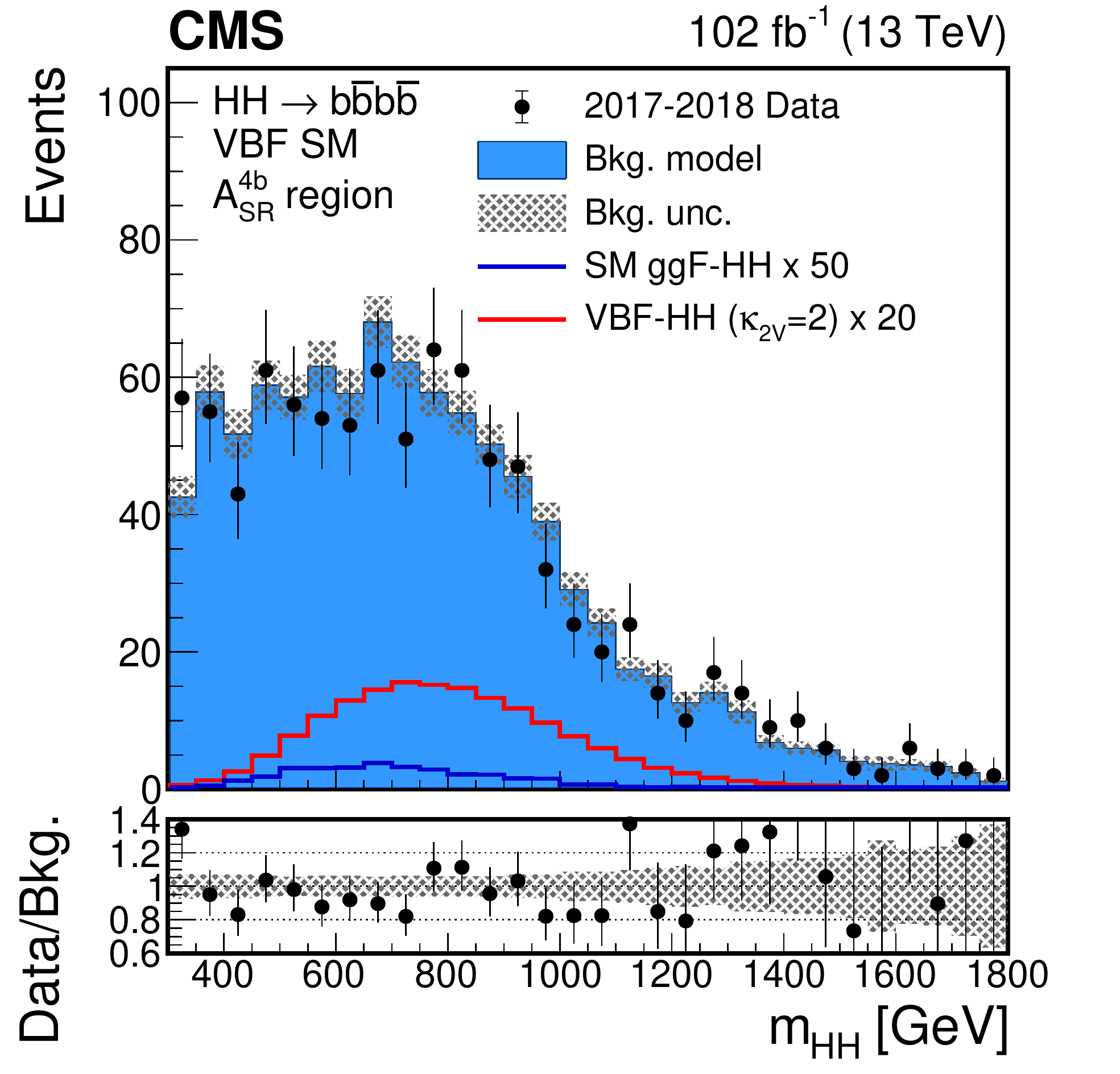}}
\caption{\label{fig:dataMCplots}
Distributions of the events observed in the \ASRfourb signal region for 2016 (top) and 2017--2018 (bottom) data.
The two leftmost columns show the BDT output in the low- and high-mass categories, and the rightmost column shows the $m_{\HH}$ distribution in the \VBF SM-like category.
}
\end{figure*}

{\tolerance=800
A multivariate BDT discriminant is trained with the \textsc{XGBoost} software~\cite{Chen:2016:XST:2939672.2939785} in the two \GGF subcategories to separate the signal from the weighted \ASRthreeb background events.
The discriminant uses as inputs $\pt(\Hfirst)$, $\pt(\Hsecond)$, $m_{\Hfirst}$, $m_{\Hsecond}$, $\abs{\Delta\eta(\Hfirst, \Hsecond)}$, $m_{\HH}$, $\pt({\HH})$, $\Delta R^{\Hfirst}(\PQb\PQb)$, $\Delta R^{\Hsecond}(\PQb\PQb)$, $\Delta R^\text{min}$, $\abs{\Delta\eta}^\text{max}$, $\sum\pt$, $N_\PQb^T$, $\sum R_e$, $\abs{\cos\theta^*}$, and $\abs{\cos\theta_\PQb^{\Hfirst}}$.
For each subcategory, a separate training is performed to separate the SM \GGF \HH signal from the weighted \ASRthreeb region data.
Since the same \ASRthreeb data are also used to model the background, this data set is divided in two equal-size subsamples. Two trainings are performed on each half and applied to the other half, and the two partial background templates are added together.
In this way, the full data set can be used for the modeling, while the BDT discriminant is not evaluated on events used for its training.
In the \VBF SM-like category, $m_{\HH}$ is used as the discriminating variable, while in the anomalous \CVV-like category, a counting experiment is performed because of the small number of expected background events.
The distributions of these variables are shown in Fig.~\ref{fig:dataMCplots}.
For the \VBF anomalous \CVV-like category in the 2016 (2017--2018) data set, 4 (13) events are observed for a total of $4.0 \pm 1.3$ ($15.0 \pm 3.4$) background and 1.5 (3.5) \VBF $\CVV = 2$ signal events expected.
\par}

A binned maximum likelihood fit is simultaneously performed in all analysis categories, where the systematic uncertainties previously discussed are introduced as nuisance parameters.
No deviation from a background-only hypothesis is observed.
Results are used to set 95\% \CL upper limits on the \HH production cross section using the modified frequentist \CLs criterion~\cite{Junk:1999kv, Read:2002hq} with the profile likelihood ratio modified for upper limits~\cite{CMS-NOTE-2011-005} as the test statistics, and making use of the asymptotic approximation~\cite{Cowan:2010js}.

\begin{figure}[!htb]
\centering
\includegraphics[width=0.9\cmsTwoWide] {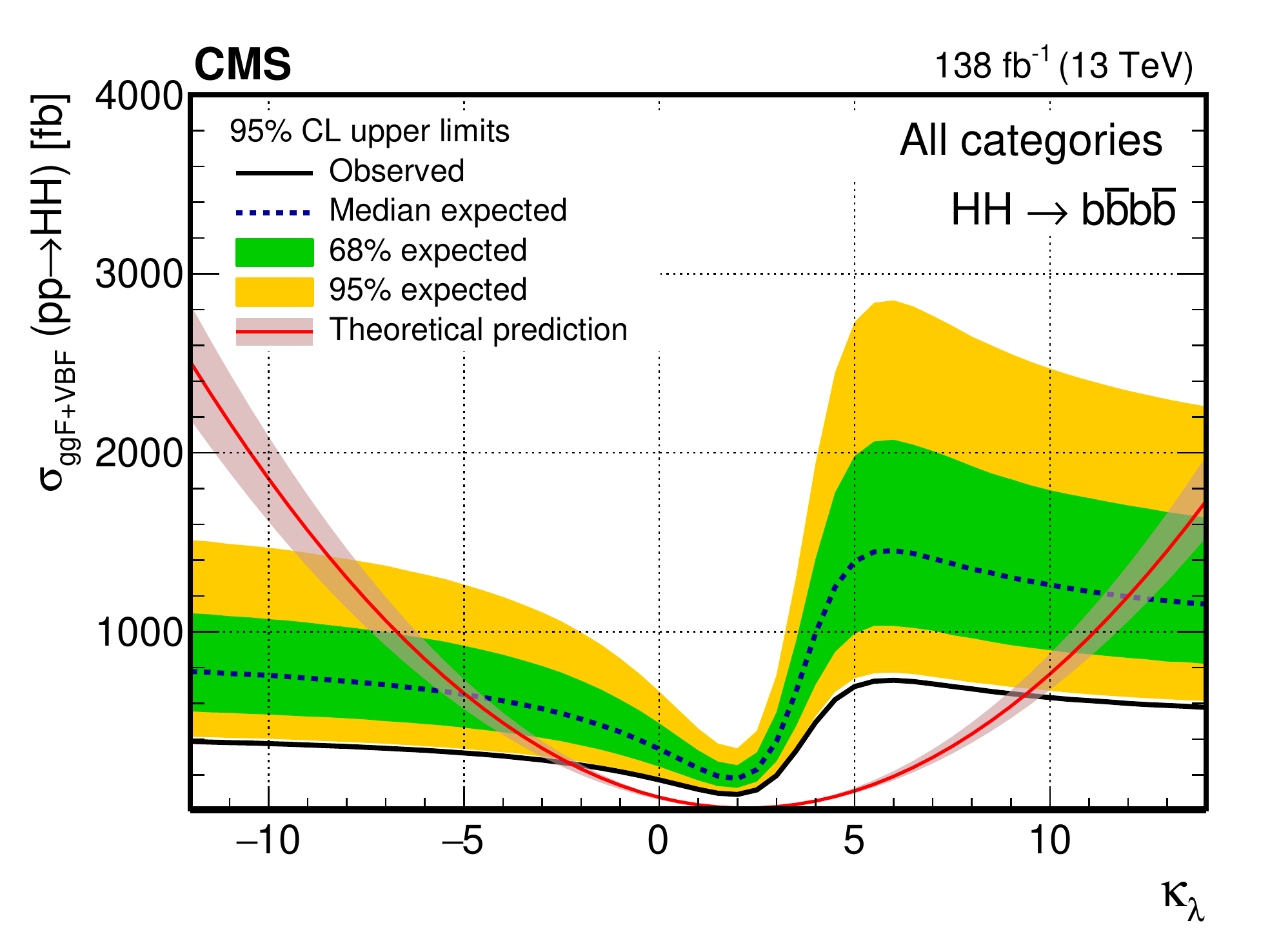}\hspace{0.1\cmsTwoWide}
\includegraphics[width=0.9\cmsTwoWide] {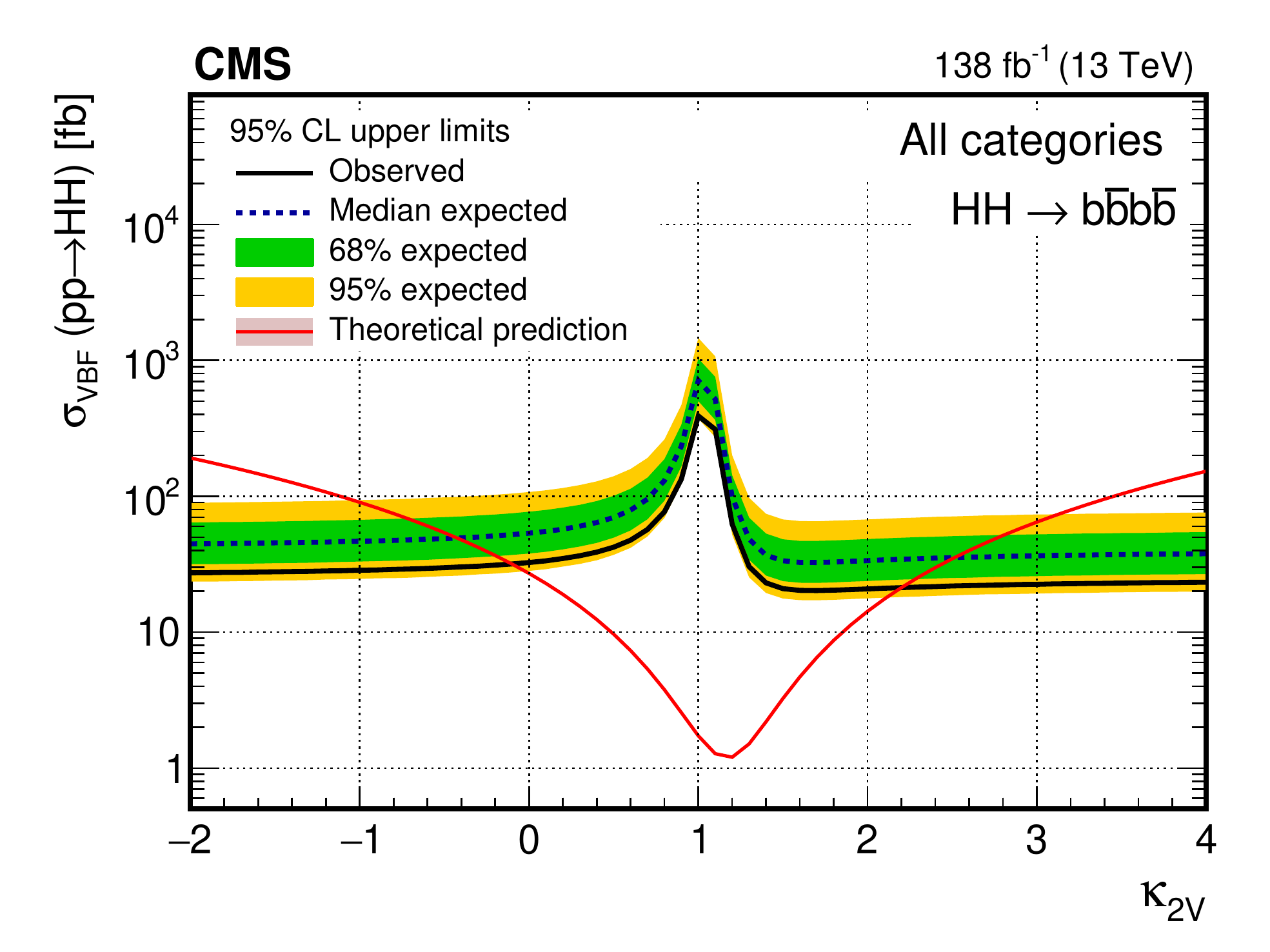}
\caption{\label{fig:ULscans}
Observed and expected 95\% \CL upper limits on the $\sigma_{\GGF+\VBF}$ \HH cross section as a function of \klambda (\cmsLeft), and on the $\sigma_{\VBF}$ \HH cross section as a function of \CVV (\cmsRight).
The green (yellow) band indicates the regions containing 68\% (95\%) of the limit values expected under the background-only hypothesis.
The red lines denote the theoretical cross section expectation assuming that other couplings are set to the SM prediction.
For the cross section limit as a function of \CVV, the \GGF \HH production is assumed to correspond to the SM prediction.
}
\end{figure}

Figure~\ref{fig:ULscans} shows the 95\% \CL cross section upper limits as functions of the \klambda and \CVV values.
The value of \klambda is observed (expected) to be in the range $-2.3 < \klambda < 9.4$ ($-5.0 < \klambda < 12.0$) at the 95\% \CL,
while the value of \CVV is observed (expected) to be in the range $-0.1 < \CVV < 2.2$ ($-0.4 < \CVV < 2.5$) at the 95\% \CL.
The total \HH production cross section, defined as the sum of the \GGF and \VBF production modes, is observed (expected) to be smaller than
120 (238)\fb, corresponding to 3.9 (7.8)
times the SM prediction, when uncertainties on the theoretical production cross section are included.
The \HH \VBF production cross section is observed (expected) to be smaller than 226 (412) times the SM prediction.
The deficit in the observed number of events localized around the BDT discriminant values of 0.85--0.9 in the \GGF high-mass category in the 2017--2018 data set, which provides the largest sensitivity to the signal, results in the observed limit to be below the expected one.
Studies of the background model for the individual BDT input variables and the absence of deficit in the 2016 data in the same high-sensitivity region, and in the other categories suggest that this under fluctuation is of statistical nature.

The intervals containing 68 and 95\% of the expected signal strength upper limits correspond to $[5.5, 12.3]$ and $[4.0, 18.7]$ ($[291, 598]$ and $[216, 846]$) for the \GGF (\VBF) production modes, and the observed limit is thus compatible with the expectation within about two standard deviations.
The sensitivity is mostly limited by the number of events in the signal and control regions of the analysis. Tabulated results are available in the HEPData record of this analysis~\cite{hepdata}.

In summary, a search for the production of Higgs boson pairs via gluon and vector boson fusion in the four \PQb quark decay channel has been presented.
The data are found to be statistically compatible with the background-only hypothesis, and an observed (expected) upper limit at the 95\% confidence level is set to
3.9 (7.8)
times the SM prediction for the combined \GGF and \VBF \HH cross section.
The value of the Higgs boson self-coupling, normalized to the SM expectation, is observed (expected) to be in the range $-2.3 < \klambda < 9.4$ ($-5.0 < \klambda < 12.0$),
and the value of the coupling of Higgs boson pairs to vector boson pairs, normalized to the SM expectation, to be in the range $-0.1 < \CVV < 2.2$ ($-0.4 < \CVV < 2.5$).
These are the most stringent observed constraints to date on the \HH production cross sections and on the \CVV coupling.

\begin{acknowledgments}
We congratulate our colleagues in the CERN accelerator departments for the excellent performance of the LHC and thank the technical and administrative staffs at CERN and at other CMS institutes for their contributions to the success of the CMS effort. In addition, we gratefully acknowledge the computing centers and personnel of the Worldwide LHC Computing Grid and other centers for delivering so effectively the computing infrastructure essential to our analyses. Finally, we acknowledge the enduring support for the construction and operation of the LHC, the CMS detector, and the supporting computing infrastructure provided by the following funding agencies: BMBWF and FWF (Austria); FNRS and FWO (Belgium); CNPq, CAPES, FAPERJ, FAPERGS, and FAPESP (Brazil); MES (Bulgaria); CERN; CAS, MoST, and NSFC (China); MINCIENCIAS (Colombia); MSES and CSF (Croatia); RIF (Cyprus); SENESCYT (Ecuador); MoER, ERC PUT and ERDF (Estonia); Academy of Finland, MEC, and HIP (Finland); CEA and CNRS/IN2P3 (France); BMBF, DFG, and HGF (Germany); GSRT (Greece); NKFIA (Hungary); DAE and DST (India); IPM (Iran); SFI (Ireland); INFN (Italy); MSIP and NRF (Republic of Korea); MES (Latvia); LAS (Lithuania); MOE and UM (Malaysia); BUAP, CINVESTAV, CONACYT, LNS, SEP, and UASLP-FAI (Mexico); MOS (Montenegro); MBIE (New Zealand); PAEC (Pakistan); MSHE and NSC (Poland); FCT (Portugal); JINR (Dubna); MON, RosAtom, RAS, RFBR, and NRC KI (Russia); MESTD (Serbia); SEIDI, CPAN, PCTI, and FEDER (Spain); MOSTR (Sri Lanka); Swiss Funding Agencies (Switzerland); MST (Taipei); ThEPCenter, IPST, STAR, and NSTDA (Thailand); TUBITAK and TAEK (Turkey); NASU (Ukraine); STFC (United Kingdom); DOE and NSF (USA).
\end{acknowledgments}

\bibliography{auto_generated}

\providecommand{\href}[2]{#2}\begingroup\raggedright\begin{thebibliography}{10}%
\makeatletter
\providecommand{\hrefCMSnoop }[0]{\@secondoftwo}%
\makeatother
\providecommand{\doi}{\texttt{doi:}\begingroup \urlstyle{tt}\Url}

\bibitem{ATLASdiscovery}
\hrefCMSnoop {}{{ATLAS Collaboration}, ``{Observation of a new particle in the
  search for the Standard Model Higgs boson with the ATLAS detector at the
  LHC}'',} \textit{ Phys. Lett. B} \textbf{ 716} (2012) 1,
  \href{http://dx.doi.org/10.1016/j.physletb.2012.08.020}{\doi{10.1016/j.physletb.2012.08.020}},
\href{http://www.arXiv.org/abs/1207.7214}{\texttt{arXiv:1207.7214}}.

\bibitem{CMSdiscovery}
\hrefCMSnoop {}{{CMS Collaboration}, ``{Observation of a new boson at a mass of
  125 GeV with the CMS experiment at the LHC}'',} \textit{ Phys. Lett. B}
  \textbf{ 716} (2012) 30,
  \href{http://dx.doi.org/10.1016/j.physletb.2012.08.021}{\doi{10.1016/j.physletb.2012.08.021}},
\href{http://www.arXiv.org/abs/1207.7235}{\texttt{arXiv:1207.7235}}.

\bibitem{Chatrchyan:2013lba}
\hrefCMSnoop {}{{CMS Collaboration}, ``{Observation of a new boson with mass
  near 125 GeV in pp collisions at $\sqrt{s}$ = 7 and 8 TeV}'',} \textit{ JHEP}
  \textbf{ 06} (2013) 081,
  \href{http://dx.doi.org/10.1007/JHEP06(2013)081}{\doi{10.1007/JHEP06(2013)081}},
\href{http://www.arXiv.org/abs/1303.4571}{\texttt{arXiv:1303.4571}}.

\bibitem{PhysRevLett.13.321}
\hrefCMSnoop {}{F.~Englert and R.~Brout, ``Broken symmetry and the mass of
  gauge vector mesons'',} \textit{ Phys. Rev. Lett.} \textbf{ 13} (1964) 321,
  \href{http://dx.doi.org/10.1103/PhysRevLett.13.321}{\doi{10.1103/PhysRevLett.13.321}}.

\bibitem{PhysRevLett.13.508}
\hrefCMSnoop {}{P.~W. Higgs, ``Broken symmetries and the masses of gauge
  bosons'',} \textit{ Phys. Rev. Lett.} \textbf{ 13} (1964) 508,
  \href{http://dx.doi.org/10.1103/PhysRevLett.13.508}{\doi{10.1103/PhysRevLett.13.508}}.

\bibitem{PhysRevLett.13.585}
\hrefCMSnoop {}{G.~S. Guralnik, C.~R. Hagen, and T.~W.~B. Kibble, ``Global
  conservation laws and massless particles'',} \textit{ Phys. Rev. Lett.}
  \textbf{ 13} (1964) 585,
  \href{http://dx.doi.org/10.1103/PhysRevLett.13.585}{\doi{10.1103/PhysRevLett.13.585}}.

\bibitem{Dawson:1998py}
\hrefCMSnoop {}{S.~Dawson, S.~Dittmaier, and M.~Spira, ``{Neutral Higgs boson
  pair production at hadron colliders: QCD corrections}'',} \textit{ Phys. Rev.
  D} \textbf{ 58} (1998) 115012,
  \href{http://dx.doi.org/10.1103/PhysRevD.58.115012}{\doi{10.1103/PhysRevD.58.115012}},
  \href{http://www.arXiv.org/abs/hep-ph/9805244}{\texttt{arXiv:hep-ph/9805244}}.

\bibitem{Borowka:2016ehy}
S.~Borowka\hrefCMSnoop {}{ {et~al.}, ``Higgs boson pair production in gluon
  fusion at next-to-leading order with full top-quark mass dependence'',}
  \textit{ Phys. Rev. Lett.} \textbf{ 117} (2016) 012001,
  \href{http://dx.doi.org/10.1103/PhysRevLett.117.079901}{\doi{10.1103/PhysRevLett.117.079901}},
  \href{http://www.arXiv.org/abs/1604.06447}{\texttt{arXiv:1604.06447}}.
  [Erratum: \DOI{10.1103/PhysRevLett.117.079901}].

\bibitem{Baglio:2018lrj}
J.~Baglio\hrefCMSnoop {}{ {et~al.}, ``{Gluon fusion into Higgs pairs at NLO QCD
  and the top mass scheme}'',} \textit{ Eur. Phys. J. C} \textbf{ 79} (2019)
  459,
  \href{http://dx.doi.org/10.1140/epjc/s10052-019-6973-3}{\doi{10.1140/epjc/s10052-019-6973-3}},
  \href{http://www.arXiv.org/abs/1811.05692}{\texttt{arXiv:1811.05692}}.

\bibitem{deFlorian:2013jea}
\hrefCMSnoop {}{D.~de~Florian and J.~Mazzitelli, ``Higgs boson pair production
  at next-to-next-to-leading order in {QCD}'',} \textit{ Phys. Rev. Lett.}
  \textbf{ 111} (2013) 201801,
  \href{http://dx.doi.org/10.1103/PhysRevLett.111.201801}{\doi{10.1103/PhysRevLett.111.201801}},
  \href{http://www.arXiv.org/abs/1309.6594}{\texttt{arXiv:1309.6594}}.

\bibitem{Shao:2013bz}
\hrefCMSnoop {}{D.~Y. Shao, C.~S. Li, H.~T. Li, and J.~Wang, ``{Threshold
  resummation effects in Higgs boson pair production at the LHC}'',} \textit{
  JHEP} \textbf{ 07} (2013) 169,
  \href{http://dx.doi.org/10.1007/JHEP07(2013)169}{\doi{10.1007/JHEP07(2013)169}},
  \href{http://www.arXiv.org/abs/1301.1245}{\texttt{arXiv:1301.1245}}.

\bibitem{deFlorian:2015moa}
\hrefCMSnoop {}{D.~de~Florian and J.~Mazzitelli, ``{Higgs pair production at
  next-to-next-to-leading logarithmic accuracy at the LHC}'',} \textit{ JHEP}
  \textbf{ 09} (2015) 053,
  \href{http://dx.doi.org/10.1007/JHEP09(2015)053}{\doi{10.1007/JHEP09(2015)053}},
  \href{http://www.arXiv.org/abs/1505.07122}{\texttt{arXiv:1505.07122}}.

\bibitem{Grazzini:2018bsd}
M.~Grazzini\hrefCMSnoop {}{ {et~al.}, ``{Higgs boson pair production at NNLO
  with top quark mass effects}'',} \textit{ JHEP} \textbf{ 05} (2018) 059,
  \href{http://dx.doi.org/10.1007/JHEP05(2018)059}{\doi{10.1007/JHEP05(2018)059}},
  \href{http://www.arXiv.org/abs/1803.02463}{\texttt{arXiv:1803.02463}}.

\bibitem{Baglio:2020wgt}
J.~Baglio\hrefCMSnoop {}{ {et~al.}, ``{$\Pg\Pg\to \PH\PH$ : Combined
  uncertainties}'',} \textit{ Phys. Rev. D} \textbf{ 103} (2021) 056002,
  \href{http://dx.doi.org/10.1103/PhysRevD.103.056002}{\doi{10.1103/PhysRevD.103.056002}},
  \href{http://www.arXiv.org/abs/2008.11626}{\texttt{arXiv:2008.11626}}.

\bibitem{Dreyer:2018qbw}
\hrefCMSnoop {}{F.~A. Dreyer and A.~Karlberg, ``Vector-boson fusion {Higgs}
  pair production at {N$^3$LO}'',} \textit{ Phys. Rev. D} \textbf{ 98} (2018)
  114016,
  \href{http://dx.doi.org/10.1103/PhysRevD.98.114016}{\doi{10.1103/PhysRevD.98.114016}},
  \href{http://www.arXiv.org/abs/1811.07906}{\texttt{arXiv:1811.07906}}.

\bibitem{Bishara:2016kjn}
\hrefCMSnoop {}{F.~Bishara, R.~Contino, and J.~Rojo, ``{Higgs pair production
  in vector-boson fusion at the LHC and beyond}'',} \textit{ Eur. Phys. J. C}
  \textbf{ 77} (2017) 481,
  \href{http://dx.doi.org/10.1140/epjc/s10052-017-5037-9}{\doi{10.1140/epjc/s10052-017-5037-9}},
  \href{http://www.arXiv.org/abs/1611.03860}{\texttt{arXiv:1611.03860}}.

\bibitem{Aaboud:2018knk}
\hrefCMSnoop {}{{ATLAS Collaboration}, ``{Search for pair production of Higgs
  bosons in the $\PQb\PAQb\PQb\PAQb$ final state using proton-proton collisions
  at $\sqrt{s} = 13$ TeV with the ATLAS detector}'',} \textit{ JHEP} \textbf{
  01} (2019) 030,
  \href{http://dx.doi.org/10.1007/JHEP01(2019)030}{\doi{10.1007/JHEP01(2019)030}},
  \href{http://www.arXiv.org/abs/1804.06174}{\texttt{arXiv:1804.06174}}.

\bibitem{Sirunyan:2018tki}
\hrefCMSnoop {}{{CMS Collaboration}, ``{Search for nonresonant Higgs boson pair
  production in the $\PQb\PAQb\PQb\PAQb$ final state at $\sqrt{s}= 13$ TeV}'',}
  \textit{ JHEP} \textbf{ 04} (2019) 112,
  \href{http://dx.doi.org/10.1007/JHEP04(2019)112}{\doi{10.1007/JHEP04(2019)112}},
  \href{http://www.arXiv.org/abs/1810.11854}{\texttt{arXiv:1810.11854}}.

\bibitem{Aaboud:2018sfw}
\hrefCMSnoop {}{{ATLAS Collaboration}, ``{Search for resonant and non-resonant
  Higgs boson pair production in the ${\PQb\PAQb\tau^+\tau^-}$ decay channel in
  $pp$ collisions at $\sqrt{s}=13$ TeV with the ATLAS detector}'',} \textit{
  Phys. Rev. Lett.} \textbf{ 121} (2018) 191801,
  \href{http://dx.doi.org/10.1103/PhysRevLett.121.191801}{\doi{10.1103/PhysRevLett.121.191801}},
  \href{http://www.arXiv.org/abs/1808.00336}{\texttt{arXiv:1808.00336}}.
  [Erratum: \DOI{10.1103/PhysRevLett.122.089901}].

\bibitem{Sirunyan:2017djm}
\hrefCMSnoop {}{{CMS Collaboration}, ``{Search for Higgs boson pair production
  in events with two bottom quarks and two tau leptons in proton-proton
  collisions at $\sqrt s$ = 13 TeV}'',} \textit{ Phys. Lett. B} \textbf{ 778}
  (2018) 101,
  \href{http://dx.doi.org/10.1016/j.physletb.2018.01.001}{\doi{10.1016/j.physletb.2018.01.001}},
  \href{http://www.arXiv.org/abs/1707.02909}{\texttt{arXiv:1707.02909}}.

\bibitem{Aaboud:2018zhh}
\hrefCMSnoop {}{{ATLAS Collaboration}, ``{Search for Higgs boson pair
  production in the ${\PQb\PAQb\PW\PW^{*}}$ decay mode at $\sqrt{s}=13$ TeV
  with the ATLAS detector}'',} \textit{ JHEP} \textbf{ 04} (2019) 092,
  \href{http://dx.doi.org/10.1007/JHEP04(2019)092}{\doi{10.1007/JHEP04(2019)092}},
  \href{http://www.arXiv.org/abs/1811.04671}{\texttt{arXiv:1811.04671}}.

\bibitem{Sirunyan:2017guj}
\hrefCMSnoop {}{{CMS Collaboration}, ``{Search for resonant and nonresonant
  Higgs boson pair production in the $\PQb\PAQb\ell\nu\ell\nu$ final state in
  proton-proton collisions at $ \sqrt{s}=13 $ TeV}'',} \textit{ JHEP} \textbf{
  01} (2018) 054,
  \href{http://dx.doi.org/10.1007/JHEP01(2018)054}{\doi{10.1007/JHEP01(2018)054}},
  \href{http://www.arXiv.org/abs/1708.04188}{\texttt{arXiv:1708.04188}}.

\bibitem{Aaboud:2018ftw}
\hrefCMSnoop {}{{ATLAS Collaboration}, ``{Search for Higgs boson pair
  production in the $\gamma\gamma\PQb\PAQb$ final state with 13 TeV $pp$
  collision data collected by the ATLAS experiment}'',} \textit{ JHEP} \textbf{
  11} (2018) 040,
  \href{http://dx.doi.org/10.1007/JHEP11(2018)040}{\doi{10.1007/JHEP11(2018)040}},
  \href{http://www.arXiv.org/abs/1807.04873}{\texttt{arXiv:1807.04873}}.

\bibitem{Sirunyan:2018iwt}
\hrefCMSnoop {}{{CMS Collaboration}, ``{Search for Higgs boson pair production
  in the $\gamma\gamma\PQb\PAQb$ final state in pp collisions at $\sqrt{s}=$ 13
  TeV}'',} \textit{ Phys. Lett. B} \textbf{ 788} (2019) 7,
  \href{http://dx.doi.org/10.1016/j.physletb.2018.10.056}{\doi{10.1016/j.physletb.2018.10.056}},
  \href{http://www.arXiv.org/abs/1806.00408}{\texttt{arXiv:1806.00408}}.

\bibitem{Aaboud:2018ewm}
\hrefCMSnoop {}{{ATLAS Collaboration}, ``{Search for Higgs boson pair
  production in the $\gamma\gamma \PW\PW^{*}$ channel using $pp$ collision data
  recorded at $\sqrt{s} = 13$ TeV with the ATLAS detector}'',} \textit{ Eur.
  Phys. J. C} \textbf{ 78} (2018) 1007,
  \href{http://dx.doi.org/10.1140/epjc/s10052-018-6457-x}{\doi{10.1140/epjc/s10052-018-6457-x}},
  \href{http://www.arXiv.org/abs/1807.08567}{\texttt{arXiv:1807.08567}}.

\bibitem{Aaboud:2018ksn}
\hrefCMSnoop {}{{ATLAS Collaboration}, ``{Search for Higgs boson pair
  production in the $\PW\PW^{(*)}\PW\PW^{(*)}$ decay channel using ATLAS data
  recorded at $\sqrt{s}=13$ TeV}'',} \textit{ JHEP} \textbf{ 05} (2019) 124,
  \href{http://dx.doi.org/10.1007/JHEP05(2019)124}{\doi{10.1007/JHEP05(2019)124}},
  \href{http://www.arXiv.org/abs/1811.11028}{\texttt{arXiv:1811.11028}}.

\bibitem{Aad:2019uzh}
\hrefCMSnoop {}{{ATLAS Collaboration}, ``{Combination of searches for Higgs
  boson pairs in $pp$ collisions at $\sqrt{s} = $13 TeV with the ATLAS
  detector}'',} \textit{ Phys. Lett. B} \textbf{ 800} (2020) 135103,
  \href{http://dx.doi.org/10.1016/j.physletb.2019.135103}{\doi{10.1016/j.physletb.2019.135103}},
  \href{http://www.arXiv.org/abs/1906.02025}{\texttt{arXiv:1906.02025}}.

\bibitem{Sirunyan:2018ayu}
\hrefCMSnoop {}{{CMS Collaboration}, ``{Combination of searches for Higgs boson
  pair production in proton-proton collisions at $\sqrt{s} = $ 13 TeV}'',}
  \textit{ Phys. Rev. Lett.} \textbf{ 122} (2019) 121803,
  \href{http://dx.doi.org/10.1103/PhysRevLett.122.121803}{\doi{10.1103/PhysRevLett.122.121803}},
  \href{http://www.arXiv.org/abs/1811.09689}{\texttt{arXiv:1811.09689}}.

\bibitem{Aad:2019yxi}
\hrefCMSnoop {}{{ATLAS Collaboration}, ``{Search for non-resonant Higgs boson
  pair production in the $\PQb\PAQb\ell\nu\ell\nu$ final state with the ATLAS
  detector in $pp$ collisions at $\sqrt{s} = 13$ TeV}'',} \textit{ Phys. Lett.
  B} \textbf{ 801} (2020) 135145,
  \href{http://dx.doi.org/10.1016/j.physletb.2019.135145}{\doi{10.1016/j.physletb.2019.135145}},
  \href{http://www.arXiv.org/abs/1908.06765}{\texttt{arXiv:1908.06765}}.

\bibitem{Sirunyan:2020xok}
\hrefCMSnoop {}{{CMS Collaboration}, ``{Search for nonresonant Higgs boson pair
  production in final states with two bottom quarks and two photons in
  proton-proton collisions at $ \sqrt{s} $ = 13 TeV}'',} \textit{ JHEP}
  \textbf{ 03} (2021) 257,
  \href{http://dx.doi.org/10.1007/JHEP03(2021)257}{\doi{10.1007/JHEP03(2021)257}},
  \href{http://www.arXiv.org/abs/2011.12373}{\texttt{arXiv:2011.12373}}.

\bibitem{ATLAS:2021ifb}
\hrefCMSnoop {}{{ATLAS Collaboration}, ``{Search for Higgs boson pair
  production in the two bottom quarks plus two photons final state in $pp$
  collisions at $\sqrt{s}=13$ TeV with the ATLAS detector}'',} 2021.
  \href{http://www.arXiv.org/abs/2112.11876}{\texttt{arXiv:2112.11876}}.
  Submitted to PRD.

\bibitem{Aad:2020kub}
\hrefCMSnoop {}{{ATLAS Collaboration}, ``{Search for the $\PH\PH \rightarrow
  \PQb\PAQb\PQb\PAQb$ process via vector-boson fusion production using
  proton-proton collisions at $\sqrt{s} = 13$ TeV with the ATLAS detector}'',}
  \textit{ JHEP} \textbf{ 07} (2020) 108,
  \href{http://dx.doi.org/10.1007/JHEP07(2020)108}{\doi{10.1007/JHEP07(2020)108}},
  \href{http://www.arXiv.org/abs/2001.05178}{\texttt{arXiv:2001.05178}}.

\bibitem{deFlorian:2016spz}
\hrefCMSnoop {}{{LHC Higgs Cross Section Working Group}, ``Handbook of {LHC
  Higgs} cross sections: 4. {D}eciphering the nature of the {Higgs} sector'',}
  \textit{ CERN} (2016)
  \href{http://dx.doi.org/10.23731/CYRM-2017-002}{\doi{10.23731/CYRM-2017-002}},
\href{http://www.arXiv.org/abs/1610.07922}{\texttt{arXiv:1610.07922}}.

\bibitem{Chatrchyan:2008zzk}
\hrefCMSnoop {}{{CMS Collaboration}, ``The {CMS} experiment at the {CERN}
  {LHC}'',} \textit{ JINST} \textbf{ 3} (2008) S08004,
  \href{http://dx.doi.org/10.1088/1748-0221/3/08/S08004}{\doi{10.1088/1748-0221/3/08/S08004}}.

\bibitem{CMS:2020cmk}
\hrefCMSnoop {}{{CMS Collaboration}, ``{Performance of the CMS Level-1 trigger
  in proton-proton collisions at $\sqrt{s} =$ 13 TeV}'',} \textit{ JINST}
  \textbf{ 15} (2020) P10017,
  \href{http://dx.doi.org/10.1088/1748-0221/15/10/P10017}{\doi{10.1088/1748-0221/15/10/P10017}},
  \href{http://www.arXiv.org/abs/2006.10165}{\texttt{arXiv:2006.10165}}.

\bibitem{Khachatryan:2016bia}
\hrefCMSnoop {}{{CMS Collaboration}, ``{The CMS trigger system}'',} \textit{
  JINST} \textbf{ 12} (2017) P01020,
  \href{http://dx.doi.org/10.1088/1748-0221/12/01/P01020}{\doi{10.1088/1748-0221/12/01/P01020}},
\href{http://www.arXiv.org/abs/1609.02366}{\texttt{arXiv:1609.02366}}.

\bibitem{CMS:2015xaf}
\hrefCMSnoop {}{{CMS Collaboration}, ``{Performance of electron reconstruction
  and selection with the CMS detector in proton-proton collisions at $\sqrt{s}
  = 8$\TeV}'',} \textit{ JINST} \textbf{ 10} (2015) P06005,
  \href{http://dx.doi.org/10.1088/1748-0221/10/06/P06005}{\doi{10.1088/1748-0221/10/06/P06005}},
\href{http://www.arXiv.org/abs/1502.02701}{\texttt{arXiv:1502.02701}}.

\bibitem{CMS:2020uim}
\hrefCMSnoop {}{{CMS Collaboration}, ``{Electron and photon reconstruction and
  identification with the CMS experiment at the CERN LHC}'',} \textit{ JINST}
  \textbf{ 16} (2021) P05014,
  \href{http://dx.doi.org/10.1088/1748-0221/16/05/P05014}{\doi{10.1088/1748-0221/16/05/P05014}},
  \href{http://www.arXiv.org/abs/2012.06888}{\texttt{arXiv:2012.06888}}.

\bibitem{CMS:2018rym}
\hrefCMSnoop {}{{CMS Collaboration}, ``{Performance of the CMS muon detector
  and muon reconstruction with proton-proton collisions at $\sqrt{s}=$ 13
  TeV}'',} \textit{ JINST} \textbf{ 13} (2018) P06015,
  \href{http://dx.doi.org/10.1088/1748-0221/13/06/P06015}{\doi{10.1088/1748-0221/13/06/P06015}},
  \href{http://www.arXiv.org/abs/1804.04528}{\texttt{arXiv:1804.04528}}.

\bibitem{CMS:2015myp}
\hrefCMSnoop {}{{CMS Collaboration}, ``{Performance of photon reconstruction
  and identification with the CMS Detector in proton-proton collisions at
  $\sqrt{s}$ = 8 TeV}'',} \textit{ JINST} \textbf{ 10} (2015) P08010,
  \href{http://dx.doi.org/10.1088/1748-0221/10/08/P08010}{\doi{10.1088/1748-0221/10/08/P08010}},
  \href{http://www.arXiv.org/abs/1502.02702}{\texttt{arXiv:1502.02702}}.

\bibitem{CMS:2014pgm}
\hrefCMSnoop {}{{CMS Collaboration}, ``{Description and performance of track
  and primary-vertex reconstruction with the CMS tracker}'',} \textit{ JINST}
  \textbf{ 9} (2014) P10009,
  \href{http://dx.doi.org/10.1088/1748-0221/9/10/P10009}{\doi{10.1088/1748-0221/9/10/P10009}},
  \href{http://www.arXiv.org/abs/1405.6569}{\texttt{arXiv:1405.6569}}.

\bibitem{Sirunyan:2017ulk}
\hrefCMSnoop {}{{CMS Collaboration}, ``{Particle-flow reconstruction and global
  event description with the CMS detector}'',} \textit{ JINST} \textbf{ 12}
  (2017) P10003,
  \href{http://dx.doi.org/10.1088/1748-0221/12/10/P10003}{\doi{10.1088/1748-0221/12/10/P10003}},
  \href{http://www.arXiv.org/abs/1706.04965}{\texttt{arXiv:1706.04965}}.

\bibitem{CMS:2018jrd}
\hrefCMSnoop {}{{CMS Collaboration}, ``{Performance of reconstruction and
  identification of $\tau$ leptons decaying to hadrons and $\nu_\tau$ in pp
  collisions at $\sqrt{s}=$ 13 TeV}'',} \textit{ JINST} \textbf{ 13} (2018)
  P10005,
  \href{http://dx.doi.org/10.1088/1748-0221/13/10/P10005}{\doi{10.1088/1748-0221/13/10/P10005}},
  \href{http://www.arXiv.org/abs/1809.02816}{\texttt{arXiv:1809.02816}}.

\bibitem{Khachatryan:2016kdb}
\hrefCMSnoop {}{{CMS Collaboration}, ``{Jet energy scale and resolution in the
  CMS experiment in pp collisions at 8 TeV}'',} \textit{ JINST} \textbf{ 12}
  (2017) P02014,
  \href{http://dx.doi.org/10.1088/1748-0221/12/02/P02014}{\doi{10.1088/1748-0221/12/02/P02014}},
  \href{http://www.arXiv.org/abs/1607.03663}{\texttt{arXiv:1607.03663}}.

\bibitem{CMS:2019ctu}
\hrefCMSnoop {}{{CMS Collaboration}, ``Performance of missing transverse
  momentum reconstruction in proton-proton collisions at $\sqrt{s} = 13$\,{TeV}
  using the {CMS} detector'',} \textit{ JINST} \textbf{ 14} (2019) P07004,
  \href{http://dx.doi.org/10.1088/1748-0221/14/07/P07004}{\doi{10.1088/1748-0221/14/07/P07004}},
\href{http://www.arXiv.org/abs/1903.06078}{\texttt{arXiv:1903.06078}}.

\bibitem{Cacciari:2008gp}
\hrefCMSnoop {}{M.~Cacciari, G.~P. Salam, and G.~Soyez, ``{The anti-\kt jet
  clustering algorithm}'',} \textit{ JHEP} \textbf{ 04} (2008) 063,
  \href{http://dx.doi.org/10.1088/1126-6708/2008/04/063}{\doi{10.1088/1126-6708/2008/04/063}},
  \href{http://www.arXiv.org/abs/0802.1189}{\texttt{arXiv:0802.1189}}.

\bibitem{Cacciari:2011ma}
\hrefCMSnoop {}{M.~Cacciari, G.~P. Salam, and G.~Soyez, ``{FastJet user
  manual}'',} \textit{ Eur. Phys. J. C} \textbf{ 72} (2012) 1896,
  \href{http://dx.doi.org/10.1140/epjc/s10052-012-1896-2}{\doi{10.1140/epjc/s10052-012-1896-2}},
\href{http://www.arXiv.org/abs/1111.6097}{\texttt{arXiv:1111.6097}}.

\bibitem{Sirunyan:2017ezt}
\hrefCMSnoop {}{{CMS Collaboration}, ``{Identification of heavy-flavour jets
  with the CMS detector in pp collisions at 13 TeV}'',} \textit{ JINST}
  \textbf{ 13} (2018) P05011,
  \href{http://dx.doi.org/10.1088/1748-0221/13/05/P05011}{\doi{10.1088/1748-0221/13/05/P05011}},
  \href{http://www.arXiv.org/abs/1712.07158}{\texttt{arXiv:1712.07158}}.

\bibitem{Bols:2020bkb}
E.~Bols\hrefCMSnoop {}{ {et~al.}, ``{Jet flavour classification using
  DeepJet}'',} \textit{ JINST} \textbf{ 15} (2020) P12012,
  \href{http://dx.doi.org/10.1088/1748-0221/15/12/P12012}{\doi{10.1088/1748-0221/15/12/P12012}},
  \href{http://www.arXiv.org/abs/2008.10519}{\texttt{arXiv:2008.10519}}.

\bibitem{Alioli:2008gx}
\hrefCMSnoop {}{S.~Alioli, P.~Nason, C.~Oleari, and E.~Re, ``{NLO vector-boson
  production matched with shower in POWHEG}'',} \textit{ JHEP} \textbf{ 07}
  (2008) 060,
  \href{http://dx.doi.org/10.1088/1126-6708/2008/07/060}{\doi{10.1088/1126-6708/2008/07/060}},
  \href{http://www.arXiv.org/abs/0805.4802}{\texttt{arXiv:0805.4802}}.

\bibitem{Nason:2004rx}
\hrefCMSnoop {}{P.~Nason, ``{A New method for combining NLO QCD with shower
  Monte Carlo algorithms}'',} \textit{ JHEP} \textbf{ 11} (2004) 040,
  \href{http://dx.doi.org/10.1088/1126-6708/2004/11/040}{\doi{10.1088/1126-6708/2004/11/040}},
  \href{http://www.arXiv.org/abs/hep-ph/0409146}{\texttt{arXiv:hep-ph/0409146}}.

\bibitem{Frixione:2007vw}
\hrefCMSnoop {}{S.~Frixione, P.~Nason, and C.~Oleari, ``{Matching NLO QCD
  computations with Parton Shower simulations: the POWHEG method}'',} \textit{
  JHEP} \textbf{ 11} (2007) 070,
  \href{http://dx.doi.org/10.1088/1126-6708/2007/11/070}{\doi{10.1088/1126-6708/2007/11/070}},
  \href{http://www.arXiv.org/abs/0709.2092}{\texttt{arXiv:0709.2092}}.

\bibitem{Alwall:2014hca}
J.~Alwall\hrefCMSnoop {}{ {et~al.}, ``{The automated computation of tree-level
  and next-to-leading order differential cross sections, and their matching to
  parton shower simulations}'',} \textit{ JHEP} \textbf{ 07} (2014) 079,
  \href{http://dx.doi.org/10.1007/JHEP07(2014)079}{\doi{10.1007/JHEP07(2014)079}},
  \href{http://www.arXiv.org/abs/1405.0301}{\texttt{arXiv:1405.0301}}.

\bibitem{Heinrich:2019bkc}
G.~Heinrich\hrefCMSnoop {}{ {et~al.}, ``{Probing the trilinear Higgs boson
  coupling in di-Higgs production at NLO QCD including parton shower
  effects}'',} \textit{ JHEP} \textbf{ 06} (2019) 066,
  \href{http://dx.doi.org/10.1007/JHEP06(2019)066}{\doi{10.1007/JHEP06(2019)066}},
  \href{http://www.arXiv.org/abs/1903.08137}{\texttt{arXiv:1903.08137}}.

\bibitem{Sjostrand:2014zea}
T.~Sj{\"o}strand\hrefCMSnoop {}{ {et~al.}, ``{An introduction to PYTHIA
  8.2}'',} \textit{ Comput. Phys. Commun.} \textbf{ 191} (2015) 159,
  \href{http://dx.doi.org/10.1016/j.cpc.2015.01.024}{\doi{10.1016/j.cpc.2015.01.024}},
  \href{http://www.arXiv.org/abs/1410.3012}{\texttt{arXiv:1410.3012}}.

\bibitem{Agostinelli:2002hh}
\hrefCMSnoop {}{{GEANT4} Collaboration, ``{{\GEANTfour}--a simulation
  toolkit}'',} \textit{ Nucl. Instrum. Meth. A} \textbf{ 506} (2003) 250,
  \href{http://dx.doi.org/10.1016/S0168-9002(03)01368-8}{\doi{10.1016/S0168-9002(03)01368-8}}.

\bibitem{Sirunyan:2020foa}
\hrefCMSnoop {}{{CMS Collaboration}, ``{Pileup mitigation at CMS in 13 TeV
  data}'',} \textit{ JINST} \textbf{ 15} (2020) P09018,
  \href{http://dx.doi.org/10.1088/1748-0221/15/09/P09018}{\doi{10.1088/1748-0221/15/09/P09018}},
  \href{http://www.arXiv.org/abs/2003.00503}{\texttt{arXiv:2003.00503}}.

\bibitem{CMS:2017wyc}
\href {https://cds.cern.ch/record/2256875}{{CMS Collaboration}, ``{Jet
  algorithms performance in 13 TeV data}'',} CMS Physics Analysis Summary
  CMS-PAS-JME-16-003, CERN, 2017.

\bibitem{Sirunyan:2019wwa}
\hrefCMSnoop {}{{CMS Collaboration}, ``A deep neural network for simultaneous
  estimation of b jet energy and resolution'',} \textit{ Comput. Softw. Big
  Sci.} \textbf{ 4} (2020) 10,
  \href{http://dx.doi.org/10.1007/s41781-020-00041-z}{\doi{10.1007/s41781-020-00041-z}},
  \href{http://www.arXiv.org/abs/1912.06046}{\texttt{arXiv:1912.06046}}.

\bibitem{Rogozhnikov:2016bdp}
\hrefCMSnoop {}{A.~Rogozhnikov, ``Reweighting with boosted decision trees'',}
  \textit{ J. Phys. Conf. Ser.} \textbf{ 762} (2016) 012036,
  \href{http://dx.doi.org/10.1088/1742-6596/762/1/012036}{\doi{10.1088/1742-6596/762/1/012036}},
\href{http://www.arXiv.org/abs/1608.05806}{\texttt{arXiv:1608.05806}}.

\bibitem{BAKER1984437}
\hrefCMSnoop {}{S.~Baker and R.~D. Cousins, ``Clarification of the use of
  chi-square and likelihood functions in fits to histograms'',} \textit{ Nucl.
  Instrum. Meth.} \textbf{ 221} (1984) 437,
  \href{http://dx.doi.org/10.1016/0167-5087(84)90016-4}{\doi{10.1016/0167-5087(84)90016-4}}.

\bibitem{Chen:2016:XST:2939672.2939785}
\hrefCMSnoop {}{T.~Chen and C.~Guestrin, ``{XGBoost}: A scalable tree boosting
  system'',} in \textit{ Proceedings of the 22nd ACM SIGKDD International
  Conference on Knowledge Discovery and Data Mining}, KDD '16, p.~785.
\newblock ACM, New York, NY, USA, 2016.
\newblock
  \href{http://dx.doi.org/10.1145/2939672.2939785}{\doi{10.1145/2939672.2939785}}.

\bibitem{Junk:1999kv}
\hrefCMSnoop {}{T.~Junk, ``{Confidence level computation for combining searches
  with small statistics}'',} \textit{ Nucl. Instrum. Meth. A} \textbf{ 434}
  (1999) 435,
  \href{http://dx.doi.org/10.1016/S0168-9002(99)00498-2}{\doi{10.1016/S0168-9002(99)00498-2}},
\href{http://www.arXiv.org/abs/hep-ex/9902006}{\texttt{arXiv:hep-ex/9902006}}.

\bibitem{Read:2002hq}
\hrefCMSnoop {}{A.~L. Read, ``Presentation of search results: the
  $\text{CL}_{\text{s}}$ technique'',} \textit{ J. Phys. G} \textbf{ 28} (2002)
  2693,
\href{http://dx.doi.org/10.1088/0954-3899/28/10/313}{\doi{10.1088/0954-3899/28/10/313}}.

\bibitem{CMS-NOTE-2011-005}
\href {https://cds.cern.ch/record/1379837}{{The ATLAS Collaboration, The CMS
  Collaboration, The LHC Higgs Combination Group}, ``Procedure for the {LHC}
  {Higgs} boson search combination in {Summer} 2011'',} Technical Report
  CMS-NOTE-2011-005, ATL-PHYS-PUB-2011-11, 2011.

\bibitem{Cowan:2010js}
\hrefCMSnoop {}{G.~Cowan, K.~Cranmer, E.~Gross, and O.~Vitells, ``Asymptotic
  formulae for likelihood-based tests of new physics'',} \textit{ Eur. Phys. J.
  C} \textbf{ 71} (2011) 1554,
  \href{http://dx.doi.org/10.1140/epjc/s10052-011-1554-0}{\doi{10.1140/epjc/s10052-011-1554-0}},
  \href{http://www.arXiv.org/abs/1007.1727}{\texttt{arXiv:1007.1727}}.
[Erratum: \DOI{10.1140/epjc/s10052-013-2501-z}].

\bibitem{hepdata}
\hrefCMSnoop {}{``{HEPD}ata record for this analysis'',} 2022.
\newblock
  \href{http://dx.doi.org/10.17182/hepdata.114358}{\doi{10.17182/hepdata.114358}}.

\bibitem{Carvalho:2015ttv}
A.~Carvalho\hrefCMSnoop {}{ {et~al.}, ``Higgs pair production: choosing
  benchmarks with cluster analysis'',} \textit{ JHEP} \textbf{ 04} (2016) 126,
  \href{http://dx.doi.org/10.1007/JHEP04(2016)126}{\doi{10.1007/JHEP04(2016)126}},
  \href{http://www.arXiv.org/abs/1507.02245}{\texttt{arXiv:1507.02245}}.

\bibitem{Buchalla:2018yce}
G.~Buchalla\hrefCMSnoop {}{ {et~al.}, ``Higgs boson pair production in
  non-linear effective field theory with full $m_t$-dependence at {NLO QCD}'',}
  \textit{ JHEP} \textbf{ 09} (2018) 057,
  \href{http://dx.doi.org/10.1007/JHEP09(2018)057}{\doi{10.1007/JHEP09(2018)057}},
  \href{http://www.arXiv.org/abs/1806.05162}{\texttt{arXiv:1806.05162}}.

\bibitem{Alwall:2007fs}
J.~Alwall\hrefCMSnoop {}{ {et~al.}, ``Comparative study of various algorithms
  for the merging of parton showers and matrix elements in hadronic
  collisions'',} \textit{ Eur. Phys. J. C} \textbf{ 53} (2008) 473,
  \href{http://dx.doi.org/10.1140/epjc/s10052-007-0490-5}{\doi{10.1140/epjc/s10052-007-0490-5}},
\href{http://www.arXiv.org/abs/0706.2569}{\texttt{arXiv:0706.2569}}.

\bibitem{Czakon:2011xx}
\hrefCMSnoop {}{M.~Czakon and A.~Mitov, ``{Top++: A program for the calculation
  of the top-pair cross-section at hadron colliders}'',} \textit{ Comput. Phys.
  Commun.} \textbf{ 185} (2014) 2930,
  \href{http://dx.doi.org/10.1016/j.cpc.2014.06.021}{\doi{10.1016/j.cpc.2014.06.021}},
  \href{http://www.arXiv.org/abs/1112.5675}{\texttt{arXiv:1112.5675}}.

\bibitem{Frederix:2012ps}
\hrefCMSnoop {}{R.~Frederix and S.~Frixione, ``{Merging meets matching in
  MC@NLO}'',} \textit{ JHEP} \textbf{ 12} (2012) 061,
  \href{http://dx.doi.org/10.1007/JHEP12(2012)061}{\doi{10.1007/JHEP12(2012)061}},
  \href{http://www.arXiv.org/abs/1209.6215}{\texttt{arXiv:1209.6215}}.

\bibitem{Campbell:2011bn}
\hrefCMSnoop {}{J.~M. Campbell, R.~K. Ellis, and C.~Williams, ``{Vector boson
  pair production at the LHC}'',} \textit{ JHEP} \textbf{ 07} (2011) 018,
  \href{http://dx.doi.org/10.1007/JHEP07(2011)018}{\doi{10.1007/JHEP07(2011)018}},
  \href{http://www.arXiv.org/abs/1105.0020}{\texttt{arXiv:1105.0020}}.

\bibitem{CMS:2015wcf}
\hrefCMSnoop {}{{CMS Collaboration}, ``{Event generator tunes obtained from
  underlying event and multiparton scattering measurements}'',} \textit{ Eur.
  Phys. J. C} \textbf{ 76} (2016) 155,
  \href{http://dx.doi.org/10.1140/epjc/s10052-016-3988-x}{\doi{10.1140/epjc/s10052-016-3988-x}},
  \href{http://www.arXiv.org/abs/1512.00815}{\texttt{arXiv:1512.00815}}.

\bibitem{CMS:2019csb}
\hrefCMSnoop {}{{CMS Collaboration}, ``{Extraction and validation of a new set
  of CMS PYTHIA8 tunes from underlying-event measurements}'',} \textit{ Eur.
  Phys. J. C} \textbf{ 80} (2020) 4,
  \href{http://dx.doi.org/10.1140/epjc/s10052-019-7499-4}{\doi{10.1140/epjc/s10052-019-7499-4}},
  \href{http://www.arXiv.org/abs/1903.12179}{\texttt{arXiv:1903.12179}}.

\bibitem{NNPDF:2014otw}
\hrefCMSnoop {}{{NNPDF} Collaboration, ``{Parton distributions for the LHC Run
  II}'',} \textit{ JHEP} \textbf{ 04} (2015) 040,
  \href{http://dx.doi.org/10.1007/JHEP04(2015)040}{\doi{10.1007/JHEP04(2015)040}},
  \href{http://www.arXiv.org/abs/1410.8849}{\texttt{arXiv:1410.8849}}.

\bibitem{NNPDF:2017mvq}
\hrefCMSnoop {}{{NNPDF} Collaboration, ``{Parton distributions from
  high-precision collider data}'',} \textit{ Eur. Phys. J. C} \textbf{ 77}
  (2017) 663,
  \href{http://dx.doi.org/10.1140/epjc/s10052-017-5199-5}{\doi{10.1140/epjc/s10052-017-5199-5}},
  \href{http://www.arXiv.org/abs/1706.00428}{\texttt{arXiv:1706.00428}}.

\bibitem{CMS-LUM-17-003}
\hrefCMSnoop {}{{CMS Collaboration}, ``{Precision luminosity measurement in
  proton-proton collisions at $\sqrt{s} =$ 13 TeV in 2015 and 2016 at CMS}'',}
  2021. \href{http://www.arXiv.org/abs/2104.01927}{\texttt{arXiv:2104.01927}}.
  Submitted to \textit{Eur. Phys. J. C}.

\bibitem{CMS-PAS-LUM-17-004}
\href {https://cds.cern.ch/record/2621960}{{CMS Collaboration}, ``{CMS
  luminosity measurement for the 2017 data-taking period at $\sqrt{s} =
  13~\mathrm{TeV}$}'',} CMS Physics Analysis Summary CMS-PAS-LUM-17-004, CERN,
  Geneva, 2018.

\bibitem{CMS-PAS-LUM-18-002}
\href {https://cds.cern.ch/record/2676164}{{CMS Collaboration}, ``{CMS
  luminosity measurement for the 2018 data-taking period at $\sqrt{s} =
  13~\mathrm{TeV}$}'',} CMS Physics Analysis Summary CMS-PAS-LUM-18-002, CERN,
  Geneva, 2019.

\bibitem{Jager:2020hkz}
B.~J{\"a}ger\hrefCMSnoop {}{ {et~al.}, ``{Parton-shower effects in Higgs
  production via vector-boson fusion}'',} \textit{ Eur. Phys. J. C} \textbf{
  80} (2020) 756,
  \href{http://dx.doi.org/10.1140/epjc/s10052-020-8326-7}{\doi{10.1140/epjc/s10052-020-8326-7}},
  \href{http://www.arXiv.org/abs/2003.12435}{\texttt{arXiv:2003.12435}}.

\end{thebibliography}\endgroup

\numberwithin{figure}{section}
\ifthenelse{\boolean{cms@external}}{}{
\clearpage\appendix
\section{Simulation details, signal systematic uncertainties, and additional figures\label{app:suppMat}}

\begin{figure*}[p]
\centering
{\includegraphics[width=\cmsTwoWide] {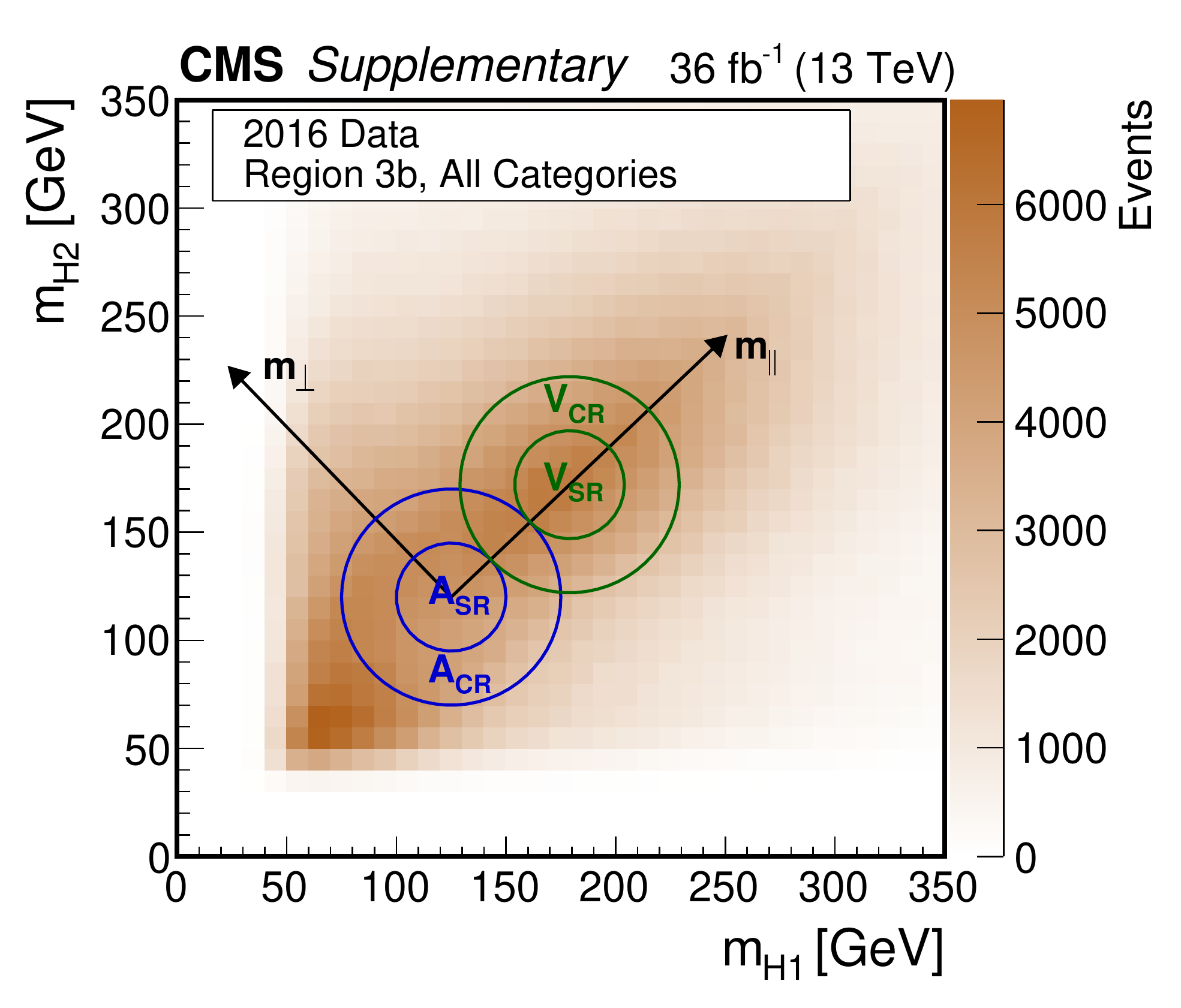}}\hspace{\fill}
{\includegraphics[width=\cmsTwoWide] {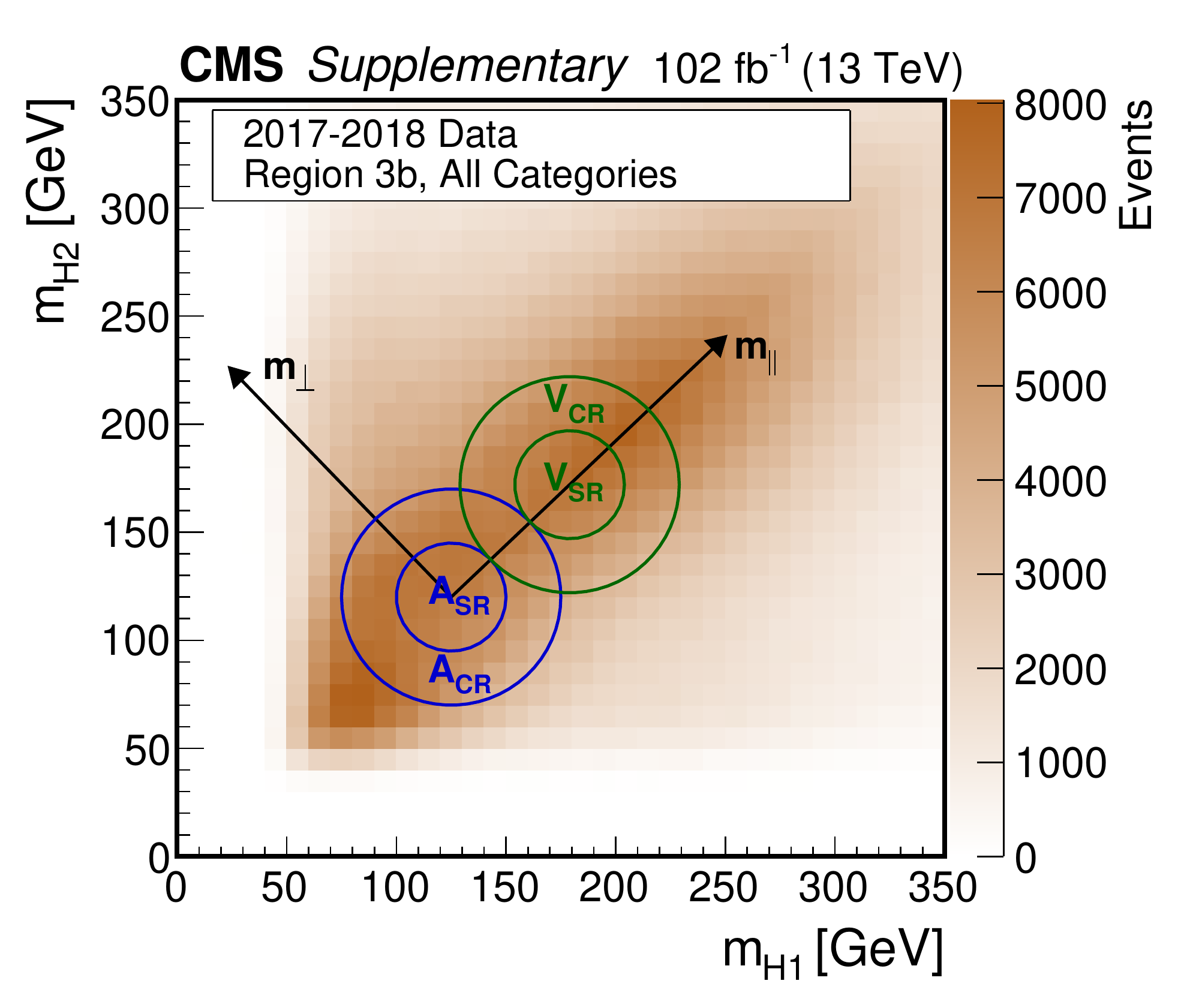}}\\
{\includegraphics[width=\cmsTwoWide] {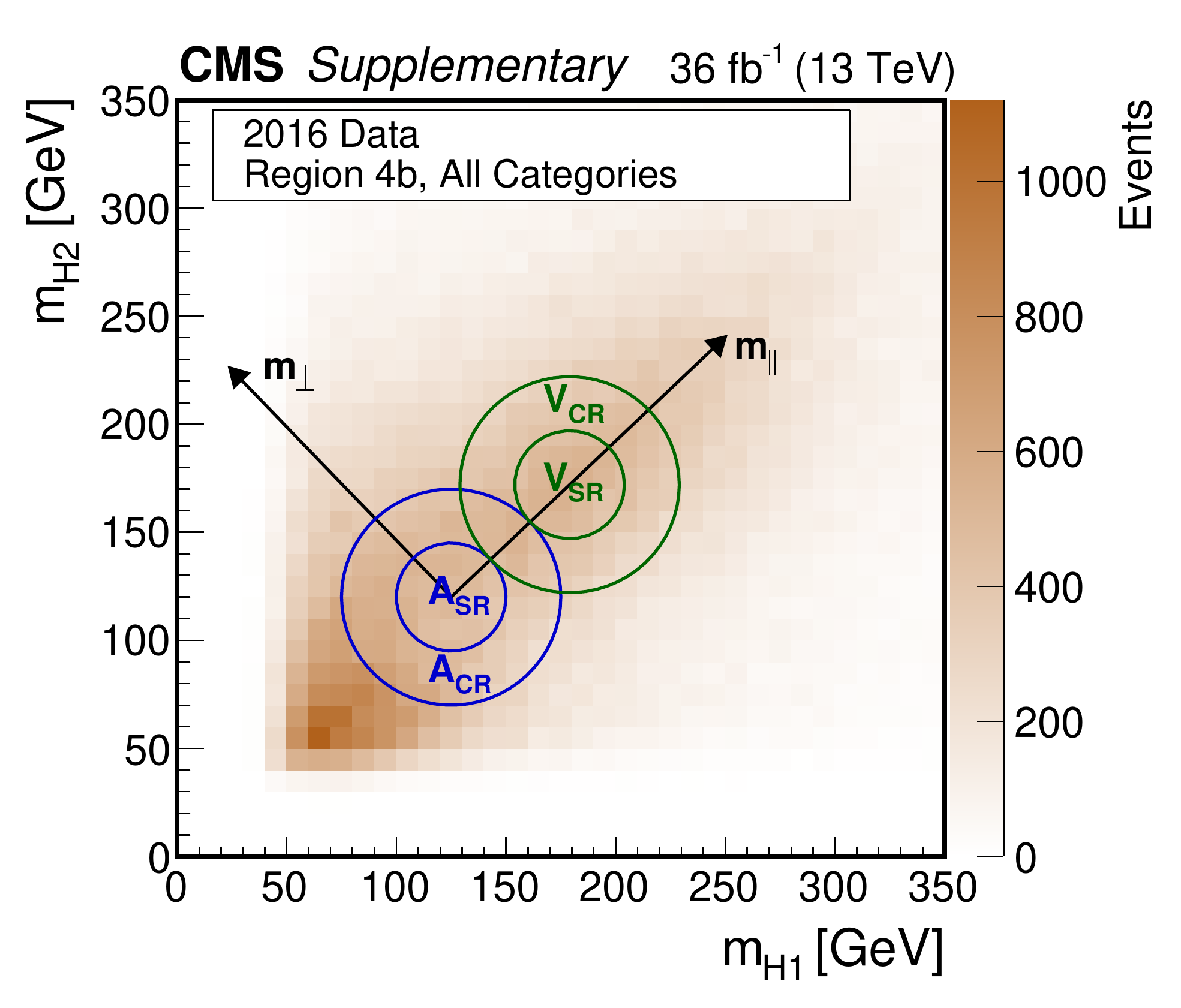}}\hspace{\fill}
{\includegraphics[width=\cmsTwoWide] {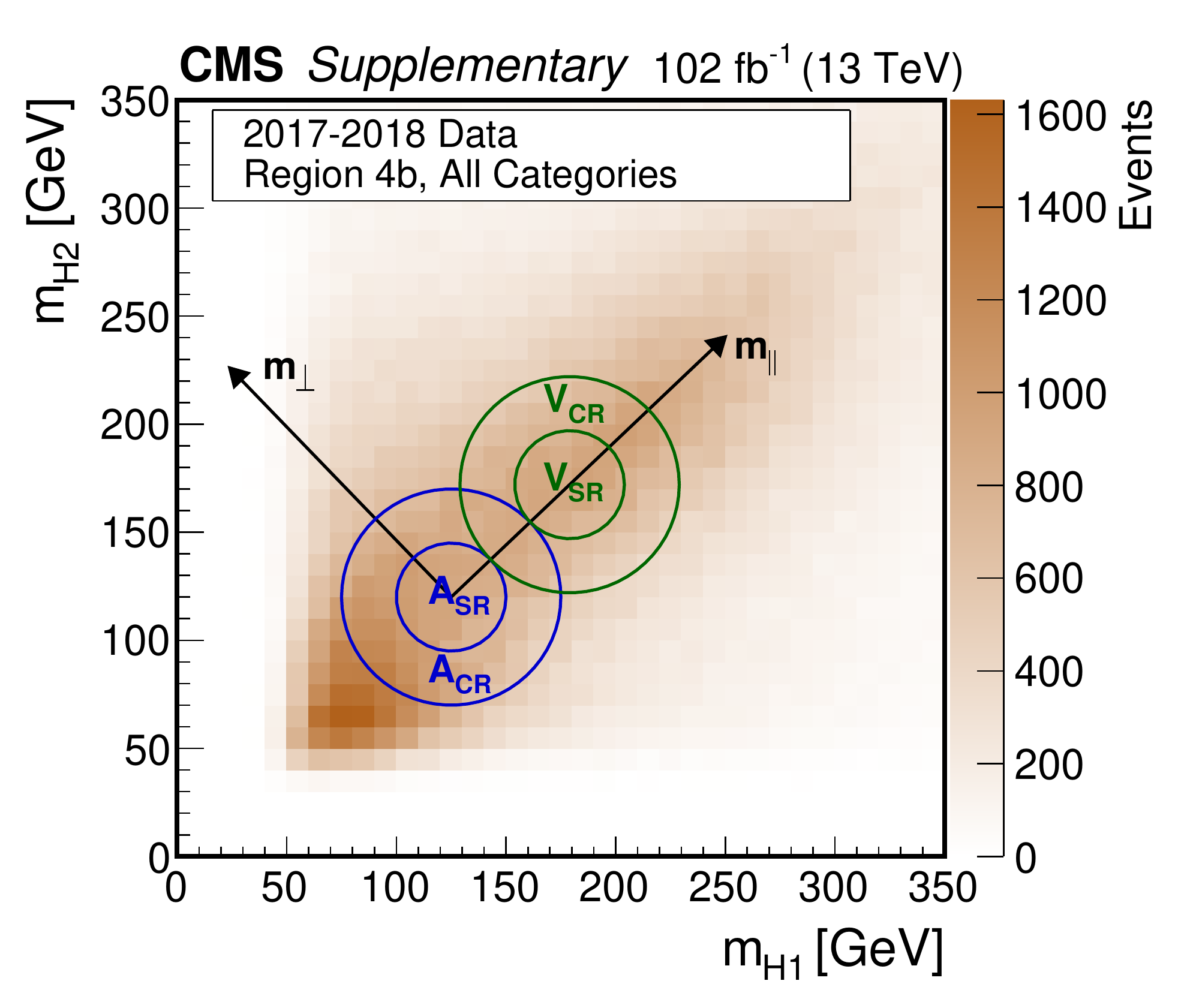}}\\
\caption{\label{fig:mH1-mH2_data}
Distribution of data events selected in the three (top row) and four (bottom row) \PQb-tagged jet categories for events recorded in 2016 (left column) and 2017 and 2018 (right column) as a function of the masses of the two Higgs boson candidates.
The blue and green circles correspond to the analysis and validation regions, respectively, denoted by the letters A and V.
The inner circle corresponds to the signal region while the ring around it defines the control region used for the background modeling. The figure also shows the definition of the variables \mparall and \mperp that are used in the analysis.
}
\end{figure*}

\begin{figure*}[p]
\centering
{\includegraphics[width=\cmsTwoWide] {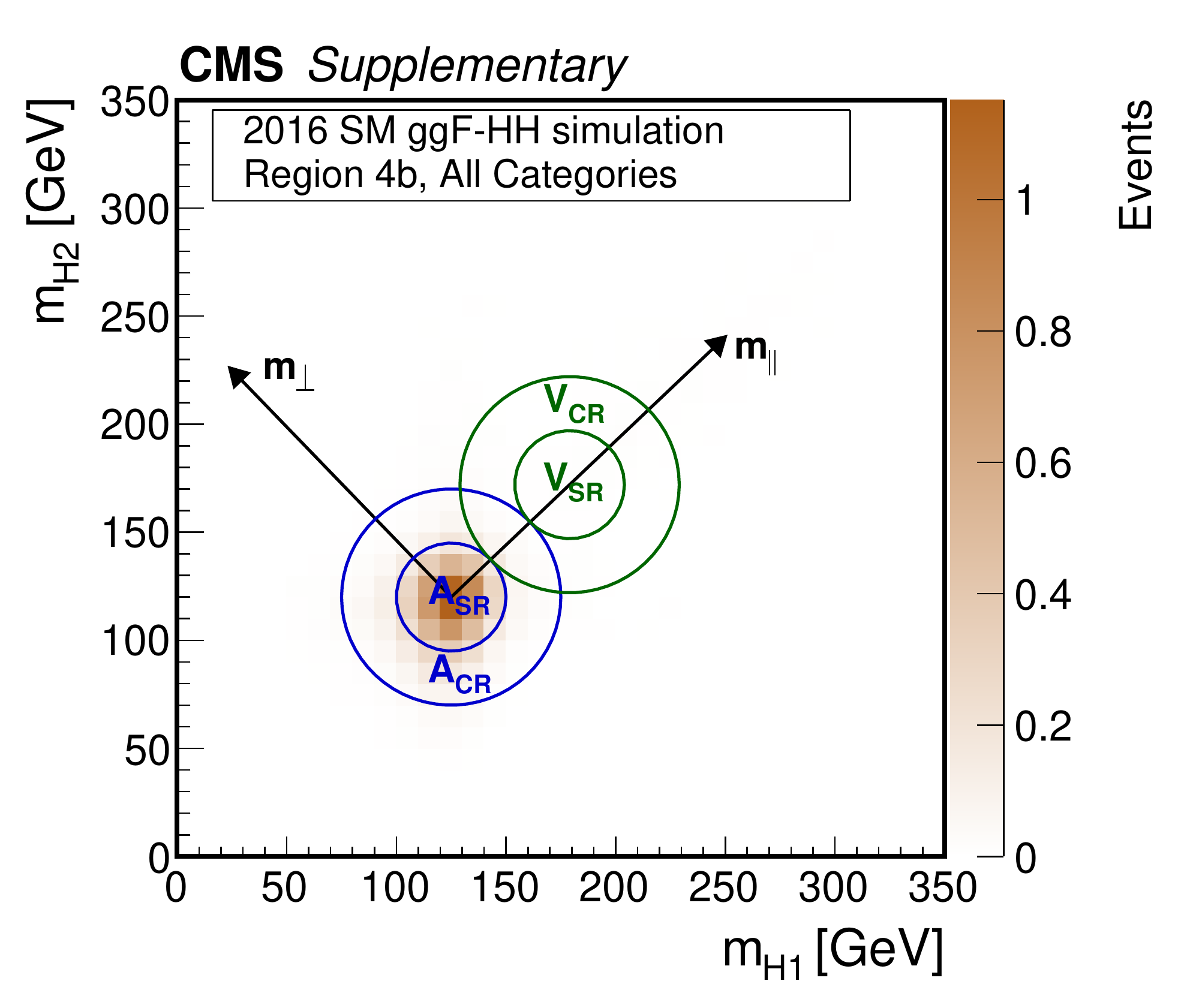}}\hspace{\fill}
{\includegraphics[width=\cmsTwoWide] {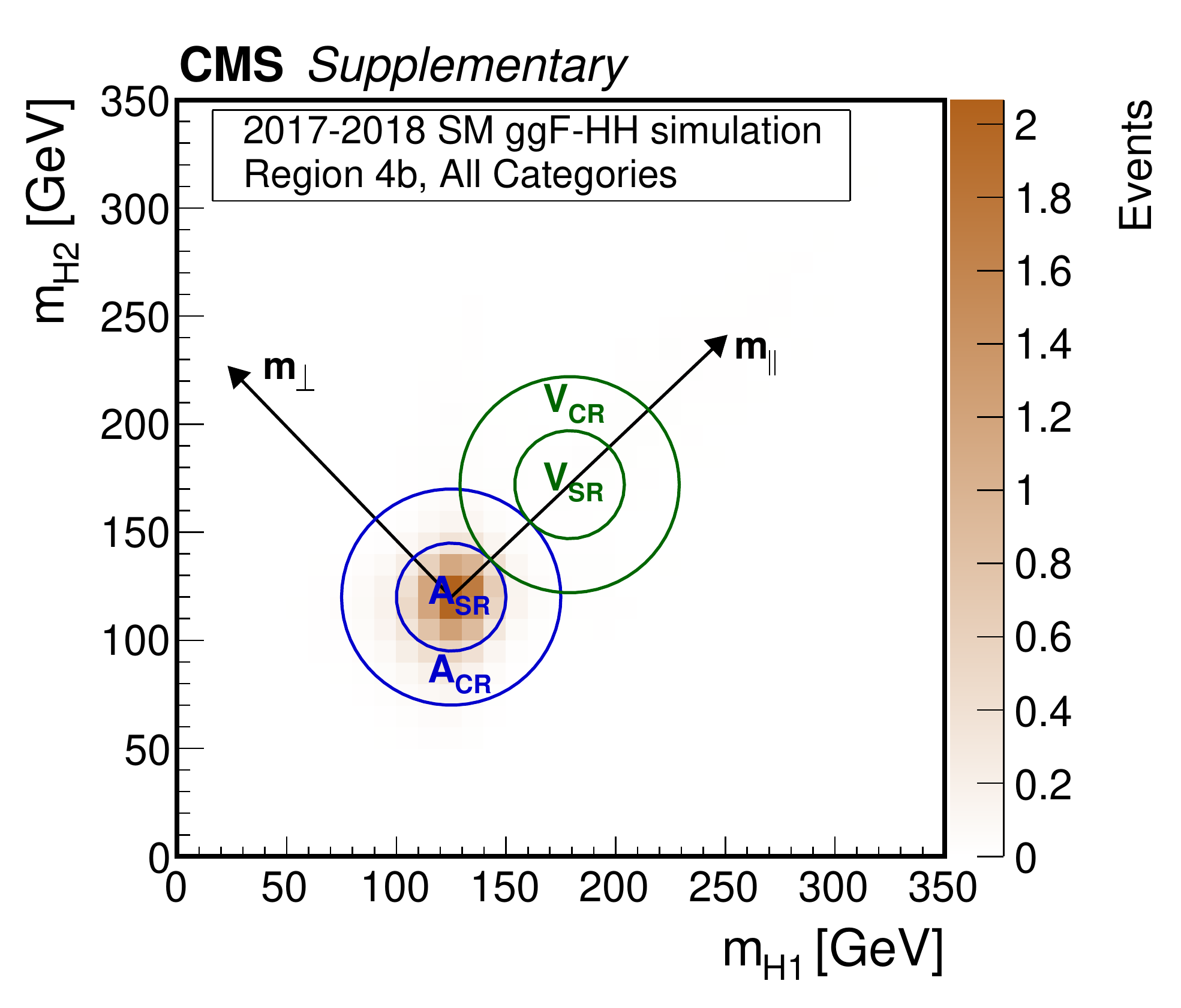}}\\
{\includegraphics[width=\cmsTwoWide] {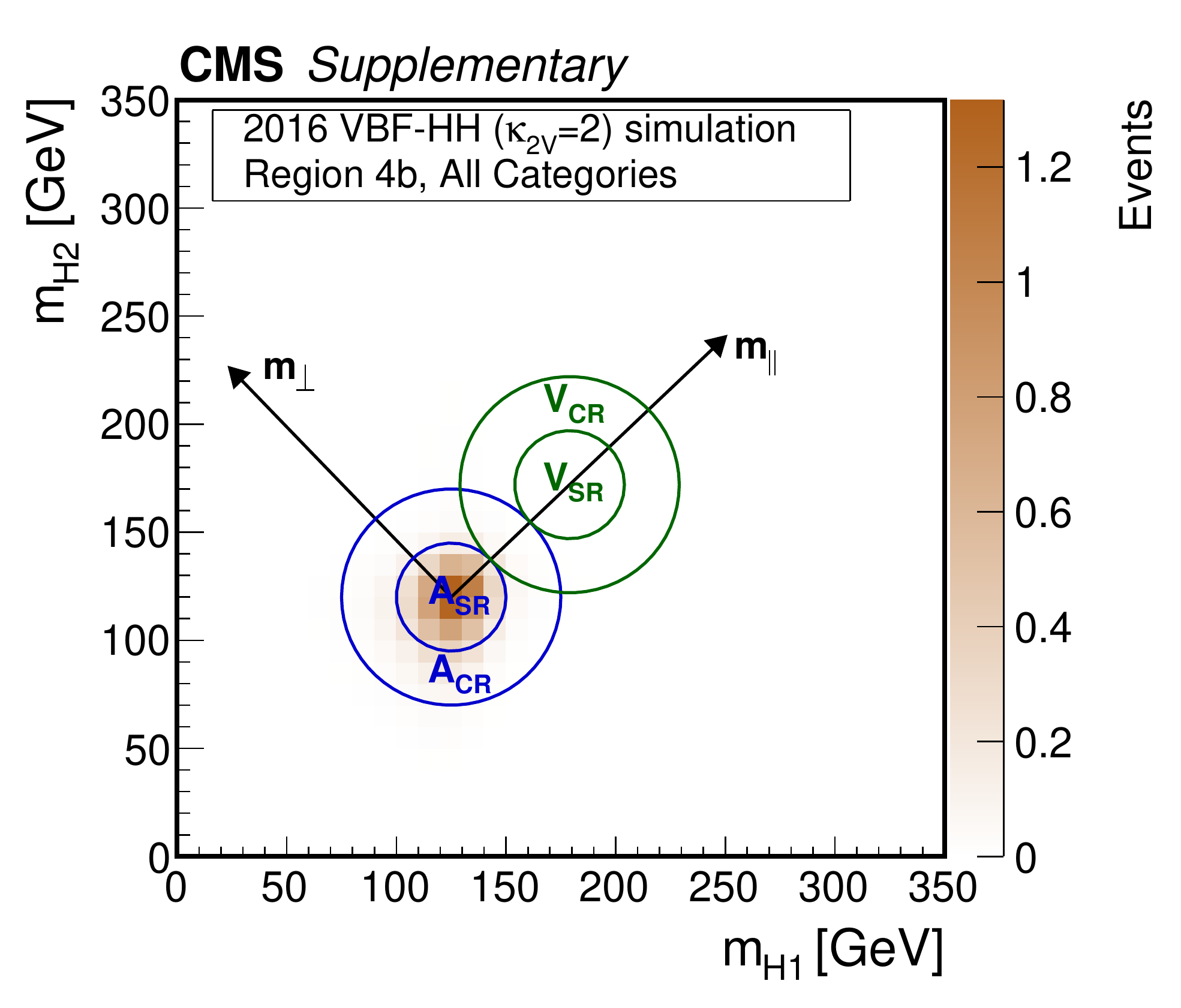}}\hspace{\fill}
{\includegraphics[width=\cmsTwoWide] {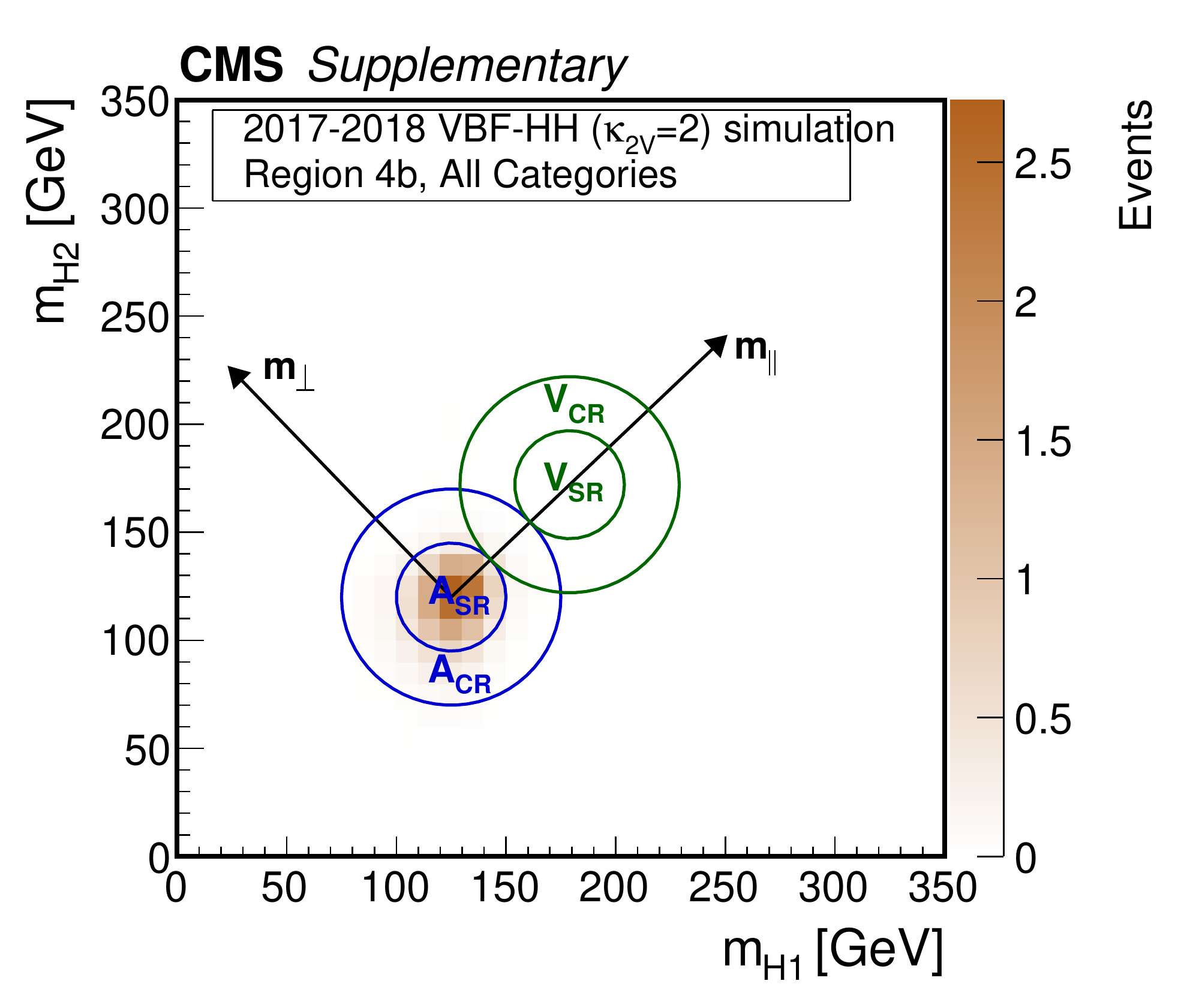}}\\
\caption{\label{fig:mH1-mH2_signal}
Distribution of simulated signal events selected for the SM \GGF production (top row) and $\CVV=2$ \VBF production (bottom row) as a function of the two Higgs boson candidates masses.
The simulations correspond to the events expected for the 2016 data taking (left column) and 2017 and 2018 (right column).
The blue and green circles correspond to the analysis and validation regions, respectively, denoted by the letters A and V.
The inner circle corresponds to the signal region while the ring around it defines the control region used for the background modeling. The figure also shows the definition of the variables \mparall and \mperp that are used in the analysis.
}
\end{figure*}

\begin{figure*}[!htb]
\centering
{\includegraphics[width=\cmsTwoWide] {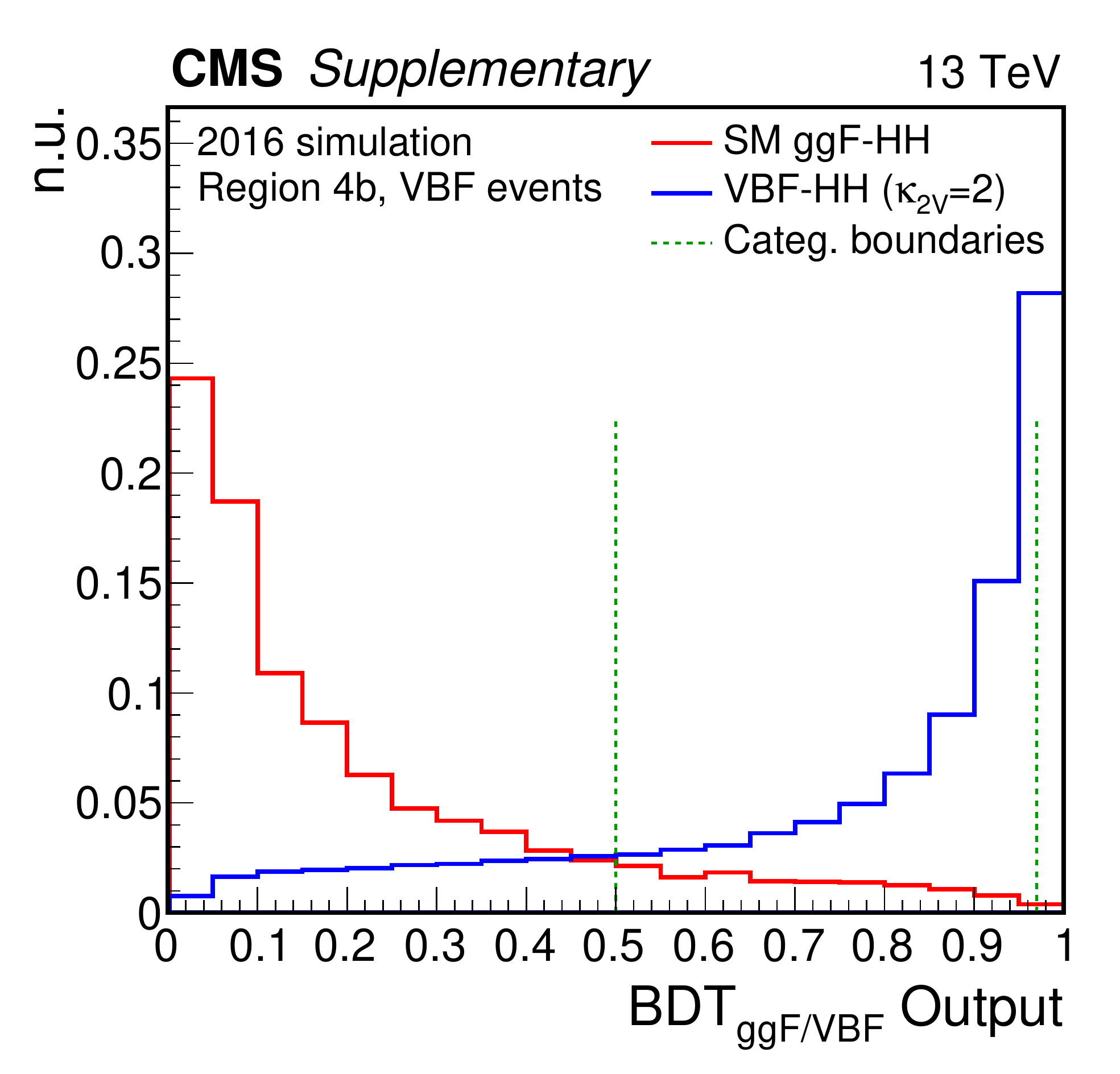}}\hspace{\fill}
{\includegraphics[width=\cmsTwoWide] {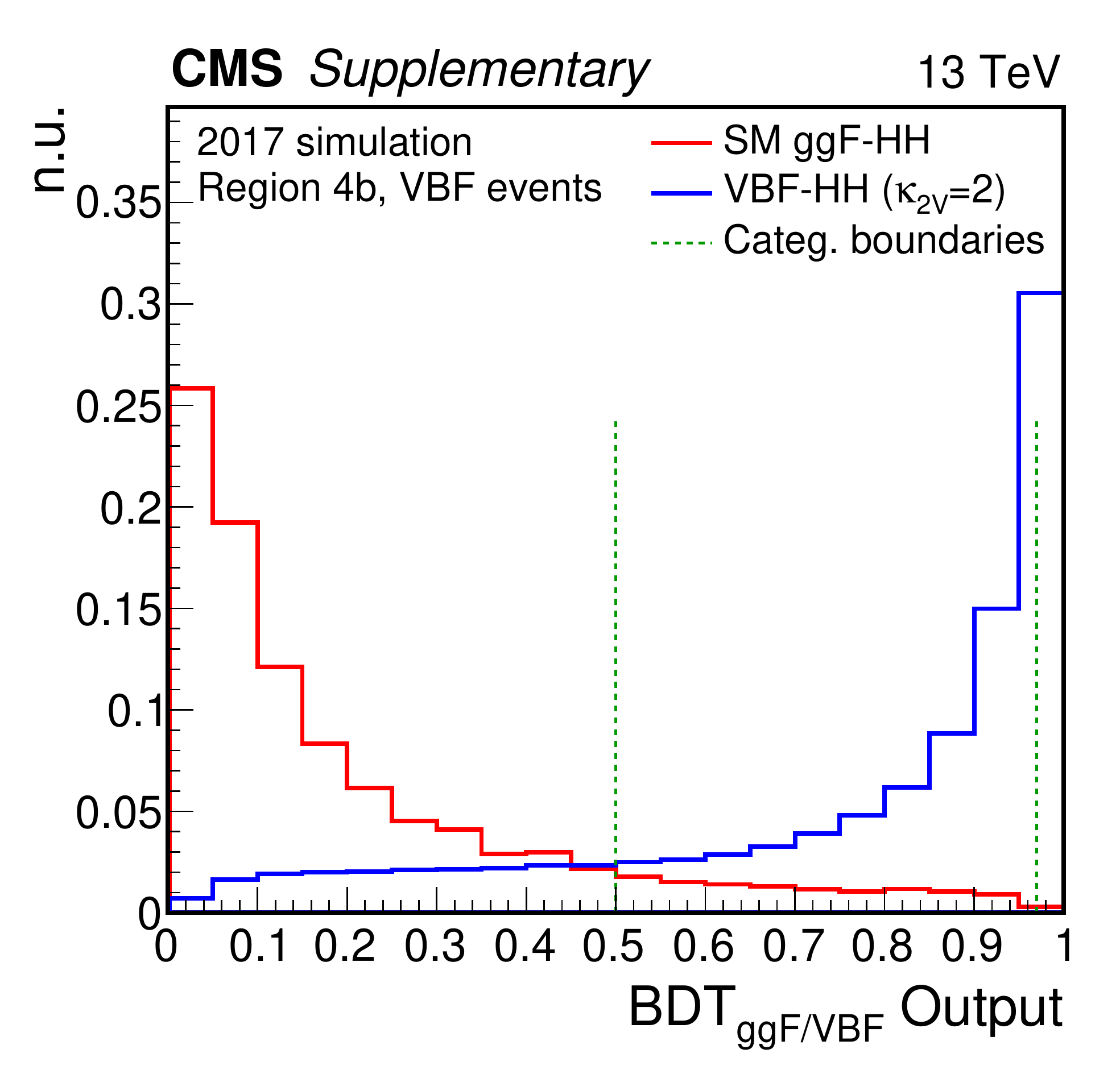}}\\
{\includegraphics[width=\cmsTwoWide] {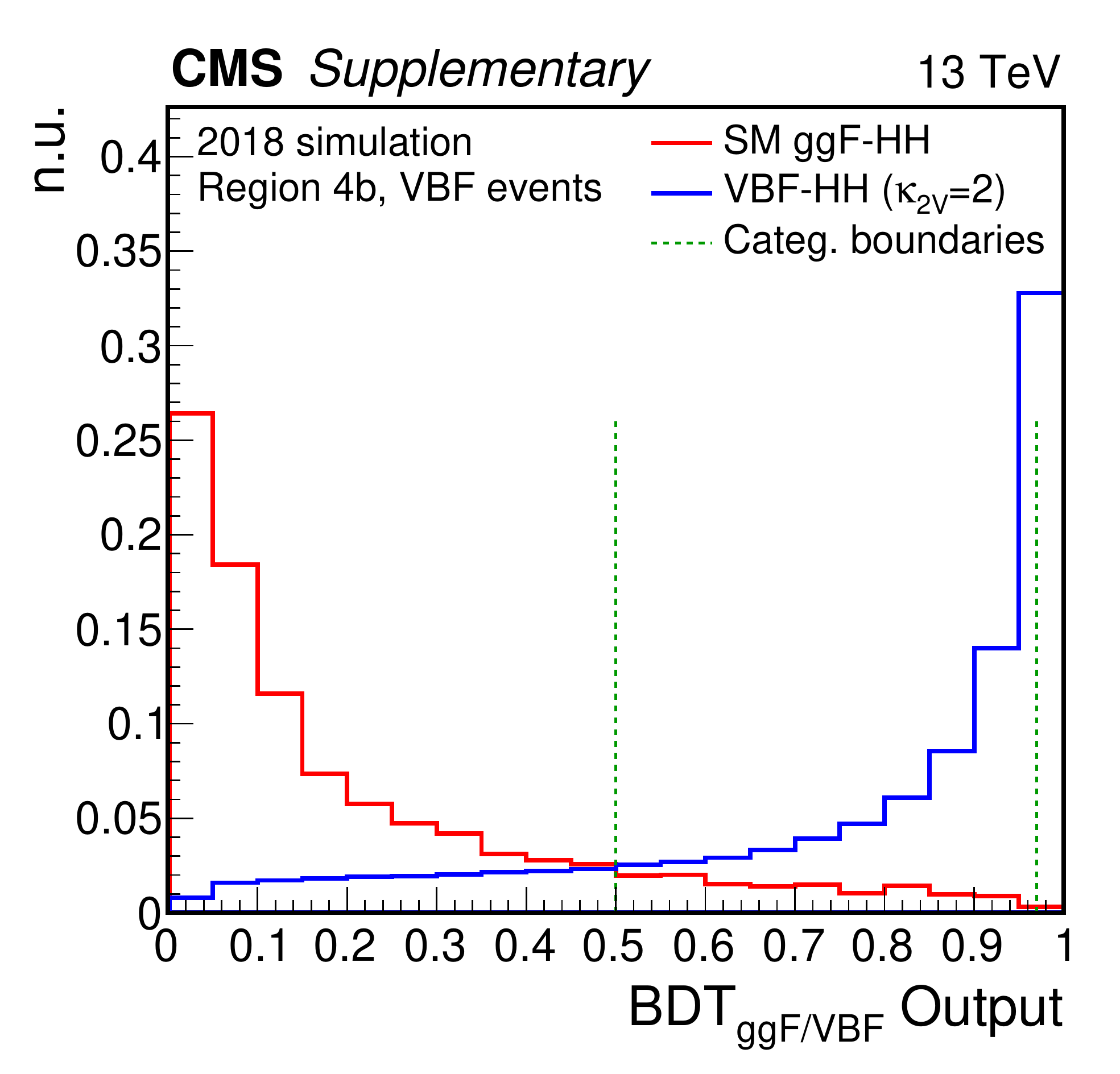}}
\caption{\label{fig:ggfKillerPlots}
Distribution of the output of the production mode BDT discriminant that is trained to separate the \GGF and the \VBF production modes, shown for the simulation of the signal processes under the 2016, 2017, and 2018 data taking conditions.
The vertical green lines denote the thresholds used for the categorization.
Events with a score below 0.5 are classified in the \GGF category.
Events with a score between 0.5 and 0.97 are classified in the \VBF SM-like category.
Events with a score above 0.97 are classified in the \VBF anomalous \CVV-like category.
}
\end{figure*}

\begin{figure*}[!htb]
\centering
{\includegraphics[width=\cmsTwoWide] {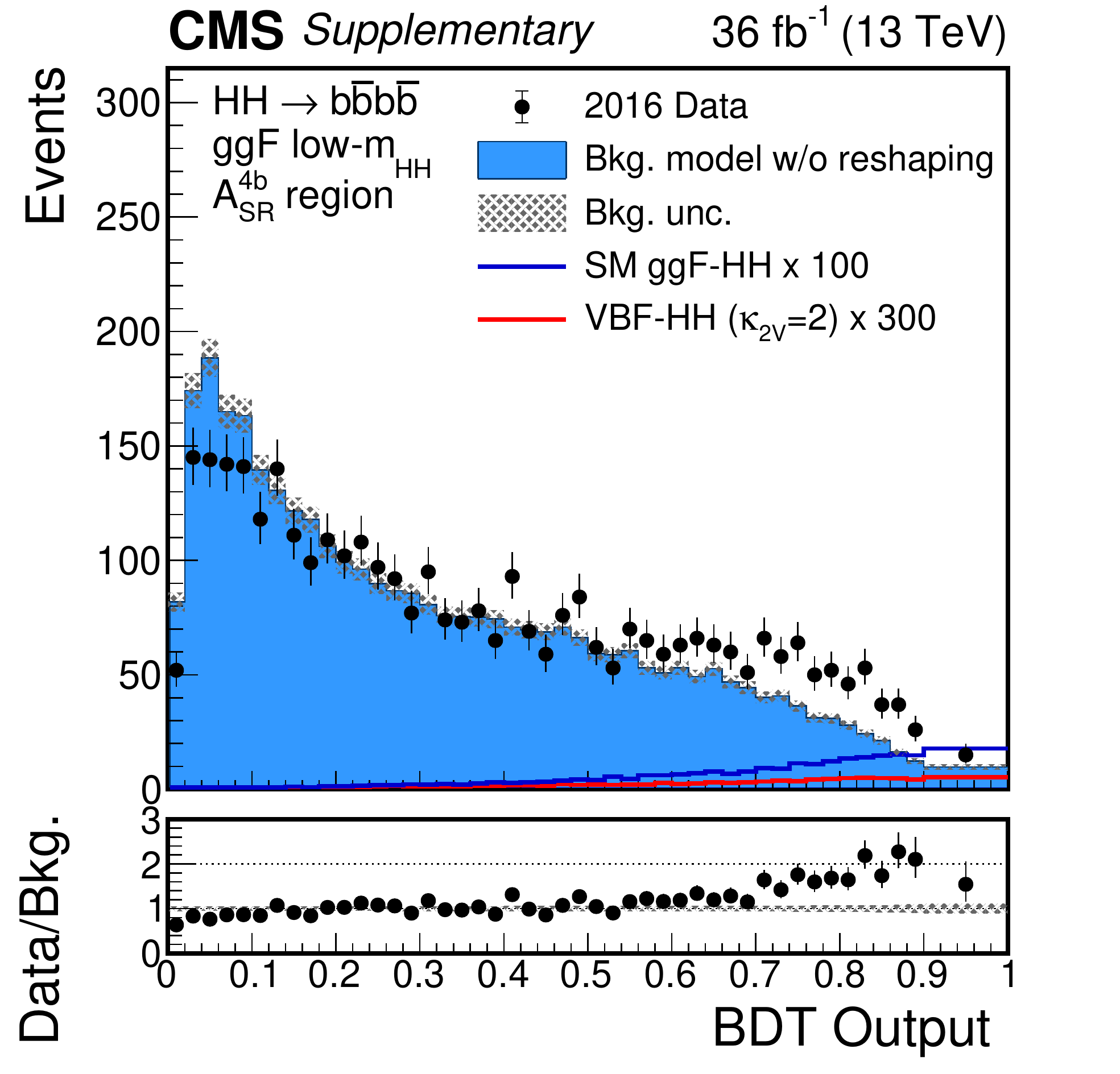}}\hspace{\fill}
{\includegraphics[width=\cmsTwoWide] {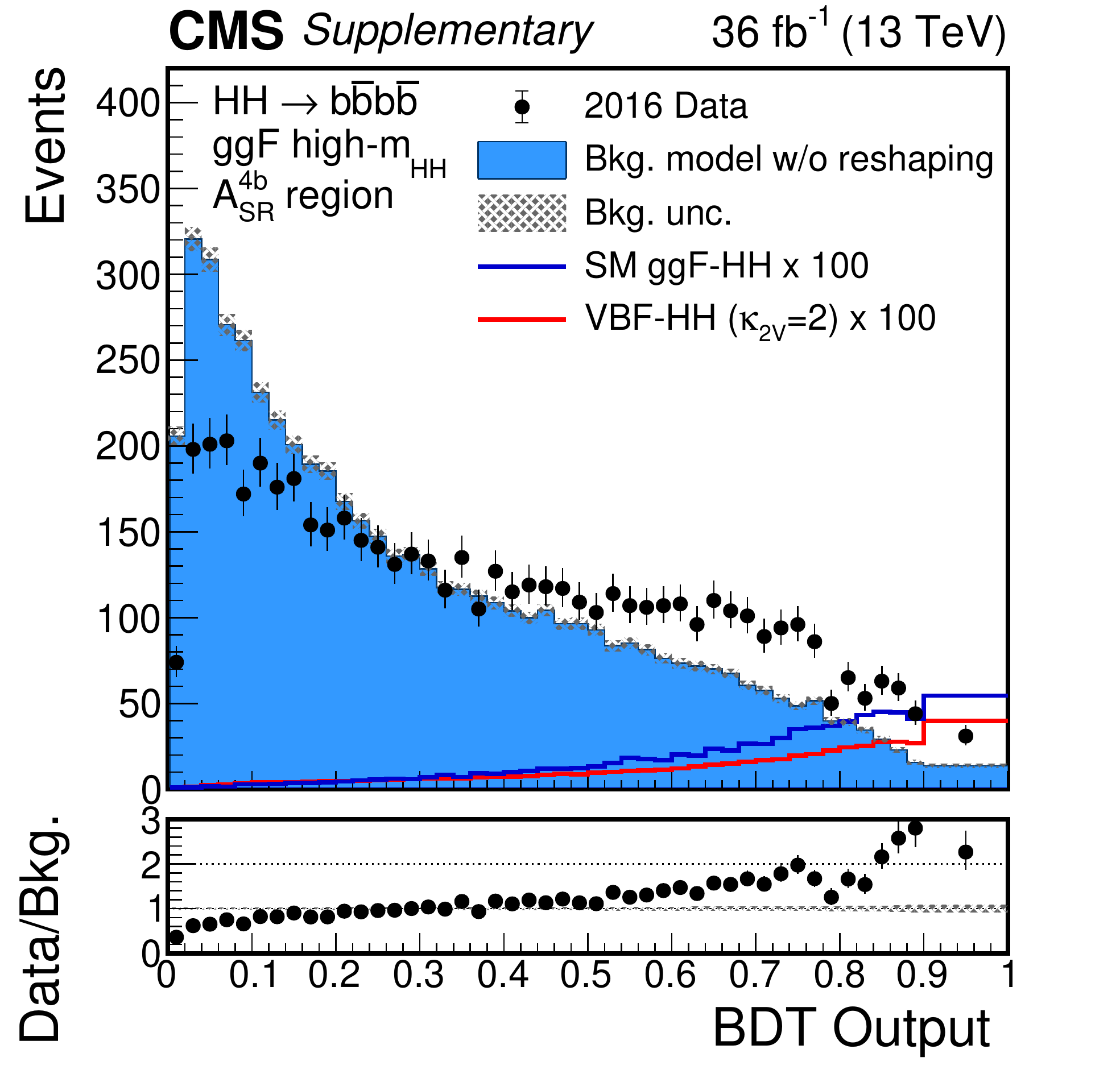}}\\
{\includegraphics[width=\cmsTwoWide] {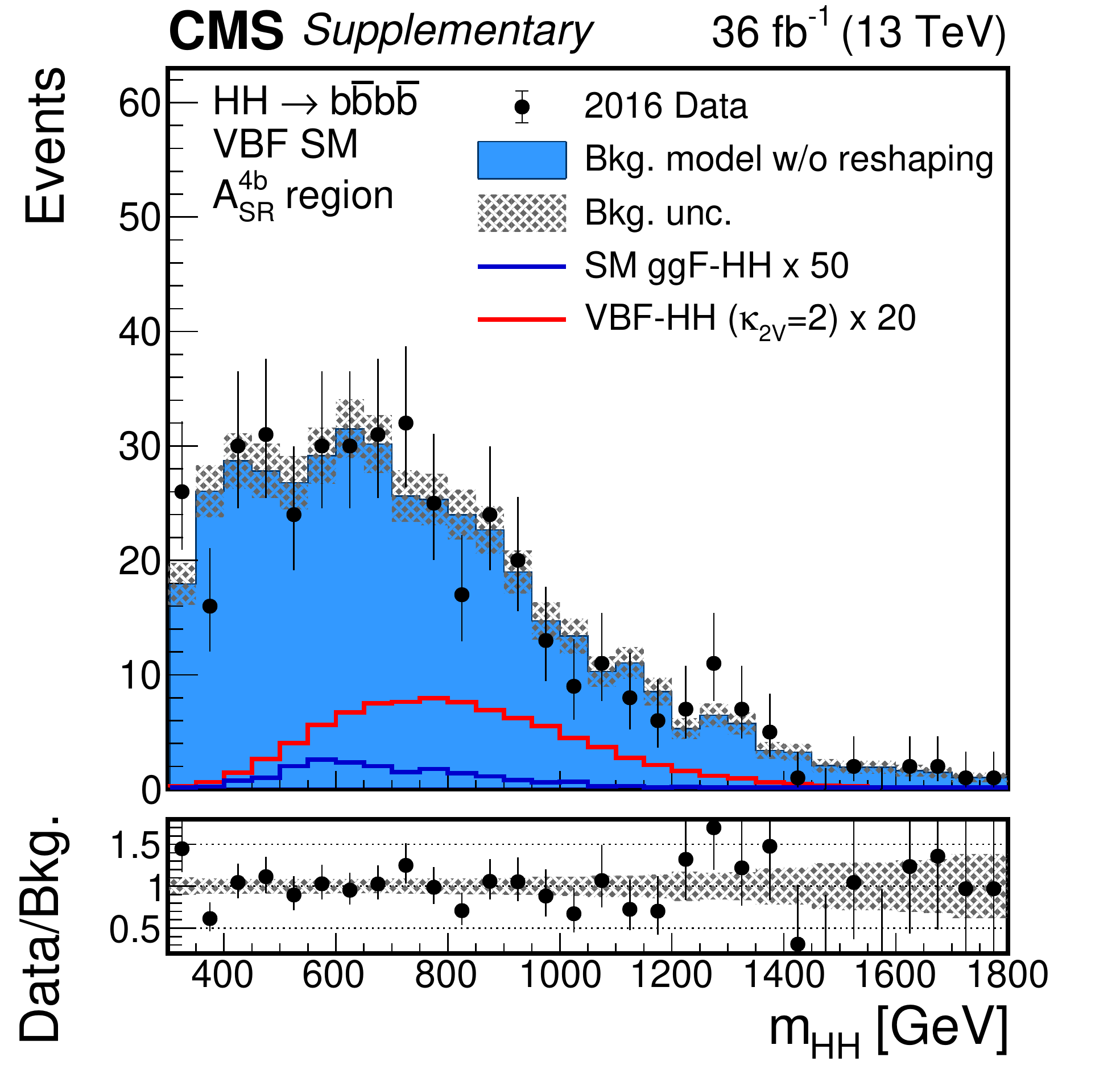}}
\caption{\label{fig:dataMCplots_2016_NoBDTR}
Prediction of the background distribution obtained directly from the \ASRthreeb region without the application of the BDT reweighting correction.
The top and bottom row show the \GGF and \VBF categories, respectively, and the data correspond to the 2016 dataset.
For the former, the output of the BDT discriminant is shown for the low-mass category on the left and for the high-mass category on the right.
For the latter, the $m_{\HH}$ distribution in the SM-like category is shown.
The anomalous \CVV-like category is not shown because no shape correction is applied for it since the overall number of observed events is used to perform a counting experiment.
Data are represented by points with error bars, while the \GGF (\VBF) signal contribution is shown in blue (red) and not stacked.
The background prediction without the shape correction is represented by the shaded blue histograms with the associated systematic uncertainties (gray dashed areas).
}
\end{figure*}

\begin{figure*}[!htb]
\centering
{\includegraphics[width=\cmsTwoWide] {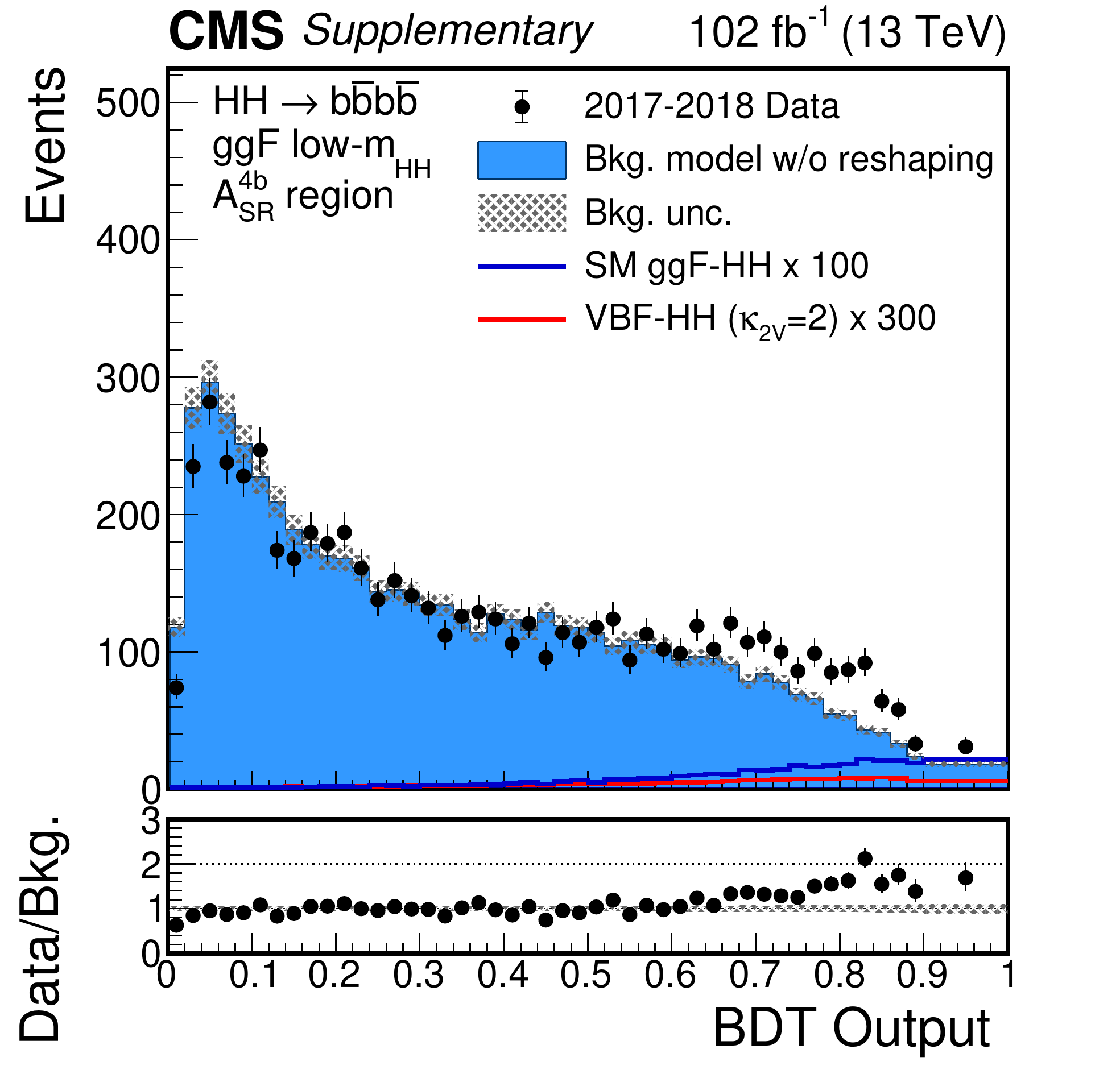}}\hspace{\fill}
{\includegraphics[width=\cmsTwoWide] {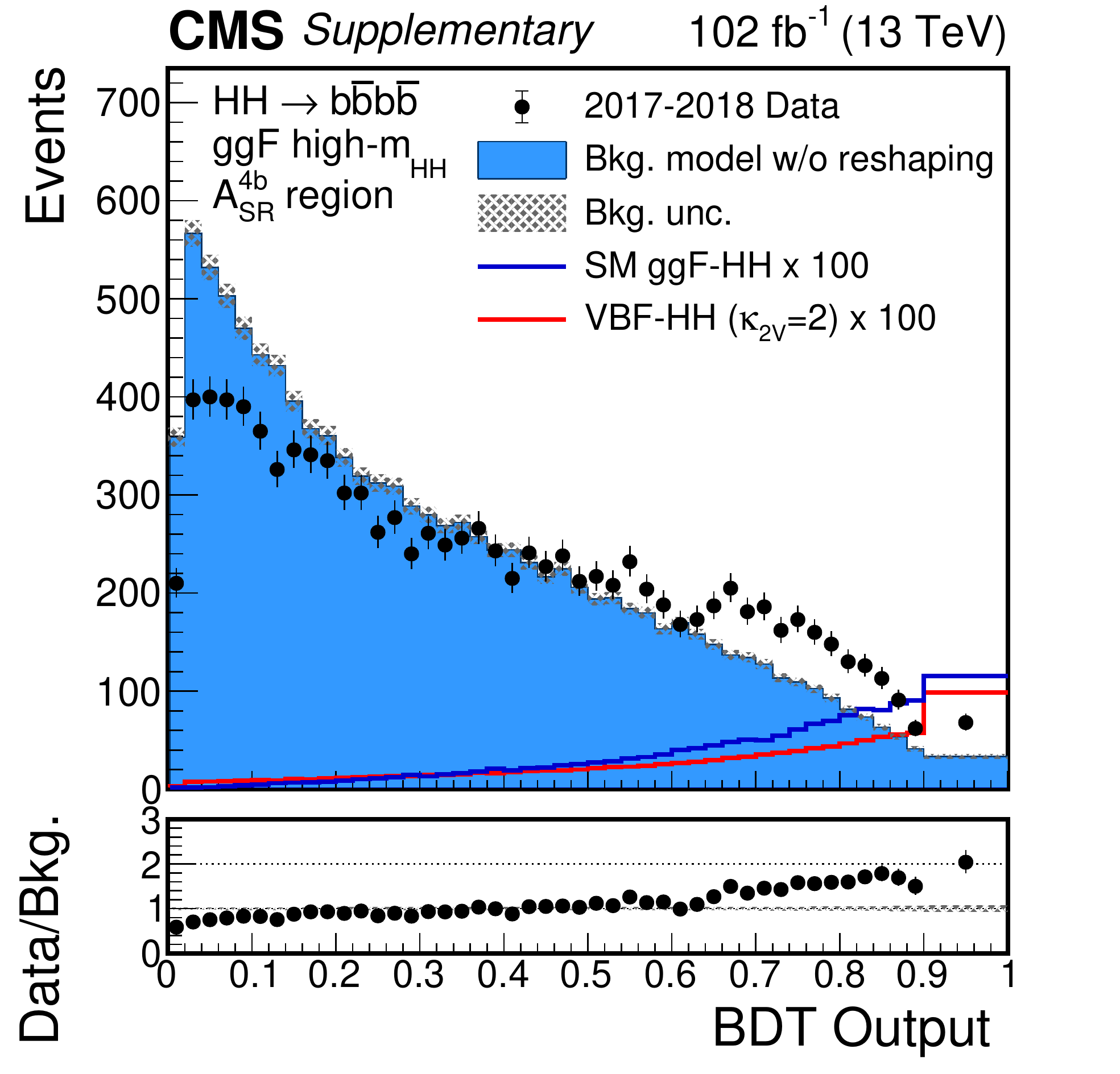}}\\
{\includegraphics[width=\cmsTwoWide] {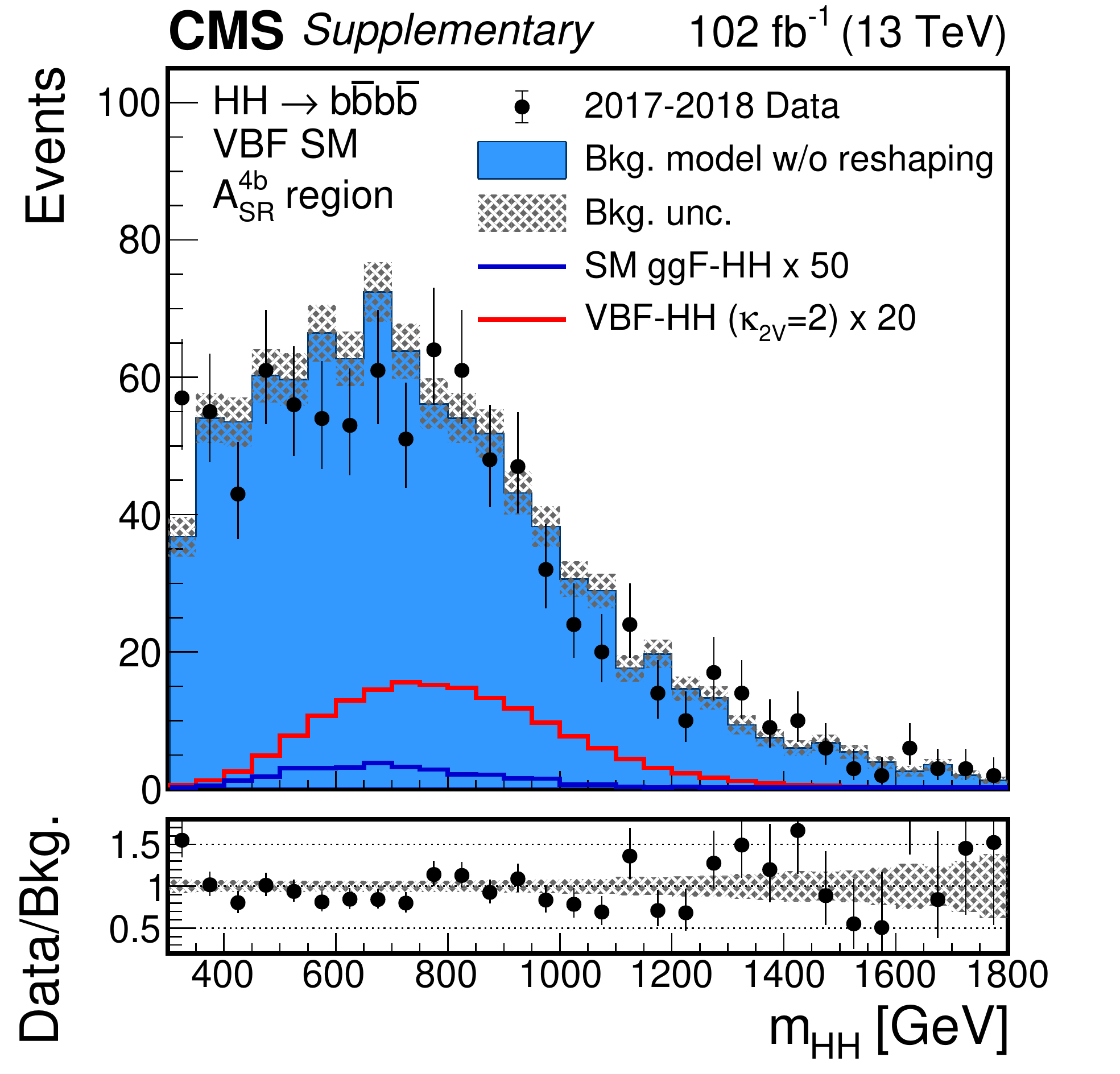}}
\caption{\label{fig:dataMCplots_20172018_NoBDTR}
Prediction of the background distribution obtained directly from the \ASRthreeb region without the application of the BDT reweighting correction.
The top and bottom row show the \GGF and \VBF categories, respectively, and the data correspond to the 2017 and 2018 datasets.
For the former, the output of the BDT discriminant is shown for the low-mass category on the left and for the high-mass category on the right.
For the latter, the $m_{\HH}$ distribution in the SM-like category is shown.
The anomalous \CVV-like category is not shown because no shape correction is applied for it since the overall number of observed events is used to perform a counting experiment.
Data are represented by points with error bars, while the \GGF (\VBF) signal contribution is shown in blue (red) and not stacked.
The background prediction without the shape correction is represented by the shaded blue histograms with the associated systematic uncertainties (gray dashed areas).
}
\end{figure*}

\begin{figure*}[!htb]
\centering
{\includegraphics[width=\cmsTwoWide] {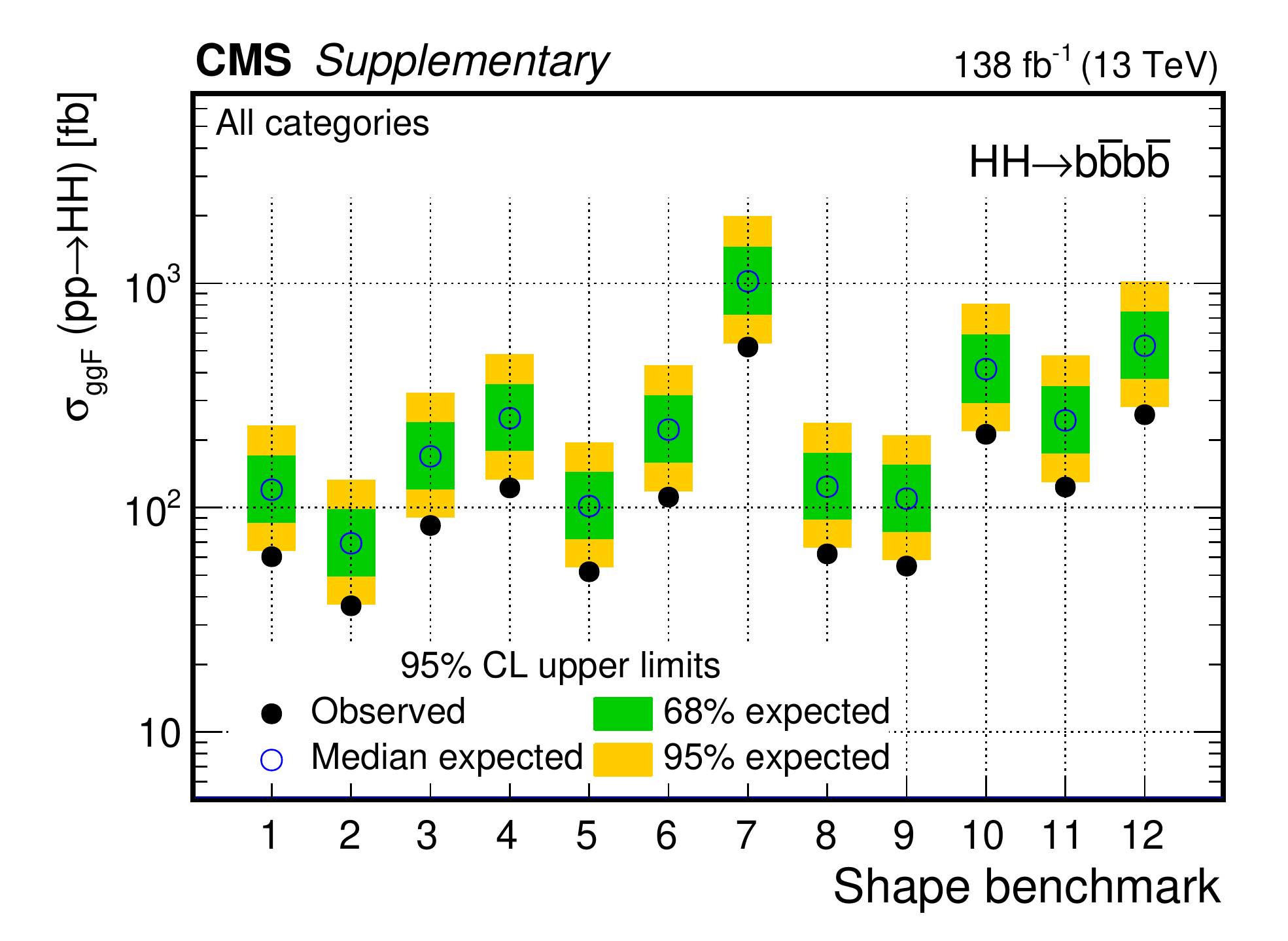}}
\caption{\label{fig:benchmarks}
Observed and expected 95\% CL upper limits on the gluon fusion Higgs boson pair production cross section for different combinations of Higgs boson couplings in the effective Lagrangian parametrization. The numbers from 1 to 12 denote the shape benchmarks defined in Ref.~\cite{Carvalho:2015ttv}, corresponding to points in the five-dimensional parameter space with characteristic kinematic properties of the \HH system. The leading order simulation is reweighted to match the $m_{\HH}$ spectrum at the next-to-leading order as predicted by simulation described in Ref.~\cite{Buchalla:2018yce}. The inner (green) band and the outer (yellow) band indicate the regions containing 68 and 95\%, respectively, of the distribution of limits expected under the background-only hypothesis.
}
\end{figure*}

\begin{figure*}[!htb]
\centering
{\includegraphics[width=\cmsTwoWide] {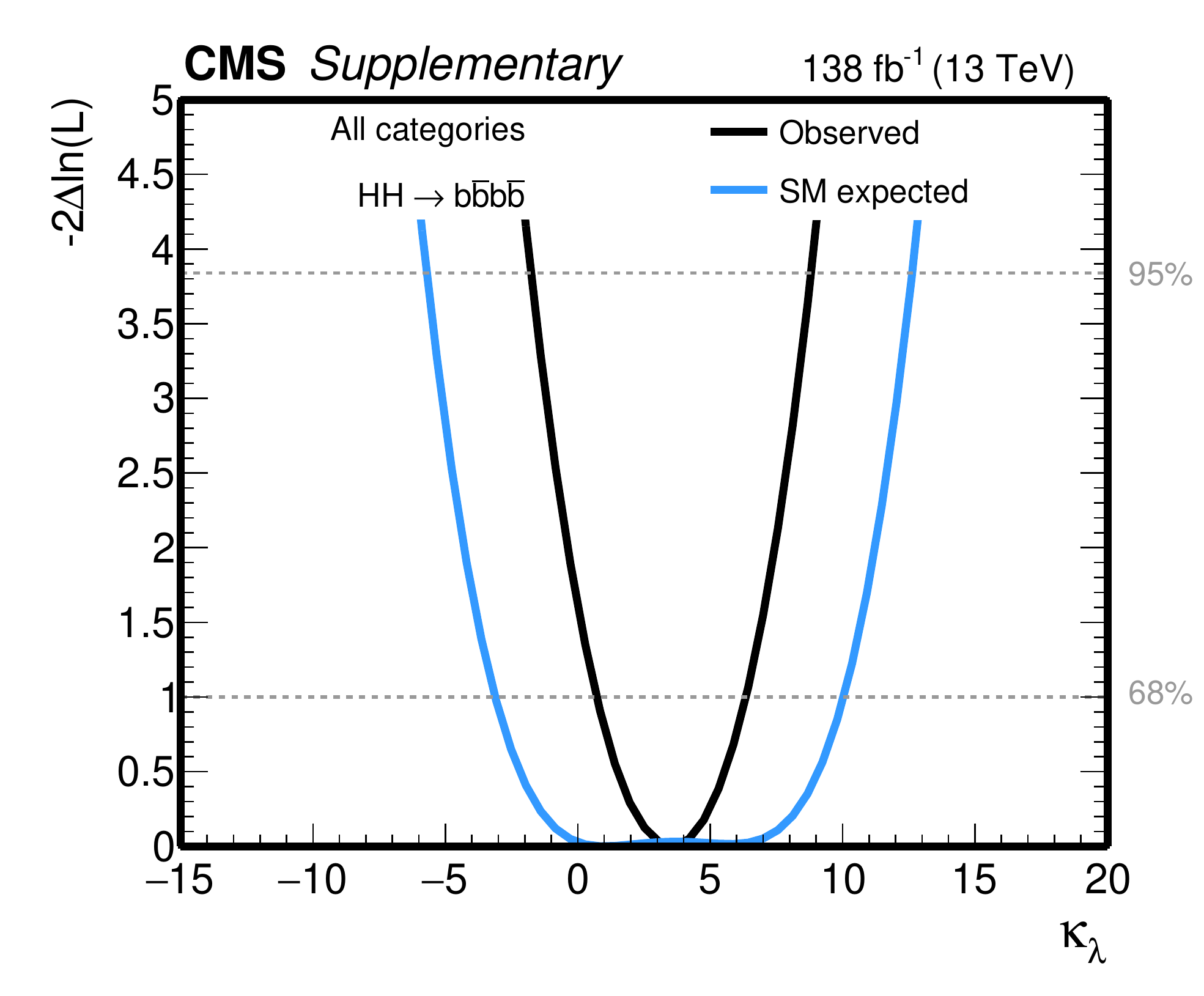}}\hspace{\fill}
{\includegraphics[width=\cmsTwoWide] {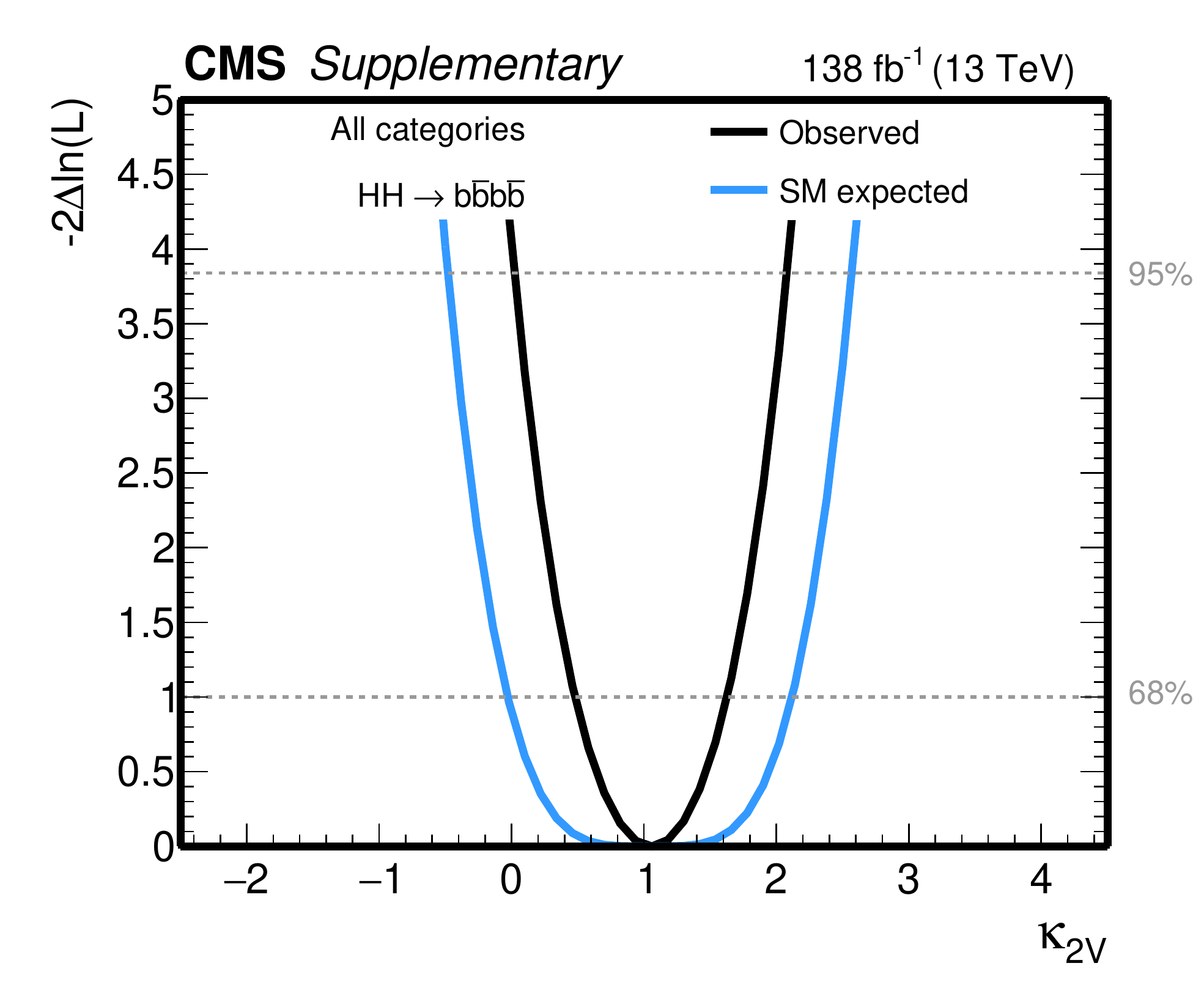}}
\caption{\label{fig:likelihood_kl_c2V_1D}
Expected and observed likelihood scan as a function of the \klambda (left) and \CVV (right) couplings.
All the couplings that are not displayed in the figure are fixed to their SM prediction.
}
\end{figure*}

\begin{figure*}[!htb]
\centering
{\includegraphics[width=\cmsTwoWide] {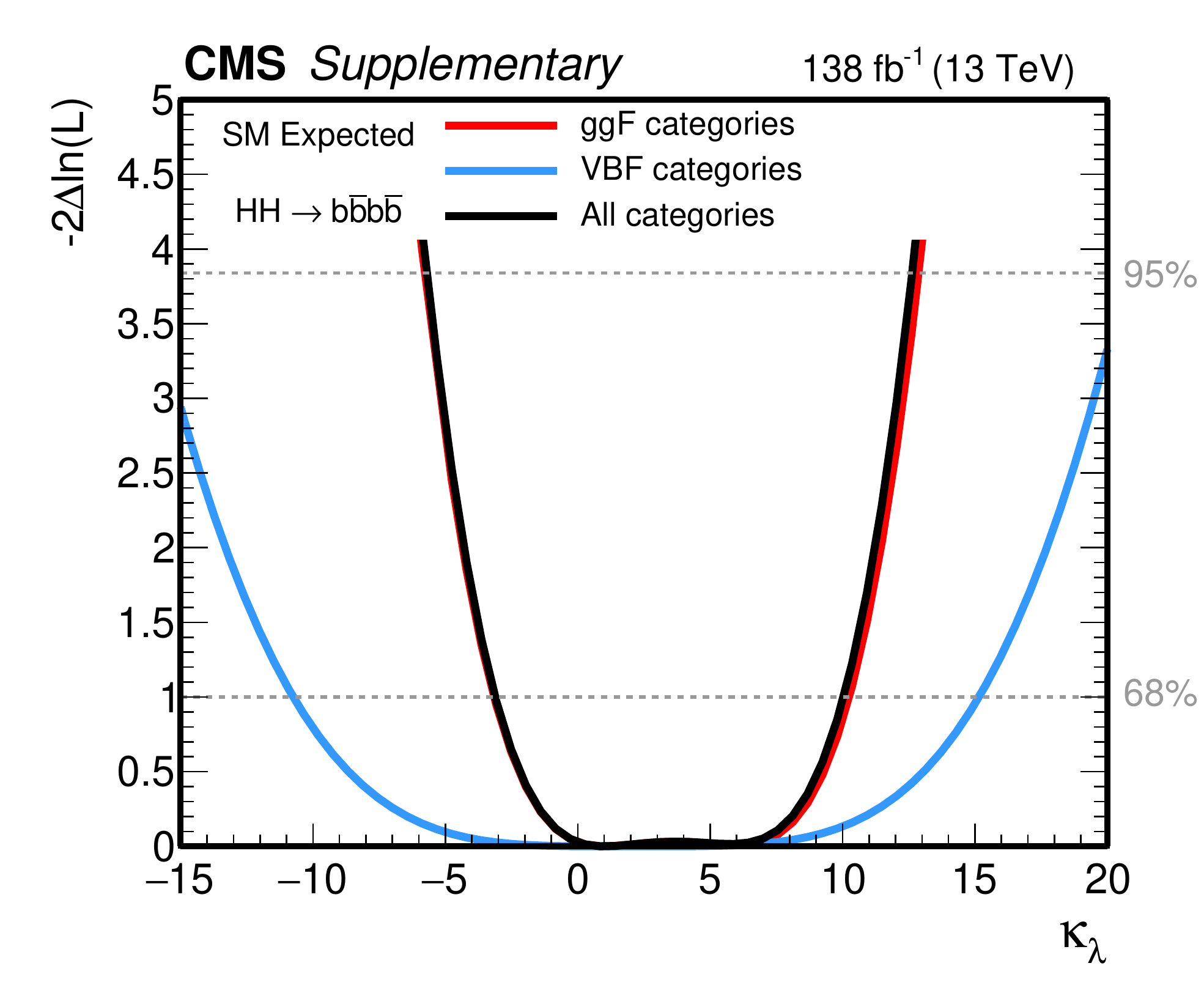}}\hspace{\fill}
{\includegraphics[width=\cmsTwoWide] {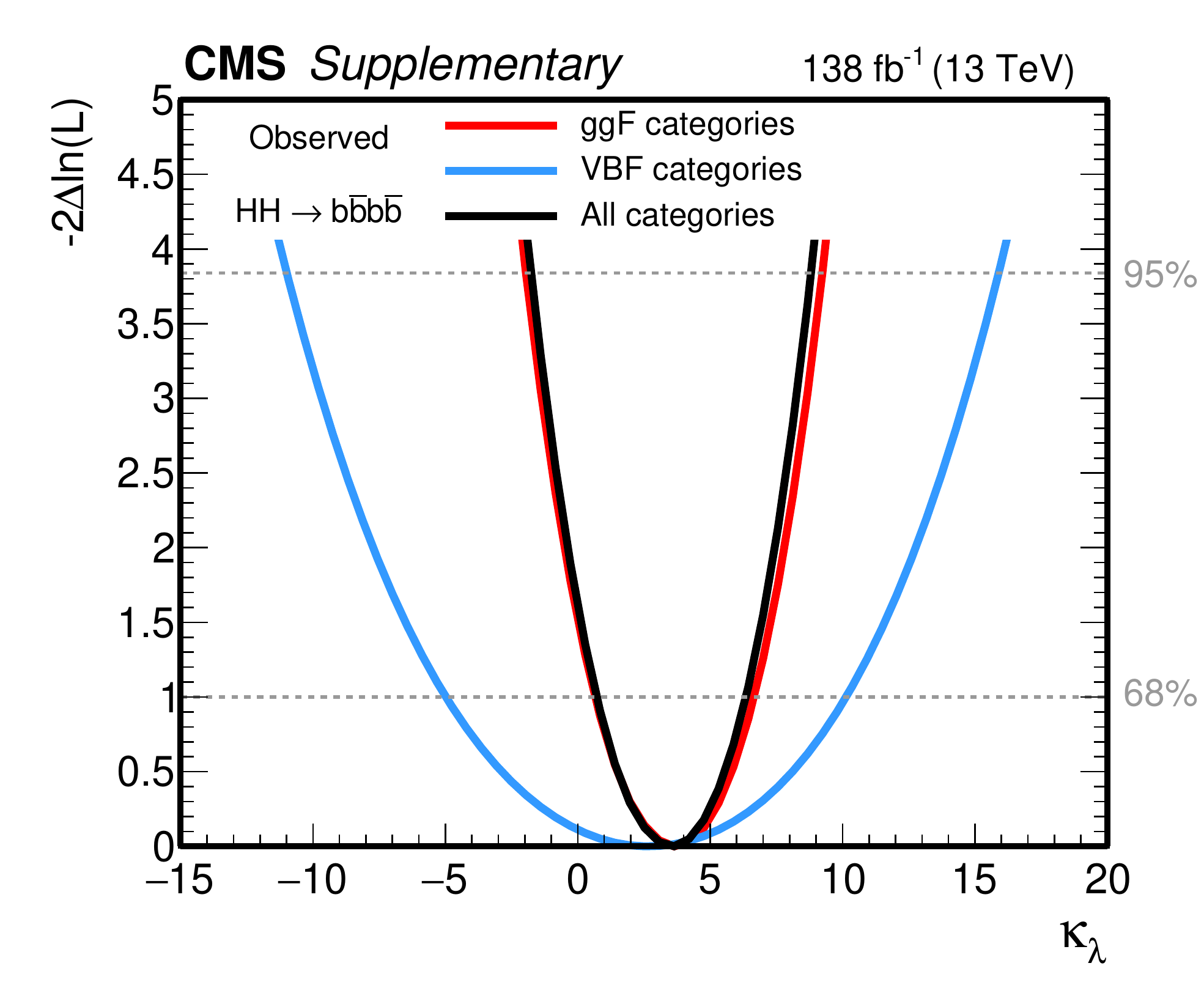}}\\
{\includegraphics[width=\cmsTwoWide] {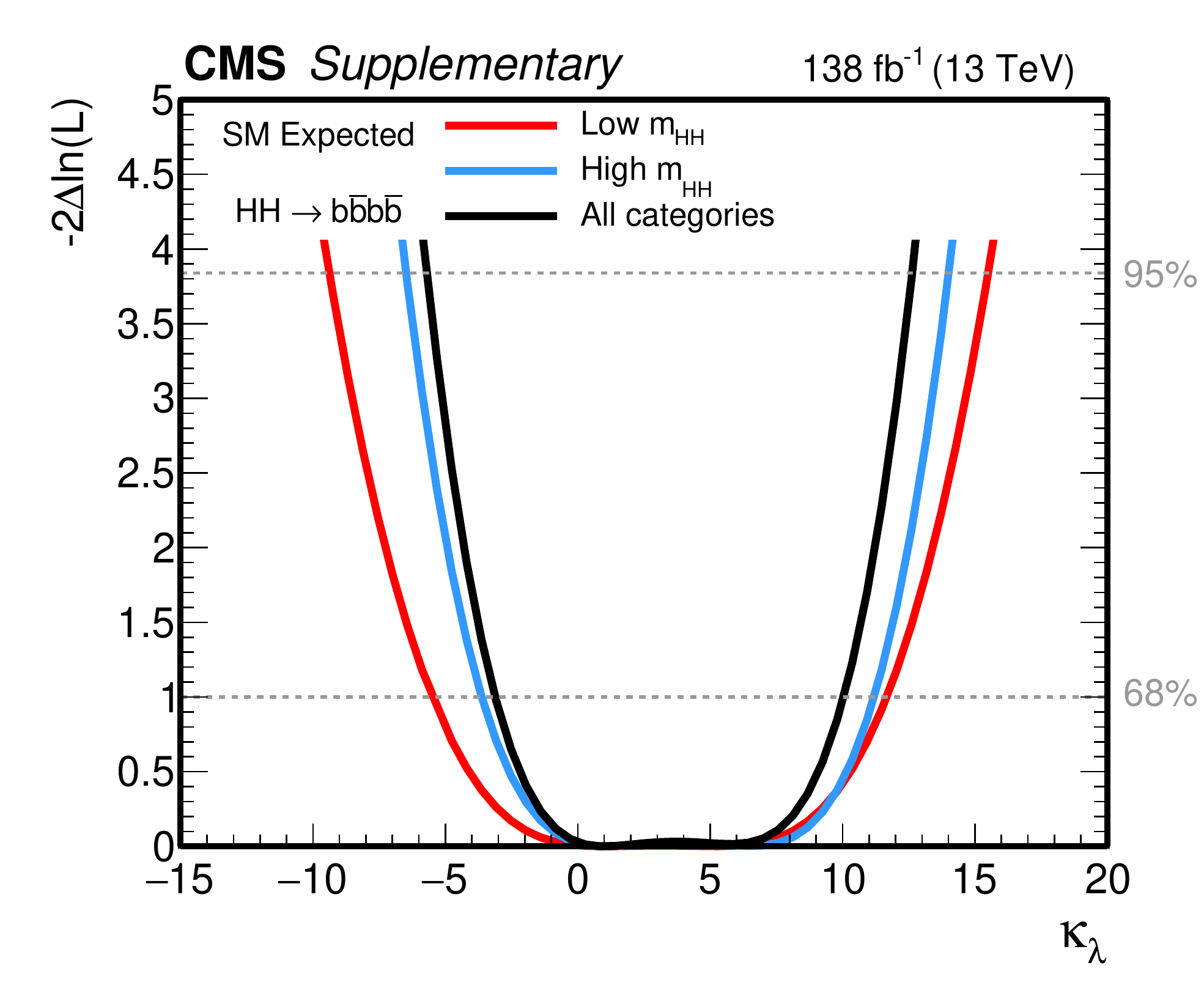}}\hspace{\fill}
{\includegraphics[width=\cmsTwoWide] {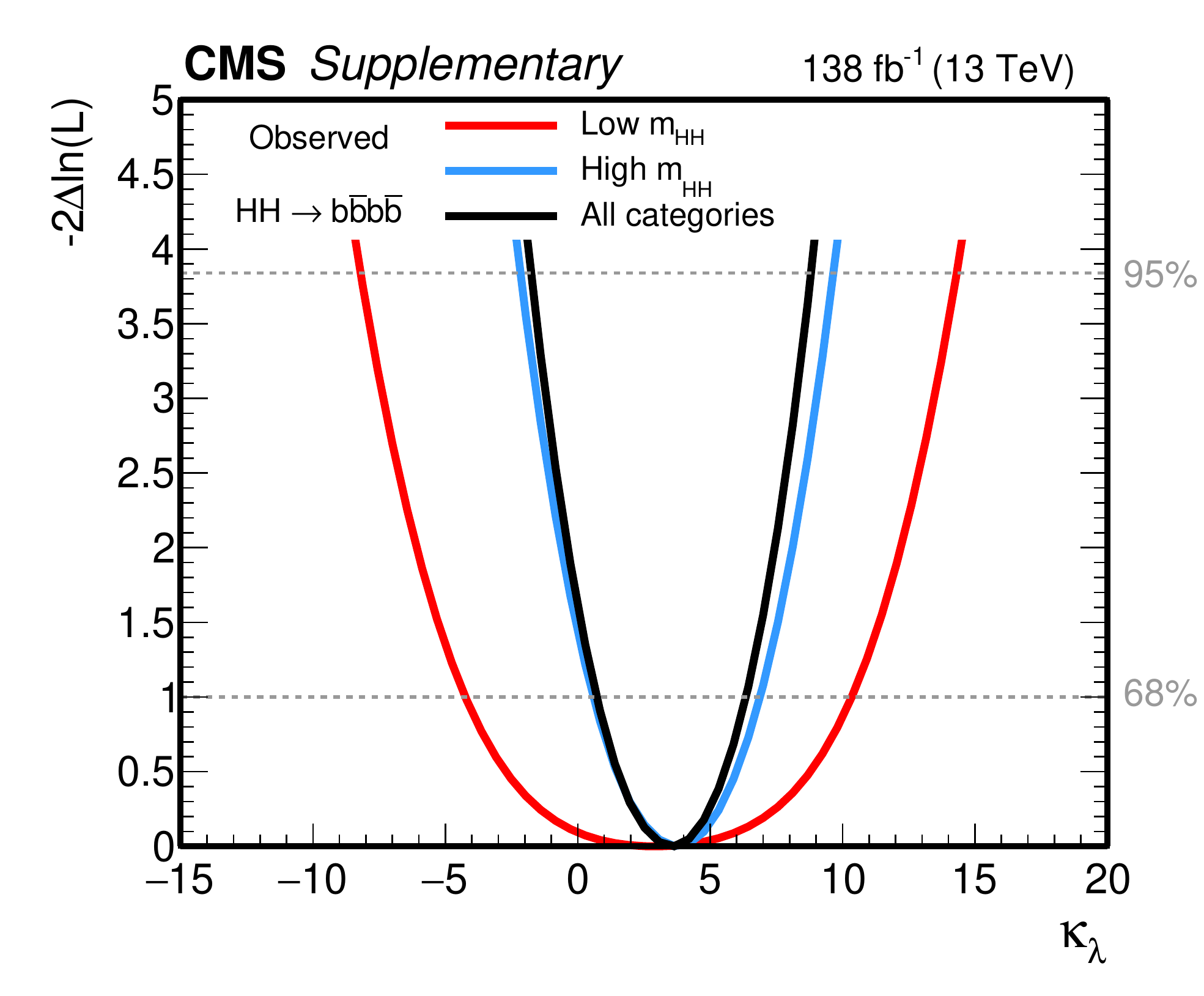}}
\caption{\label{fig:likelihood_kl_c2V_1D_categs}
Expected (left column) and observed (right column) likelihood scans as a function of the \klambda couplings, assuming all the other couplings to be fixed to their SM prediction.
The top row shows the contribution of the \GGF (red) and \VBF (blue) categories and their combination (black).
The bottom row shows the contribution of the low- (red) and high-mass (blue) \GGF categories, and the overall combination of all the \GGF and \VBF categories (black).
For all the curves, the 2016, 2017 and 2018 datasets are combined.
}
\end{figure*}

\begin{figure*}[!htb]
\centering
{\includegraphics[width=\cmsTwoWide] {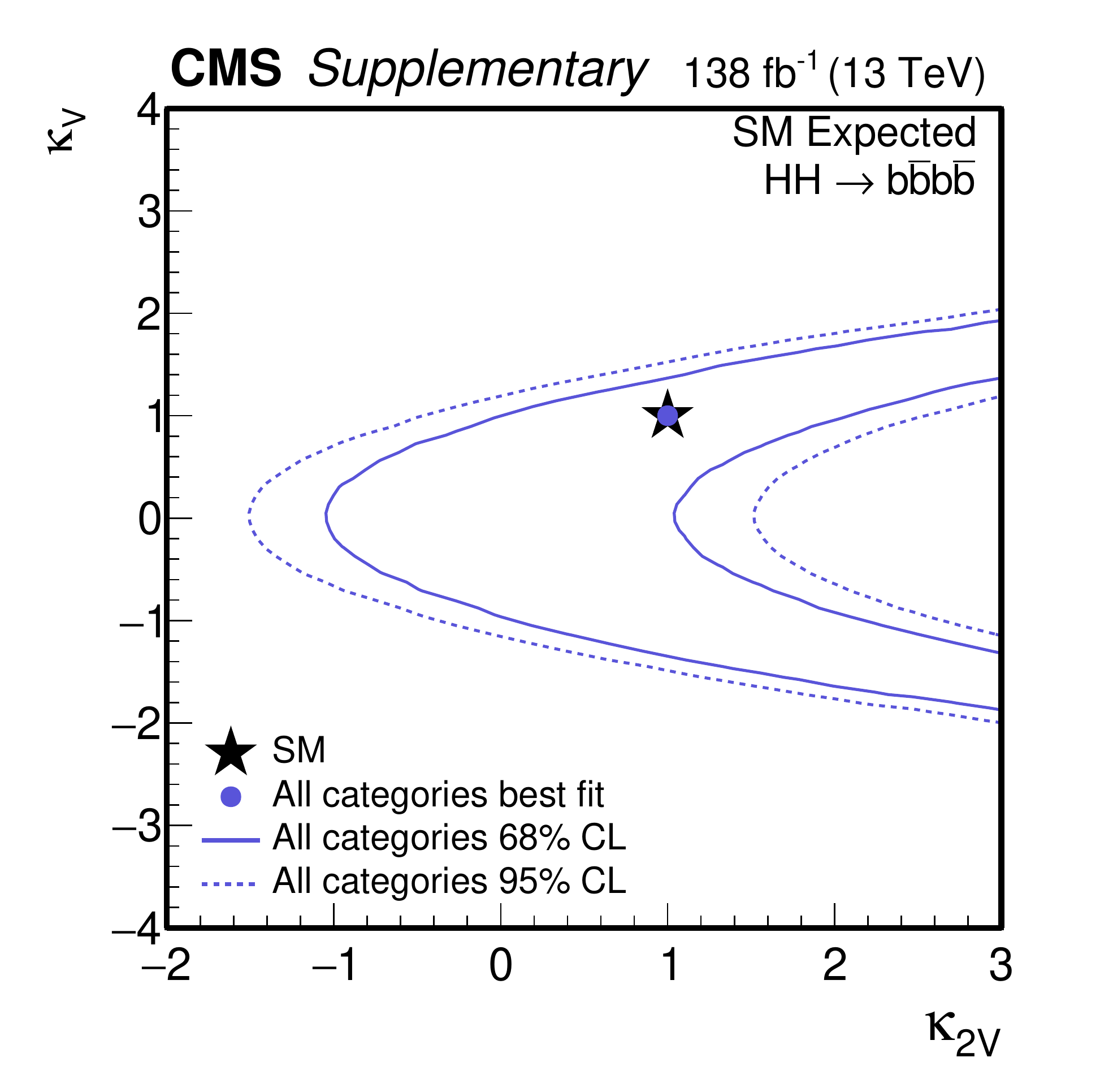}}\hspace{\fill}
{\includegraphics[width=\cmsTwoWide] {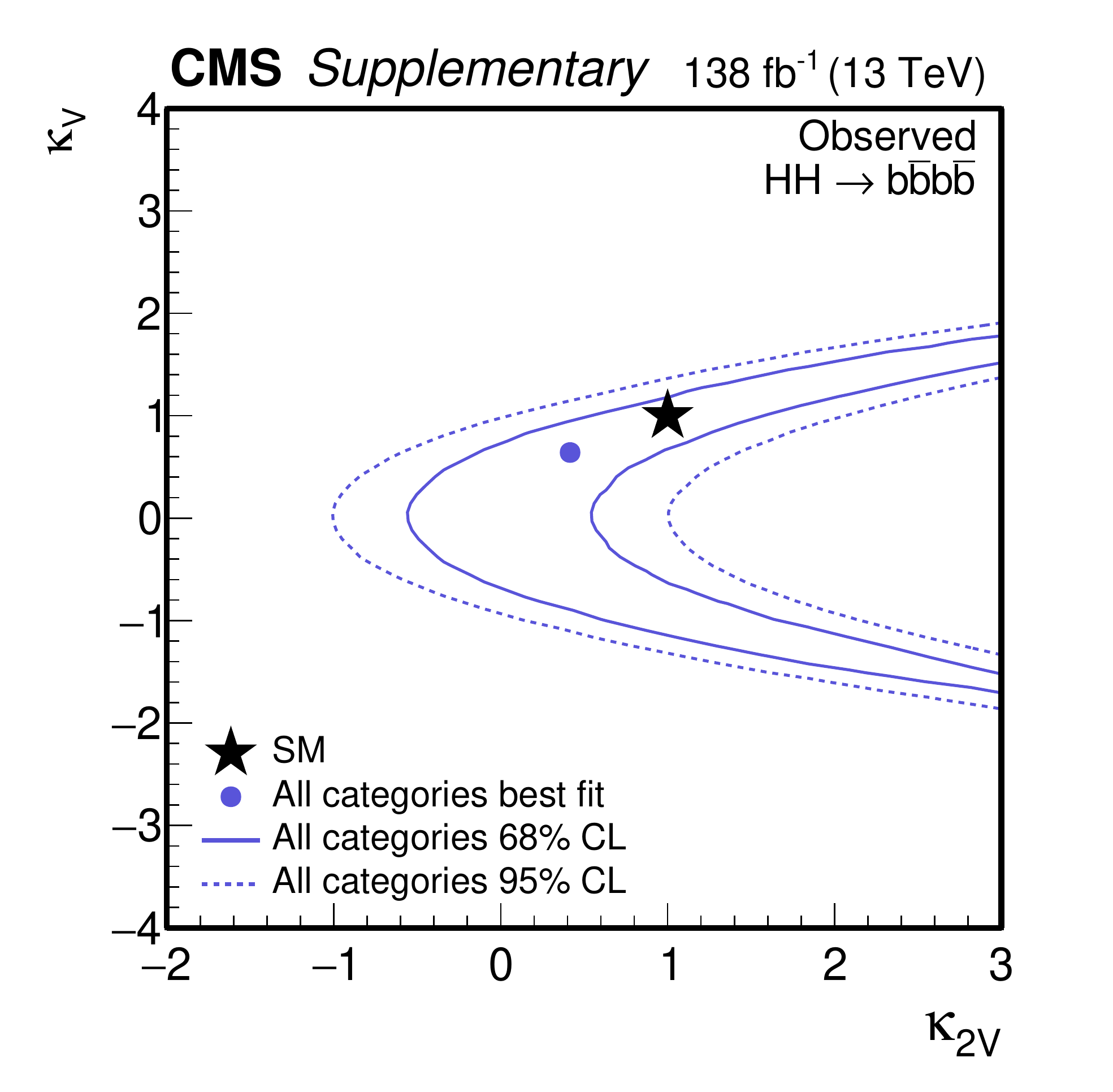}}
\caption{\label{fig:likelihood_c2V_cV_2D}
Expected (left) and observed (right) regions allowed at the 68 and 95\% CL as a function of the \CVV and \CV couplings, assuming the \ktop and \klambda couplings to be fixed to 1.
The dot denotes the best-fit hypothesis while the star marker represents the SM prediction.
}
\end{figure*}

\begin{figure*}[!htb]
\centering
{\includegraphics[width=\cmsTwoWide] {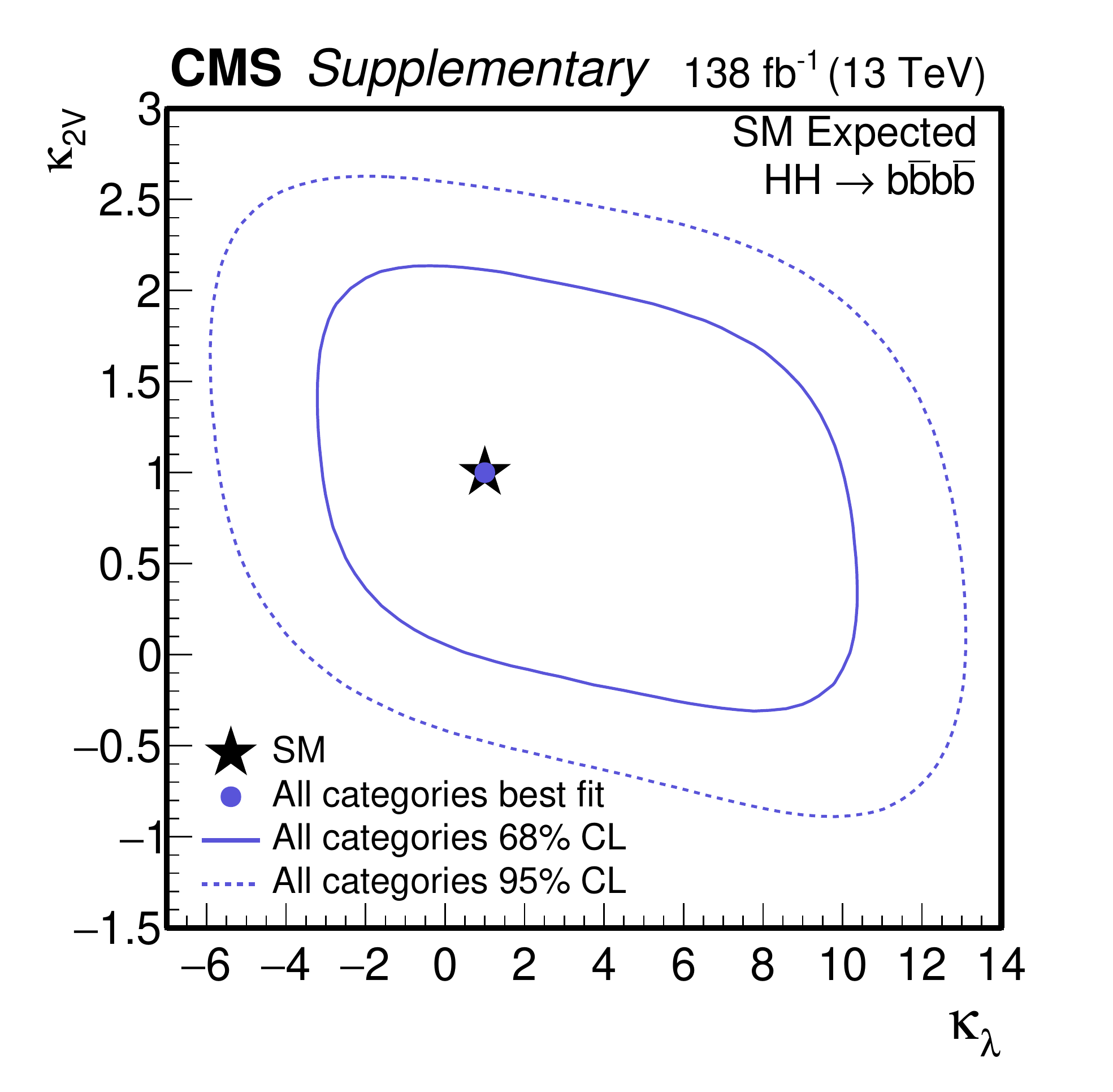}}\hspace{\fill}
{\includegraphics[width=\cmsTwoWide] {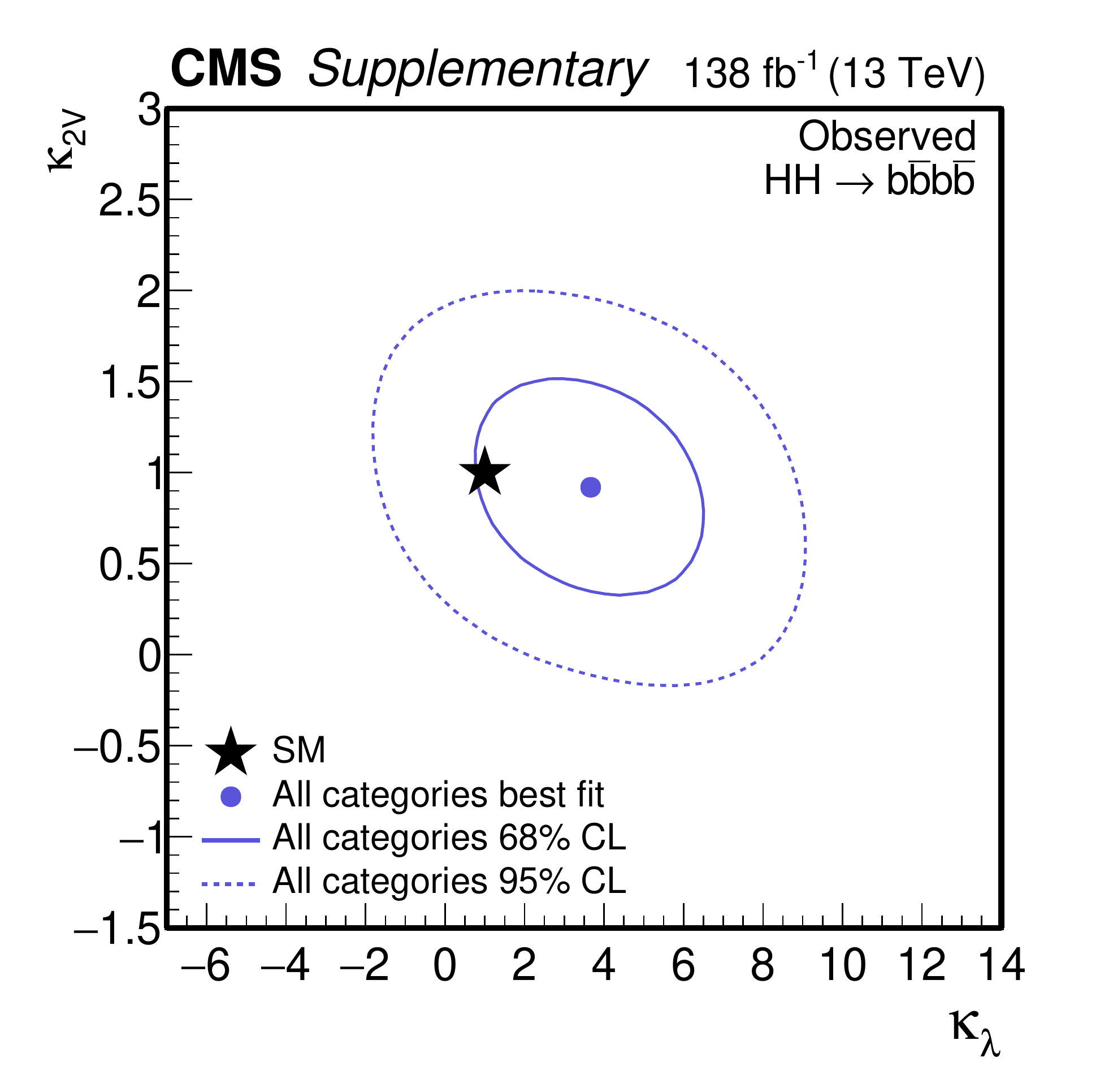}}
\caption{\label{fig:likelihood_c2V_kl_2D}
Expected (left) and observed (right) regions allowed at the 68 and 95\% CL as a function of the \CVV and \klambda couplings, assuming the \ktop and \CV couplings to be fixed to 1.
The dot denotes the best-fit hypothesis while the star marker represents the SM prediction.
}
\end{figure*}

\begin{figure*}[!htb]
\centering
{\includegraphics[width=\cmsTwoWide] {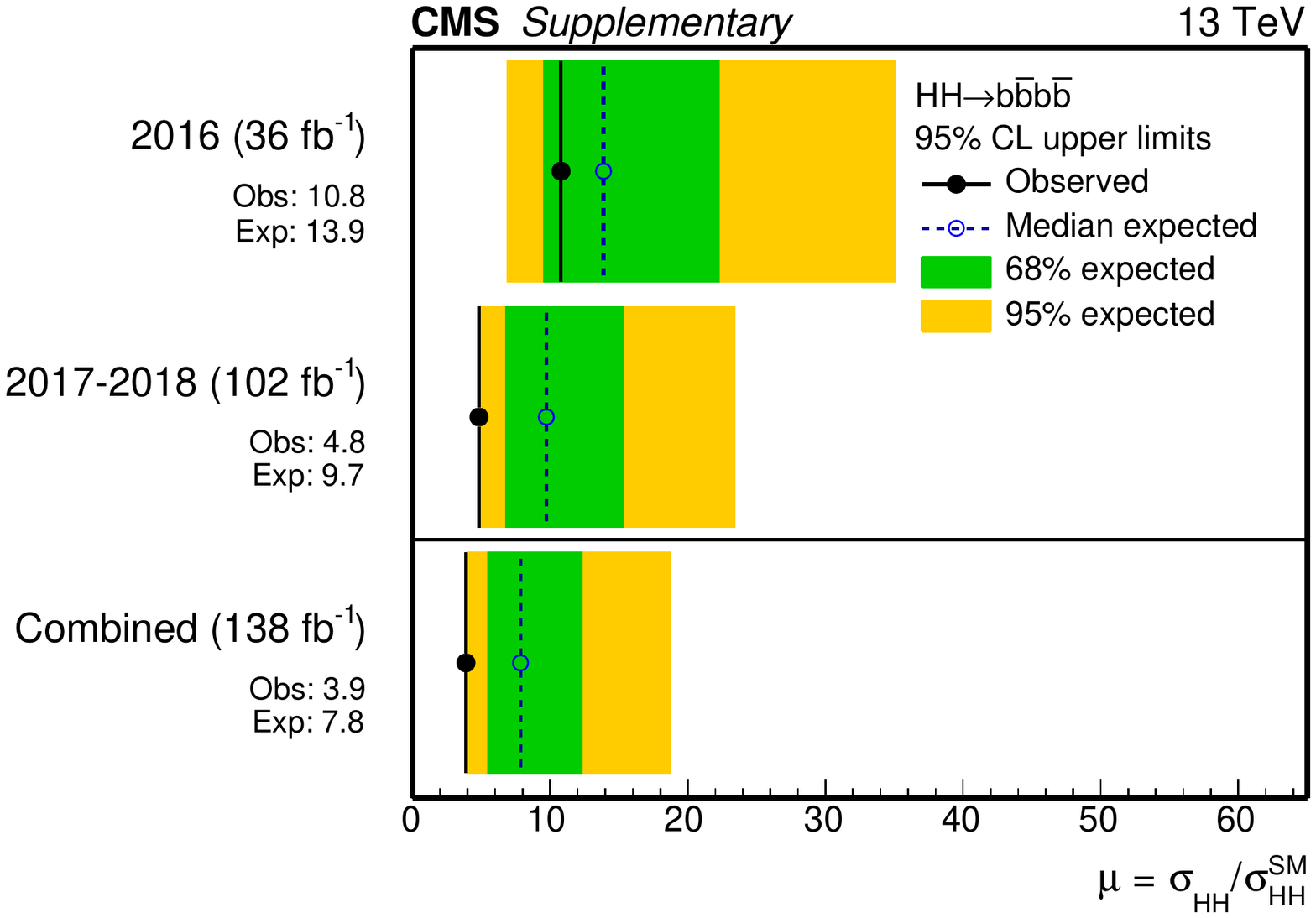}}
\caption{\label{fig:limitsbyyear}
Observed and expected 95\% CL limits on the standard model Higgs boson pair (HH) production signal strength ($\mu$) by year: 2016 (top row), 2017-2018 (center row) and Combined (bottom row).}
\end{figure*}

\begin{figure*}[!htb]
\centering
{\includegraphics[width=\cmsTwoWide] {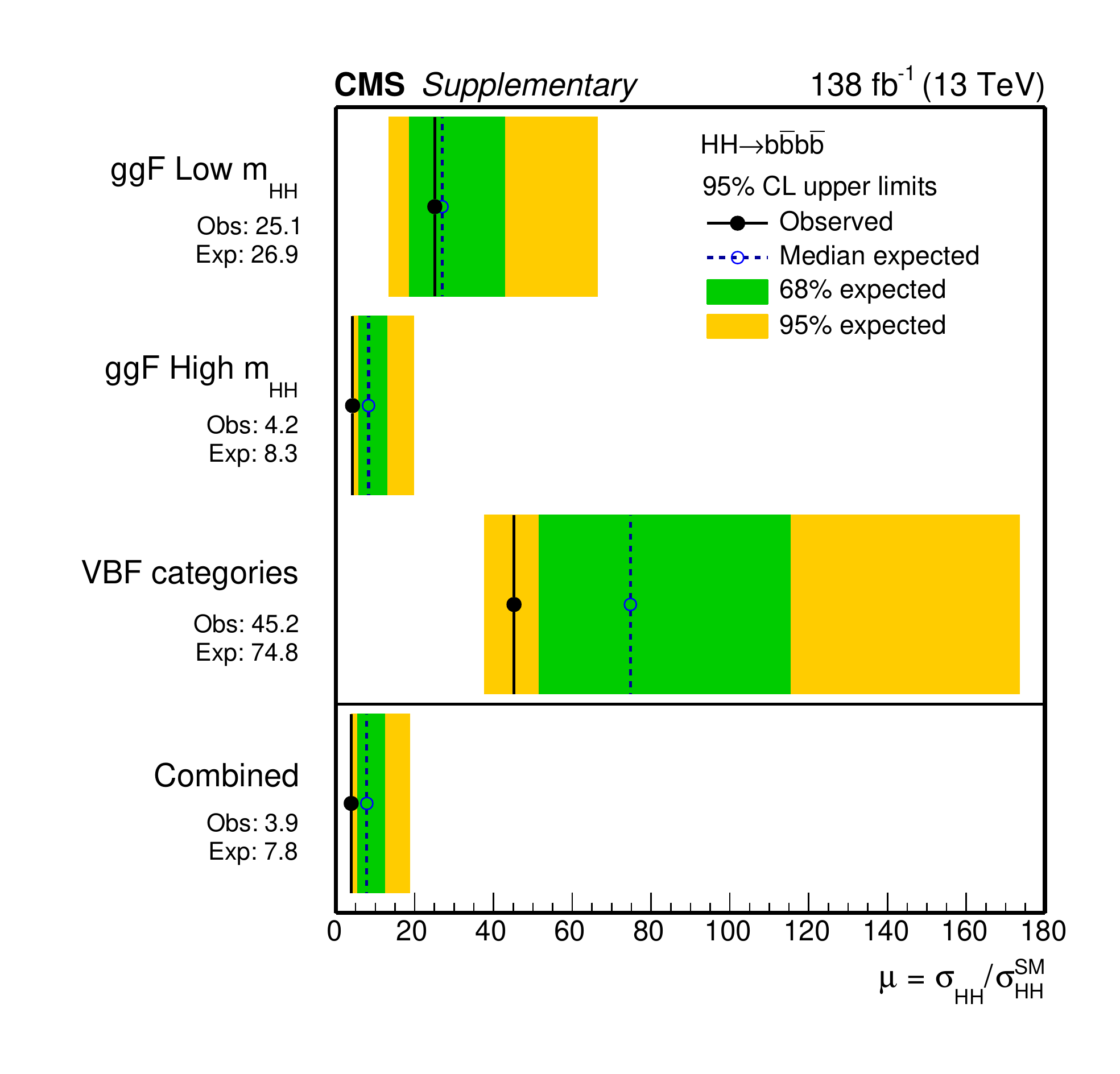}}
\caption{\label{fig:limitsbycateg}
Observed and expected 95\% CL limits on the standard model Higgs boson pair (HH) production signal strength ($\mu$) by category: ggF Low $m_{\HH}$ (first row), ggF High $m_{\HH}$ (second row), VBF categories (third row), and Combined (fourth row).}
\end{figure*}

\begin{figure*}[!htb]
\centering
{\includegraphics[width=\cmsTwoWide] {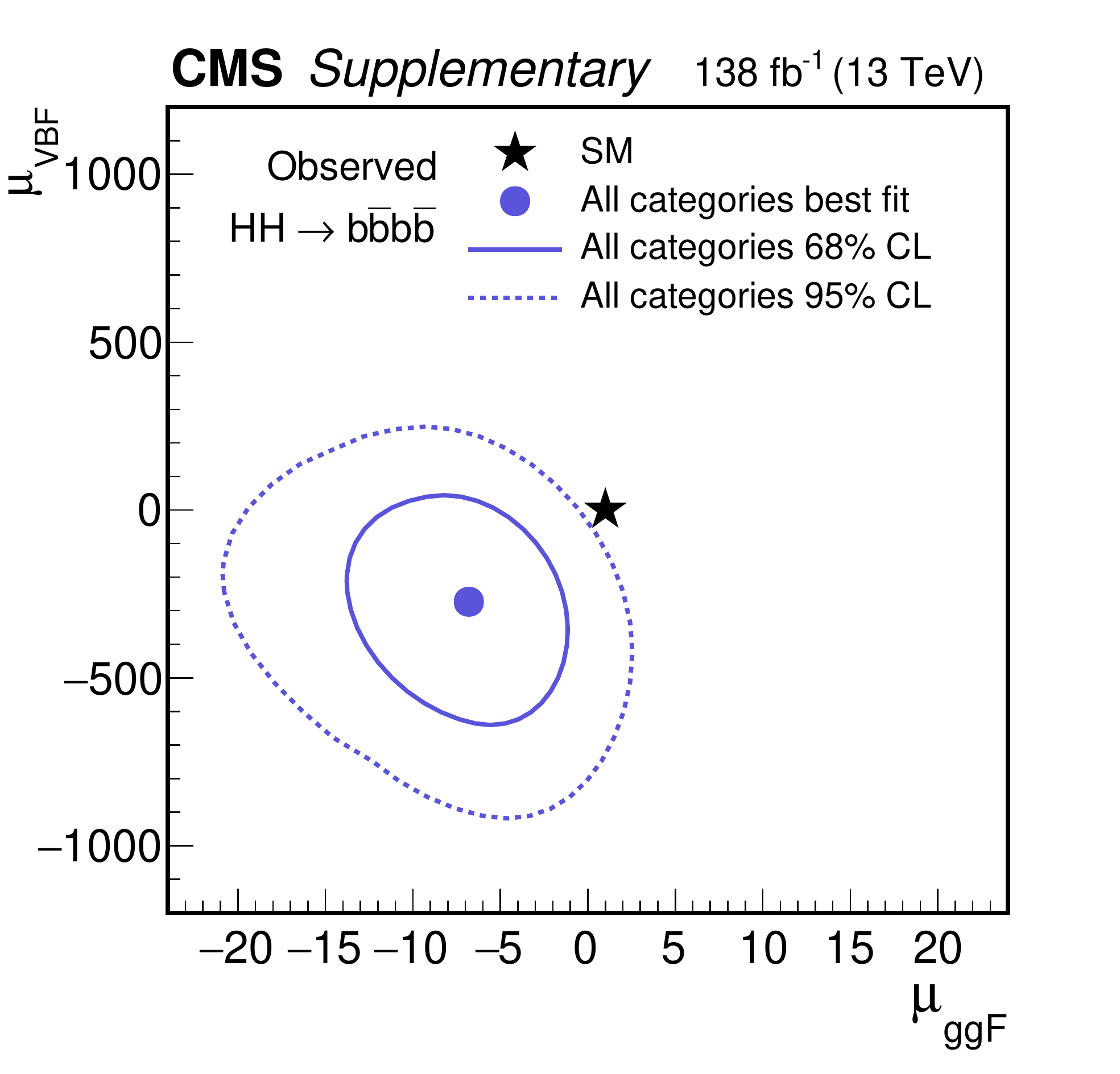}}
\caption{\label{fig:plot_2d_mu_obs}
Signal strength of the SM ggF and VBF production modes. The solid (dashed) line denotes the contour corresponding to the 68\% (95\%) confidence levels. The circle represents the best-fit point while the star marker represents the SM expectation. The signal strength is treated as a free parameter that is not bound to be positive, and the best-fit point takes a negative value as a consequence of the deficit in the observed number of the events discussed in the Letter.}
\end{figure*}

\begin{figure*}[!htb]
\centering
{\includegraphics[width=\cmsTwoWide] {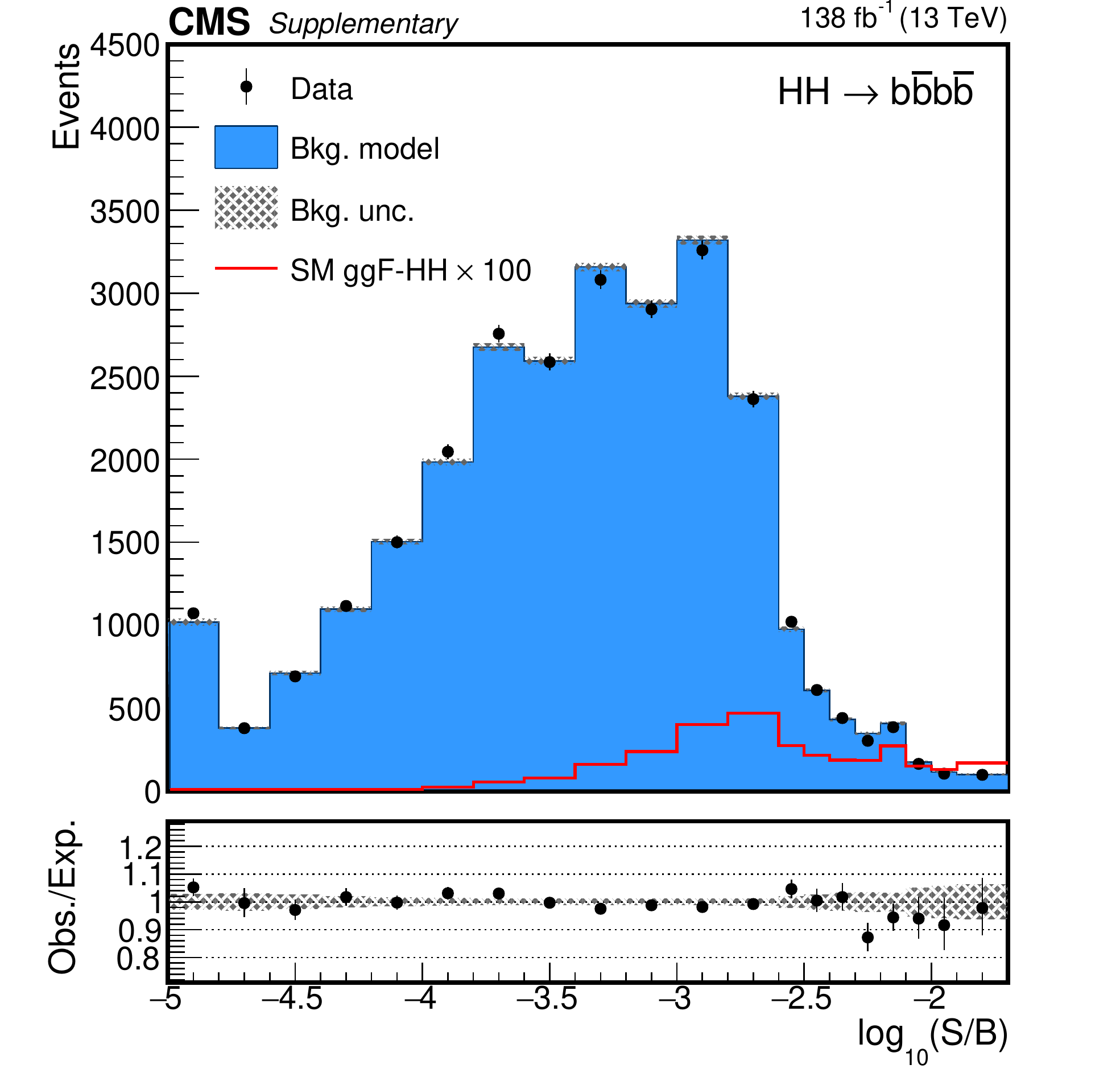}}
\caption{\label{fig:log10sb}
Distribution of the ratio of the expected signal to background ratio for events selected in the categories of the analysis. The leftmost bin also includes events that are below the represented x-axis range.}
\end{figure*}

Signal processes from \GGF \HH production are simulated at NLO accuracy in QCD with \POWHEG~2.0 for \klambda values of 1, 2.45, and 5.
Samples for \VBF \HH production are generated at LO accuracy in QCD using \MGvATNLO~2.6.5 for six different sets of $(\CV,\CVV,\klambda)$ couplings
corresponding to $(1,1,\{0,1,2\})$, $(1,\{0,2\},1)$, and $(1.5,1,1)$, where each number inside brackets corresponds to one alternative choice that defines one set.

QCD multijet processes are generated with \MGvATNLO at LO with MLM merging~\cite{Alwall:2007fs}. The production of pairs of top quarks is simulated at NLO with \POWHEG, and normalized to the theoretical cross section at NNLO precision in QCD, including resummation of NNLL soft gluon terms~\cite{Czakon:2011xx}. The production of pairs of \PZ bosons decaying to hadrons with up to one jet emitted at the matrix element level is simulated with \MGvATNLO at NLO with FxFx merging~\cite{Frederix:2012ps} and normalized to the cross section at the same precision~\cite{Campbell:2011bn}.

For all simulations, the generators are interfaced with \PYTHIA~8.226 (2016) and 8.230 (2017--2018)~\cite{Sjostrand:2014zea} for parton showering, fragmentation, and modeling of the underlying event using the CUETP8M1 tune for 2016 and the CP5 tune for 2017--2018~\cite{CMS:2015wcf,CMS:2019csb}.
The NNPDF3.0 NLO~\cite{NNPDF:2014otw} and NNPDF3.1 NNLO~\cite{NNPDF:2017mvq} parton distribution function (PDF) sets are used for the 2016 and 2017--2018 samples, respectively.
The CMS detector response is modeled with \GEANTfour~\cite{Agostinelli:2002hh}.

The total 2016--2018 integrated luminosity has an uncertainty of 1.6\%, partially correlated for the three years~\cite{CMS-LUM-17-003,CMS-PAS-LUM-17-004,CMS-PAS-LUM-18-002}. 
The uncertainty on the jet energy scale and resolution is accounted for by modifying the corresponding jet properties within the uncertainties associated to their independent measurement, and affects the shape of the signal distributions and the associated yields by up to 5\%, partially correlated for the three years.
The trigger efficiency in data is measured in an orthogonal data set enriched in multijet \ttbar events and compared to the same efficiency obtained from the simulation to derive a correction for the simulated signal events that depends on \HT and on the \pt and \textsc{DeepJet} discriminant of the selected jets. The uncertainties of these measurements are considered as shape and normalization systematic uncertainties, uncorrelated for the three years, and range between 5 and 40\% depending on the coupling and the category, with the largest values for the low-mass \GGF category in regions of the phase space with little separation from the background. The impact of this uncertainty on the sensitivity is of about 1\%.
The \PQb-tagging efficiency and the distribution of the \textsc{DeepJet} discriminant are measured in \ttbar and QCD multijet events, and independent uncertainties in this measurement for each year are propagated to the simulated events and range between 5 and 12\%.
For data recorded in 2016 and 2017, a gradual shift in the timing of the inputs of the ECAL Level-1 trigger in the forward endcap region ($\abs{\eta} > 2.4$) led to an inefficiency whose impact is evaluated using an unbiased data sample, and an uncertainty of 1--2\% on the corresponding correction is applied.
The uncertainty associated with the correction of the pileup distribution is considered in the analysis by varying the assumed interaction cross section by 4.6\%.
Uncertainties arising from the theoretical modeling of the signal and related to the factorization and renormalization scales (2--8\%), PDF (1--12\%), and parton shower parameters (6--13\%) used for the sample generation are considered.
For the \VBF production mode, two alternative models of signal generation that enable and disable a dipole-ordered parton shower in \PYTHIA~\cite{Jager:2020hkz} are compared and their difference in the predicted number of signal events ranges from 2 to 13\% and is considered to be a systematic uncertainty.
Uncertainties in the $\HH\to\bbbb$ branching fraction and, when quoting a limit relative to the standard model prediction, theoretical uncertainties on the total \HH cross section prediction are also considered.

}\cleardoublepage \section{The CMS Collaboration \label{app:collab}}\begin{sloppypar}\hyphenpenalty=5000\widowpenalty=500\clubpenalty=5000\cmsinstitute{Yerevan~Physics~Institute, Yerevan, Armenia}
A.~Tumasyan
\cmsinstitute{Institut~f\"{u}r~Hochenergiephysik, Vienna, Austria}
W.~Adam\cmsorcid{0000-0001-9099-4341}, J.W.~Andrejkovic, T.~Bergauer\cmsorcid{0000-0002-5786-0293}, S.~Chatterjee\cmsorcid{0000-0003-2660-0349}, K.~Damanakis, M.~Dragicevic\cmsorcid{0000-0003-1967-6783}, A.~Escalante~Del~Valle\cmsorcid{0000-0002-9702-6359}, R.~Fr\"{u}hwirth\cmsAuthorMark{1}, M.~Jeitler\cmsAuthorMark{1}\cmsorcid{0000-0002-5141-9560}, N.~Krammer, L.~Lechner\cmsorcid{0000-0002-3065-1141}, D.~Liko, I.~Mikulec, P.~Paulitsch, F.M.~Pitters, J.~Schieck\cmsAuthorMark{1}\cmsorcid{0000-0002-1058-8093}, R.~Sch\"{o}fbeck\cmsorcid{0000-0002-2332-8784}, D.~Schwarz, S.~Templ\cmsorcid{0000-0003-3137-5692}, W.~Waltenberger\cmsorcid{0000-0002-6215-7228}, C.-E.~Wulz\cmsAuthorMark{1}\cmsorcid{0000-0001-9226-5812}
% ...
\vskip\cmsinstskip
% ...
1:~Also at TU Wien, Wien, Austria\\
% ...
\end{sloppypar}
\end{document}